\documentclass[article]{svmono}

\usepackage[greek,english.german]{babel}
\usepackage{marvosym}
\usepackage{graphicx,caption}        

\usepackage{booktabs}
\usepackage{threeparttable}

\def\begmarg{\par \begingroup  \leftskip2.5em \rightskip2em \large}
\def\endmarg{\par \endgroup }

\def\3H{{3\over2}}
\def\simlt{\lower.5ex\hbox{$\; \buildrel < \over \sim \;$}}
\def\simgt{\lower.5ex\hbox{$\; \buildrel > \over \sim \;$}}

\def\degs{$^{\circ}$}
\def\mins{$^{\prime}$}

\def\secspt{$\buildrel{\prime\prime}\over .$}

\def\degspt{$\buildrel{\circ}\over .$}
\def\hs{$^{\rm h}$}
\def\ms{$^{\rm m}$}
\def\mspt{$\buildrel{\rm m}\over .$}

\def\simlt{\lower.5ex\hbox{$\; \buildrel < \over \sim \;$}}
\def\simgt{\lower.5ex\hbox{$\; \buildrel > \over \sim \;$}}

\def\bcol{\begin{column}}
\def\ecol{\end{column}}
\def\bcols{\begin{columns}[]}
\def\ecols{\end{columns}}

\usepackage[german,short]{isodate}
\usepackage{everypage}

\begin{document}

\pagestyle{plain}
\ \ 
\vspace{1cm}

\begin{center}
{\huge {\bf }}
\bigskip
  
{\LARGE {\bf PANNEKOEK'S GALAXY}}
\vspace{1cm}

\noindent
{\Large Pieter C. van der Kruit},\\
{\large Kapteyn Astronomical Institute, University of Groningen,}\\
{\large P.O. Box 800, 9700AV Groningen, the Netherlands}.\\
{\large vdkruit@astro.Rug.nl; www.astro.rug.nl/$\sim$vdkruit}
\vspace{2cm}

\end{center}

{\large

\noindent
\vspace{2cm}

\noindent
This paper has been accepted for publication by
the {\it Journal of Astronomical History and
Heritage}. This preprint version has been produced in \LaTeX.

\newpage

\noindent
{\Large {\bf Abstract}}\\
\bigskip

Antonie (`Anton’) Pannekoek (1873–1960), the founder of the Astronomical Institute of the University of Amsterdam, is remembered as one of the initiators of the field of stellar atmospheres. A second, maybe equally significant part of his research and legacy did not concern stellar, but Galactic astronomy. He has been awarded an honorary doctorate by Harvard University and the Gold Medal of the Royal Astronomical Society (UK) for his total research effort. From his longterm interest in viewing and mapping the Milky Way he became convinced that the sidereal system was built up of clouds of stars, condensations of various sizes in a smooth but low-density  stratum of stars. In addition there were dark clouds together with streaks with little or no extinction in between. So, he took the opposite view of Jacobus C. Kapteyn, concerning the stellar distribution, which the latter regarded  in first approximation to be smooth and uniform in longitude so that star counts as a function of apparent magnitude depended primarily on latitude.

Pannekoek’s research into the structure of the stellar system consisted of various parts. He first looked at bright clouds of the Milky Way, such as in Cygnus, and assuming these were isolated structures he estimated their distance from their contribution to star counts by stars far out in the bright tail of the luminosity function. He found values of tens of kpc, which would mean their distribution was similar in extent to that of  Shapley’s globular cluster system. Later he had to reduce his distance by a factor over two, and later still to admit he had to retract the method altogether. He developed a rigorous method of estimating distances of dark clouds from modeling star counts off and on the cloud, preceding Wolf’s quick and dirty method that became very widely applied. He should have received more credit for this. 

During the 1920s it became clear that extinction was a major effect and that the Galaxy rotated around a distant center. Pannekoek unsuccessfully looked for the large central mass, that Oort found had to be in the center of rotation in the Galaxy, by estimating masses of some bright star clouds in the Milky Way. He also pioneered the mapping of the local structure in the Galaxy from spectroscopic parallaxes, using the fact —following Kapteyn’s study of `helium stars’ — that for B- and A-stars and K-giants the intrinsic dispersion in absolute magnitude is small. The distributions of different spectral types are substantially different. He tried without success to explain Kapteyn’s Star Streams from the gravitational pull of the irregular distribution of matter in the solar neighborhood.

As a teenager he was fascinated by the appearance of the Milky Way and he started producing sketches of the Milky Way by drawing curves of equal (surface) brightness or in modern terms isophotes. In the course of his career this resulted in isophotal maps of both the northern and the southern Milky Way, first from visual observations, later from photographic surface photometry using out-of-focus exposures. The care and patience with which he visually produced his maps is admirable. For the photographic work he enlisted help from Wolf in Heidelberg and Vo\^ute at Lembang, who both provided extensive plate material. Unlike the care that has been employed  for photographic surface photometry in more recent times, a wide range of photographic emulsions and developers was used and calibration of the characteristic curve was done on separate plates `of the same type, {\it if possible’}! (my italics). In this paper I compare Pannekoek’s resulting isophotal maps with detailed photographic surface photometry of the southern Milky Way in the 1980s and 1990s by the group in Bochum and to the almost  all-sky mapping by the Pioneer 10 spacecraft, free of zodiacal light, from  beyond the asteroid belt. This shows Pannekoek’s visual and photographic maps to be surprisingly accurate, in spite of all the non-uniformity in observers for the first category and  in  exposure, emulsion, etc.,  with magnitude scale and zero point errors of a few tenths of a magnitude, except in the southern Coalsack where he quoted a surface brightness that is much too faint.

The legacy of Pannekoek in the area of Galactic research consists of his mapping of the structure of the nearby part of the Galaxy, the distances of dark clouds, and isophotal maps of the Milky Way. His other contributions turned out inconclusive or wrong as a result of his conviction, resulting from his many years of observing and mapping the Milky Way, that the nearby distribution is characterized primarily by more or less isolated clouds of stars and by dust restricted to isolated dark clouds and streaks.

In one appendix I show that the wide spread notion that the center of the Galaxy is located in the constellation Sagittarius is not so compelling as usually thought and could with equal justification by replaced by Scorpius. A second appendix provides short curriculae vitae of a few Dutch PhDs that have been referred to in this paper.
\bigskip


\newpage

\section{Introduction}\label{sect:Introduction}

In 2016 the Anton Pannekoek Astronomical Institute of the University of Amsterdam organized a symposium on the famous founder of the institute and gave this the title {\it ‘Anton Pannekoek (1873–1960): Ways of viewing science and society’}. Pannekoek (see Fig.~\ref{fig:Pannekoek1}) was a prominent astronomer: in 1951 he was awarded the probably most prestigious astronomical honor, the Gold Medal of the UK Royal Astronomical Society (in 1936 he had already been awarded an honorary doctorate by Harvard University). I knew rather little about him, although I had read some years earlier his autobiographic notes {\it Herinneringen} [Recollections] (Pannekoek, 1982), `written by candlelight’ in October 1944, when he was almost 72 years old (‘for my son and grandson’), which has been published with some further essays in 1982. I must admit I only had read his `Astronomical  Reminiscences’, which are completely separate from those as a prominent Marxist and communist. Unfortunately, I was notified of this meeting in Amsterdam only a few weeks before it took place and had other commitments by then, which prevented me to my regret from attending. However, in 2019 the proceedings appeared, available as a very commendable open access electronic publication (Tai, van der Steen \&\ van Dongen, 2019). This publication gives a fascinating account of Pannekoek’s remarkable personality, which made him not only a prominent astronomer, but also a dedicated and influential communist. I read it superficially at the time and resolved to have a closer look at it when other matters and obligations were out of the way.

When more recently studying the book more carefully I was impressed by various papers,  particularly the studies of his astronomical research, that emphasized the innovative accomplishments of Pannekoek and stressed his obvious importance for astronomy in general and for Dutch and Amsterdam astronomy in particular. One was the remarkable role he played in founding the field of the study of the structure of stellar atmospheres and the understanding of the physics behind stellar spectra. With my background in structure of galaxies I was particularly interested in his work on our Galaxy, which I knew in general terms without appreciating the details of it. First there was his fascination of the structure and appearance of the Milky Way, which led to extensive efforts to capture this in  detailed maps. The contributions in the proceedings discussed in general terms how he arrived at these, but I wondered -- in fact had wondered for some time -- how he had calibrated the luminosity scale and the zero-point, how his surface brightnesses would compare with modern studies, and what new insight into Galactic structure he had deduced from this. A second issue that aroused my interest was triggered by two remarks. One was made by  my Amsterdam colleague and former director of the Anton Pannekoek Astronomical Institute Edward P.J. van den Heuvel and another by authoritative historian of astronomy and expert in this area Robert W. Smith.

\begin{figure}[t]
\sidecaption[t]
\includegraphics[width=0.54\textwidth]{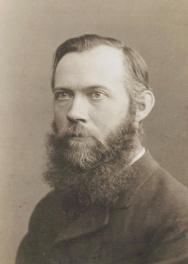}
\caption{\normalsize Pannekoek in 1908 or earlier. He had obtained a PhD in Leiden in 1902 with van de Sande Bakhuyzen on \textit{Untersuchungen \"uber den Lichtwechsel Algols}, and following his Marxist sympathies had moved to Germany in 1906 to teach socialist theories and write articles on this. He still published some astronomical work through his former supervisor. During World War I he taught physics and chemistry at various secondary schools in the Netherlands, and from 1916 as privaat-docent (unpaid appointment) history of astronomy at Leiden University. This photograph comes from the {\it Album Amicorum}, presented to H.G. van de Sande Bakhuyzen on the occasion of his retirement as Professor of Astronomy and Director of Leiden Observatory in 1908 (after Leiden Observatory, 1908). In the public domain under Creative Commons CC.}
\label{fig:Pannekoek1}
\end{figure}

Van den Heuvel (2019; p.40) wrote:
\begmarg
In the previous year, Pannekoek had published an important paper in the {\it Monthly Notices of the Royal Astronomical Society} on the Earth’s distance from the centre of the Milky Way. By studying the distribution of star clouds in the Galaxy, Pannekoek calculated a distance to the galactic centre of 60,000 lightyears in the direction of the constellation Sagittarius.
\endmarg

The ‘important paper’ referred to is Pannekoek (1919b), which has the eye-catching title {\it The distance to the Milky Way}. These remarks immediately raised some questions, since after all we now know that the Galactic center is hidden from view by interstellar extinction: What would Pannekoek in 1919 have had in mind when referring to the center of the Galaxy? How did he determine the distribution of star clouds, and how actually did he derive the distance to the  Galactic center from this? How could he possibly measure distances of tens of kpc in the first place? How did he determine that the center is in the direction of Sagittarius? If indeed an important paper, how influential has it actually been in the debate about the size of the Galaxy?
 
In his chapter in the book, Smith (2019; p.126) wrote:
\begmarg
Pannekoek indeed treated each of the  different parts of the Galaxy separately to derive the changes in star density with distance. By this route, he ended-up siding with Shapley. Some of the bright parts of the Milky Way, Pannekoek calculated, were some 40,000 to 60,000 parsecs distant, and ‘the starry masses of the Galaxy are spread over space as far as the remotest [globular] clusters, and clearly both belong to one system.’ Although Shapley’s arguments in favour of an eccentric position of the Sun in the Galactic system were, ‘contrary to the common view’, Pannekoek reckoned that ‘Shapley’s result is wholly in accordance with the aspect of the Milky Way’. 
\endmarg

Smith here also referred to the Pannekoek (1919b)  paper. Note that both van den Heuvel and Smith quote a values of 40,000 and 60,000, but the former with  unit lightyears and the latter parsecs. Checking Pannekoek’s paper confirms that Smith has got it right. Smith’s paper makes clear that Pannekoek did not directly measure the distance to the center, nor had he determined it was in the direction of Sagittarius. Pannekoek  found distances up to 60 kpc for parts of the Milky Way. He concluded therefore that it was part of the same structure as Shapley’s system of globular clusters (Shapley, 1918; Shapley \&\ Shapley, 1919), which had its center at (old) longitude 325\degs\ at a distance of 20 kpc. Harlow Shapley (1885--1972) used pulsating stars (cluster variables) in globular clusters at apparent magnitudes 13 to 14 or fainter to arrive at distances of a few tens of kpc, so what did Pannekoek use (60 kpc  corresponds to a distance modulus of 13.9 magnitudes)?  Shapley assigned the center of his globular cluster system to the constellation Sagittarius and others followed, such as Jan Hendrik Oort (1900--1992) when he found the center of rotation to be at longitude 323\degs\ (Oort, 1927). In fact, Shapley could equally well have chosen Scorpius as the constellation harboring the center of his system of globular clusters, and so could Oort for his rotation center. I address this sideline in Appendix A.

Since that 2016 meeting in Amsterdam the comprehensive PhD thesis of Chaokang Tai (2021) was defended successfully  in which the work of Pannekoek was examined in much detail. Pannekoek’s analysis  and determination of the distances of the star clouds in Cygnus and Aquila  has been described there and this treatment is much more illuminating (p.79):
\begmarg
Pannekoek constructed a model of a single star cluster placed in an otherwise uniform system. For this model, he could compute a theoretical star count, which could be fitted to observed star counts by adjusting the distance and size of the theoretical cluster. 
\endmarg
So, Pannekoek viewed the cloud as an isolated conglomeration of stars for which the brightest stars contributed to the star counts and used this to estimate the distance. Yet, it still is not described what assumptions he had made to find distances of tens of kpc while we now know we cannot look out that far, really no more than some kpc, at those positions. He clearly assumed an absence of interstellar extinction, but even then ending up with a distance of 60 kpc is surprising.

The only other relevant recent reference to Pannekoek’s 1919b paper that I could find (using the {\it SAO/NASA Astrophysics Data System} (ADS); Lattis (2014) mentions it only in passing), is a review by Virginia Trimble (1995)  on the 1920  ‘Great Debate’ between Harlow Shapley and Heber Doust Curtis (1872--1942), in which she noted (p.1139):
\begmarg
Anton Pannekoek (1919b) agreed with Shapley in placing the Sun far off-center but in a smaller galaxy ($R_\odot$=40-60,000 LY, $d$=80-120,000 LY).
\endmarg
The distance $R_{\odot}$ to the Galactic center she must have taken from Pannekoek’s statement (p.507)
\begmarg
The  results we have arrived at here are in accordance with these later views, as they place some bright parts of the Milky Way at a distance of 40-60,000 parsecs
\endmarg
and the diameter $d$ by simply doubling that. However, this means she has also confused lightyears and parsecs.

I had recently encountered Pannekoek in  a study concerning the Kapteyn Astronomical Laboratory under his successor Pieter J. van Rhijn (van der Kruit, 2022), when discussing the PhD thesis study of van Rhijn’s student Broer Hiemstra  (1938) on the distances of dark clouds. Pannekoek had done some important work in this area (Pannekoek 1921, 1942), but this is not mentioned, let alone discussed, in the Amsterdam proceedings. 
\bigskip

The first part of this article will be devoted to finding out how actually Pannekoek concluded on these remarkable and enormous distances of 40 to 60 kpc and further work on the structure of the Stellar System. I will then address Pannekoek’s work on the distances of dark clouds in the Milky Way. In the second part, I will turn to another topic. He also spent quite some time in drawing up detailed maps of the structure of the Milky Way on the sky, observing both visually and photographically, which intrigued me as an investigator in the late 1970s and 1980s of nearby galaxies extensively using the technique of photographic surface photometry. How he obtained these isophotal maps is indeed discussed in some detail in Tai, van der Steen \&\  van Dongen (2019) and Tai (2021), but no comparison with more recent maps is made to evaluate the accuracy of the features and surface brightnesses. I will compare his results to the photographic surface photometry of the southern Milky Way by the Bochum group under Theodor Schmidt-Kaler and particularly to the almost full-sky maps obtained with the Pioneer 10 spacecraft. The Pioneer 10 and 11 spacecraft had in the 1970s mapped the Galactic background on their way to Jupiter when beyond the asteroid belt the sky was free of zodiacal light. 

Pannekoek did only devote a part of his scientific research effort on the study of the structure of the stellar system we live in. As van den Heuvel wrote in the proceedings of the Amsterdam symposium (van den Heuvel, 2019; p.45):
\begmarg
In the 1920s and 1930s, Pannekoek was the pioneer in numerically calculating the structure of stellar atmospheres, and the spectra produced by these atmospheres. This put him on the map as an international expert in the physics of stellar atmospheres[...]
\endmarg
Nowadays Pannekoek is remembered as one of the founders of this major part of astrophysics. 
\bigskip

A note of notation: I will use the convention to capitalize Galaxy and Galactic when this refers to our Galaxy and use lower case when concerning external galaxies. For consistency with other coordinate systems  galactic in combination with longitude and latitude will not be capitalized. In quotations I will leave things as they are.

\section{The distance to the Milky Way}

It is good advice to examine at the start  the circumstances under which developments took place and in this case to ask what the state of astronomy was at the time considered. Virginia Trimble (1995) has described the state of the study of the structure of the sidereal system (and many related and unrelated matters) at the time Pannekoek set out to determine the `distance to the Milky Way’ (that is 1919) extensively in her  inimitable way as the background against which the Great Debate between Curtis and Shapley in 1920 took place. Issues here were -- to quote the phrasing from Trimble’s abstract -- `the existence of external galaxies (on which Curtis had been more nearly correct)’ and `the size of the Milky Way (on which Shapley had been more nearly correct)’.

\begin{figure}[t]
\sidecaption[t]
\includegraphics[width=0.54\textwidth]{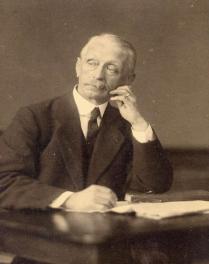}
\caption{\normalsize  Jacobus Cornelius Kapteyn (1851--1922) on an undated photograph. From his appearance compared to other photographs I would date this as some time around 1910 or at age of around sixty. Courtesy Kapteyn Astronomical Institute and University Museum Groningen.}
\label{fig:Kapteyn}
\end{figure}

Kapteyn (see Fig.~\ref{fig:Kapteyn}) had taken the  approach of turning counts of stars as a function of apparent magnitudes and observations of proper motions into a three-dimensional distribution (Kapteyn \&\ van Rhijn, 1920).  In his `first attempt’ he had primarily looked at the dependence on galactic latitude, because the large-scale dependence on longitude was in first approximation very small. This resulted in a symmetric model, flattened into an oblate spheroid, with the Sun not too far from the center  (Kapteyn, 1922). He also ignored, following Harlow Shapley, interstellar extinction, even though he had hypothesized its effect was wavelength dependent and he had found the corresponding reddening with distance (Kapteyn, 1909a). Shapley (1916) had obtained color indices for 1300 stars in the globular cluster Messier 13 in Hercules and found their color range to be quite similar to that of stars in the solar neighborhood. From the apparent magnitudes of the individual stars Shapley inferred that the distance of the Hercules cluster was ‘not less than’ 10 kpc (currently it is believed to be 6.8 kpc). Shapley noted that if Kapteyn’s value for the reddening were correct, the bluest color index (phot—vis) he would have observed should be 2.5. For M-stars he should have observed a color index of 5! This was magnitudes redder than he actually observed. Shapley concluded from this (p.15):
\begmarg
If we grant, on the basis of our data, a color excess of a tenth of a magnitude, and attribute it all to space
absorption, the value of the coefficient can not then exceed +0.0001 mag., an amount completely negligible in dealing with the ordinary isolated stars. In the light of this result we are probably justified in assuming that the non-selective absorption in space (obstruction) is also negligible.
\endmarg
Kapteyn had to conclude, as he had always kept open as a possibility, that what he  had seen was not extinction but a change in the mix of stars with distance.
\bigskip

It has been emphasized before in both Tai, van der Steen \&\ van Dongen (2019) and Tai (2021) that Pannekoek had deliberately taken an approach to the problem of the structure of the sidereal system different from Kapteyn. The latter had in first approximation ignored all detailed structure treating star counts in the first place as a function only of apparent magnitude and galactic latitude. Pannekoek had argued that the appearance of the Milky Way on the sky did not warrant such an approximation of a smooth ellipsoid and that the system should be considered as consisting primarily of an agglomeration of star clouds. To quote from Pannekoek (1910; p.241, his italics):
\begmarg
[...] the structure of the Universe [in Kapteyn’s approach] is determined as a figure of revolution, a flat disk, its axis at right angles with the Milky Way.  [...] This conclusion, however, is {\it in direct opposition to the appearance of the galaxy}. We see the galaxy as a belt of more or less circular masses, patches and drifts designating a totally different structure.
\endmarg

\begin{figure}[t]
\begin{center}
\includegraphics[width=0.88\textwidth]{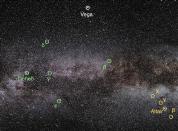}
\end{center}
\caption{\normalsize Part of the Milky Way from the constellation of Cygnus (left) to that of Aquila from the \textit{ESO Milky Way Panorama} (ESO, 2009), corresponding to longitudes about 86\degs\ to about 46\degs. The bright stars Deneb ($\alpha$ Cygni), Vega ($\alpha$ Lyrae) and Altair ($\alpha$ Aquilae) have been identified by name. Stars in Cygnus are in green and in Aquila in yellow. Pannekoek’s Cygnus cloud is between $\beta$ and $\gamma$ Cygni and the Aquila Stream includes the bright patch near $\chi$ Aql. In the space between Cygnus and Aquila lie the faint constellations Vulpecula and Sagitta. As for an orientation: the line between $\gamma$ Cygni and Vega is close to east-west, so north is to the top-left and east to the bottom-left. Credit: ESO/S. Brunier.}
\label{fig:CygAqu}
\end{figure}

Pannekoek (1919b) treated two regions in Cygnus and Aquila/Sagitta as isolated enhancements in stellar density in space with a dimension along the line-of-sight quite similar to the size that was displayed on the sky. These obvious, bright parts would have fascinated him in his extensive visual mapping of the Milky Way that I will discuss below, and would in his view have to be major aspects of the stellar distribution in the sidereal universe.
It is instructive to examine what these regions look like on the sky (see Fig.~\ref{fig:CygAqu}). I have taken the relevant area from the photographic survey {\it The Milky Way Panorama} produced by Serge Brunier and the  European Southern Observatory (ESO, 2009), based on photographs with a digital camera from the Paranal and La Silla Observatories in Chili and from La Palma, Canary Islands, on the occasion of the International Year of Astronomy 2009\footnote{\normalsize The center of the Galaxy is not precisely in the center of the ESO \textit{Milky Way Panorama} (ESO, 2009). In this paper I use the TIFF version which measures 6000 by 3000 pixels. My best estimate of its position is 3050 pixels from left and 1475 pixels from top.}. Fig.~\ref{fig:CygAqu} shows the Milky Way between longitudes about 86\degs\  to about 46\degs, encompassing the constellations Cygnus, Sagitta and Aquila. The view in latitude is chosen to include the bright star Vega for orientation for those familiar with the constellations.  Compare this figure to the text in Pannekoek (1919b; p.500):
\begmarg
The parts chosen were the southern part of the great bright elliptic patch between $\beta$ and $\gamma$ Cygni, the poorer regions east of it, and a part of the Eastern stream between 10\degs\ and 20\degs declination, in Aquila and Sagitta, including the bright patch of $\chi$ Aquilae.
\endmarg

According to Fig.~\ref{fig:CygAqu} these are parts of more or less elliptical clouds on the sky with long axes of 15\degs\ to 20\degs\ parallel to the Milky Way. Pannekoek viewed them as well-defined, distinct star clouds in which all but the very brightest stars were not visible individually. The crucial assumption is that these are isolated structures with well-defined extents on the sky and along the line-of -sight and are not affected by interstellar extinction. In our modern view these are both questionable. Extinction is everywhere at low latitudes; in e.g. Fig.~\ref{fig:CygAqu} it manifests itself as a lane of dust in the form of a sort of wave from left to right curving down below the Cygnus cloud and then up above the Aquila cloud. So the appearance of two clouds really is primarily shaped by dust rather than variations in stellar space density.

In his paper in 1910, Pannekoek had collected star count material and had concluded from that (Pannekoek, 1910; p.257, his italics):
\begmarg
[...] It appears that with the stars as far as the 13.9 magnitude we only just reach into the great star clusters forming the Milky Way, and in order to ascertain more about their structures and distances we have to go on to still lower magnitudes. That is why we do not venture here a comparison between our numbers and the light of the Milky Way. [...] What has been found here indicates {\it that no organic relation exists between the great mass of stars of the 9-th magnitude, and perhaps  as far as the 11-th, and the star-clusters forming the Milky Way.}
\endmarg

In this context it becomes clear what Pannekoek’s (see Fig.~\ref{fig:Pannekoek})  starting point in 1919 was. His work concerned the bright Cygnus patch and the Aquila stream, that he already had studied in the 1910 paper just quoted. He assumed that the stellar density distribution in such clouds was a Gaussian and that the extent along the line-of-sight was not greater than that on the sky (he quoted 15\degs\ to 20\degs\ diameter, so an extent a quarter to a third of the distance). He then concluded this can be ignored in the modeling since the range of absolute magnitudes far exceeds that produced in apparent magnitudes by such a relatively small range in distances. In Pannekoek's view, the stars in the clouds could not be counted due to the distance, except for the very brightest ones. The general stellar distribution is made up of stars uniformly distributed (`stars of the 9-th magnitude, and perhaps  as far as the 11-th’). This is the general component in the stellar distribution that  is seen as the star counts as made by Kapteyn and collaborators. The stars in the Cygnus and Aquila Clouds then manifest themselves as a small increase on top of these star counts. The way Pannekoek investigated that then is through the run of the gradient (the first derivative) in the counts as a function of apparent magnitude.

\begin{figure}[t]
\sidecaption[t]
\includegraphics[width=0.64\textwidth]{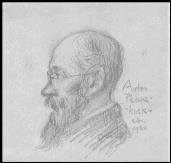}
\caption{\normalsize Pannekoek in 1920. After having been appointed part-time at the Municipal University of Amsterdam and after the Prime Minister in 1918 had blocked an appointment at Leiden Observatory as adjunct-director because of his Marxist sympathies, he was appointed full-time lecturer in Amsterdam and extra-ordinary professor in 1925. In 1931 he became full professor. Photograph of a drawing by an unknown artist, provided by the University Museum Utrecht to the public domain in  Wikimedia (commons. wikimedia.org/wiki/File: Anton Pannekoek 1920.jpg).}
\label{fig:Pannekoek}
\end{figure}

Pannekoek then did not treat different parts of the Galaxy separately to derive the changes in {\it star density  with distance}, as Smith wrote, he rather looked for changes in the {\it gradients in the numbers of star per solid angle with apparent magnitude} to derive distances of two star clouds of 40 and 60  kpc. He never calculated densities as a function of distance! His approach was fundamentally different, so the method is entirely misunderstood, or at least misrepresented by Smith. Nor has  he {\it calculated a distance to the galactic centre of [40 and 60 kpc] in the direction of the constellation Sagittarius}, as van den Heuvel has described the procedure. He only {\it derived distances for two clouds}. Pannekoek (1919b;  p.507) discussed the distribution of star clouds in the Milky Way ($\lambda$ is old galactic longitude):
\begmarg
[...], it must be emphasized that Shapley's result is wholly in accordance with the aspect of the Milky Way. For the hemisphere with centre at $\lambda$=325, where judging by the abundance of globular clusters our Universe reaches farthest, we observe in the direction of Sagittarius a series of round or irregular and rather small bright patches and clouds, which probably lie at a great distance. At the opposite side of the heavens we have at $\lambda$=100 to 140 the faint region of Perseus, where the Galactic light nearly vanishes; on both sides of it the light band gradually becomes brighter. 
\endmarg 

He concluded therefore that the distribution of star clouds in the Milky Way {\it supported} Shapley’s very eccentric location of the Sun in the Galaxy and the direction of the Galactic center in Sagittarius. 
\bigskip

Pannekoek (1910, 1912, 1919a) based his work on star counts he had collected in the directions of these bright Cygnus and Aquila complexes, and of a faint region in Cygnus (`between the branches’). These came from a number of sources, namely the {\it Bonner Durchmusterung} up to magnitudes 9 or so, star counts by one Epstein from Frankfurt am Mainz, counts on photographic plates of the {\it Carte du Ciel} and William Herschel’s star gauges. For the first he used three cuts in apparent magnitudes, in the other cases he adopted values for the limiting magnitudes. I will look into this in detail in the discussion section. The comparison of the slopes (gradients) of the star counts in the Aquila and the Cygnus faint end led him to the conclusion that at levels near magnitude 14 or so the counts reach into the background star clouds. The 1912-paper shows how with photographic exposures of a great variety of lengths the magnitude scales and hence gradients can be calibrated. In the 1919a-paper he divided up the Aquila field and calculated gradients per magnitude interval in the counts with magnitude  and concluded that (p.1333; his italics):

\begmarg
\noindent
The first gradient [between the brightest magnitudes]  is larger than the others.\\
Here then is manifest {\it the influence of the distinct galactic condensations, which  therefore is perceptible in the gradients only after the 13.5th magnitude.}
\endmarg

From his star count material Pannekoek found that at bright magnitudes the decrease in star numbers per magnitude interval (the gradient or first derivative) changed similarly towards the star cluster and the general Kapteyn star counts towards the Galactic poles. This reflects the run with distance of star density in the `central cluster’ (the local system around the Sun). But towards the star cluster this gradient increased for stars beyond magnitude 13, reflecting the `influence of the remote star cluster’. This was the case both towards the bright Cygnus and Aquila regions, but was absent in the faint Cygnus field. He then modeled the effect of a distant cloud, assuming it has a (in modern terms)  `luminosity function’ of a Gaussian shape (Kapteyn, 1902), with parameters as redetermined in the PhD thesis of Kapteyn’s student Willem Johannes Adriaan Schouten (1893--1971; see Schouten, 1918)\footnote{\normalsize Schouten  supported the approach to solving the inversion of the equations of statistical astronomy by Karl Schwarzschild (1875--1916) and vehemently criticized that of (Ritter) Hugo Hans von Seeliger (1849--1924). The to his taste too antagonistic tone in the thesis embarrassed Kapteyn, who refused to publish the thesis in the \textit{Publications of the Astronomical Laboratory at Groningen}. See Paul (1993) and van der Kruit (2015, 2021a). For a brief c.v. of Schouten, see Appendix B. Schouten's thesis and other theses under Kapteyn that remained unpublished are available electronically as part of Kapteyn’s publications on the dedicated Website accompanying van der Kruit (2015).}.

Pannekoek’s argument why the extent of the star cloud along the line-of-sight can be ignored can then be understood as follows. He approximated both  the spatial distribution of the stars in the cloud  and that over absolute magnitude as a Gaussian and the resulting star counts in apparent magnitude thus becomes the convolution of two Gaussians. This is another Gaussian with a dispersion (standard deviation) equal to the two original ones added in quadrature. The dispersion in magnitudes squared of the luminosity function he then found is at least 30 times that for the spatial distribution expressed as distance moduli, and thus also in magnitude, so the range in distances can be ignored. He then calculated a small number of models of counts of cluster stars superposed on either Kapteyn’s (1908) counts in the Galactic poles (as improved by van Rhijn and further in Schouten’s  thesis) or that in the faint region in Cygnus. He then looked for the following signature in the star counts (Pannekoek, 1919b, p504):
\begmarg
[...] after the normal decrease the gradient reaches a minimum, increases during
an interval of about 4 magnitudes -- just the interval in which the transition from minority to majority occurs -- and then the gradient of $\log N$, now due almost solely to cloud-stars, decreases again.
\endmarg
Now the apparent magnitude where the signature in the counts appeared in the form of a minimum in the gradients, where the cloud started to contribute to the counts. This he then compared to his model calculations. Then (p.504; Pannekoek’s italics):
\begmarg
For the Cygnus patch the minimum lies at m=11.5, for the Aquila stream somewhat below the 12th magnitude. These values correspond respectively to $\rho$ = 18 and 19. {\it So we find for the parallax of these parts of the Millky Way 0\secspt000025 (for the bright Cygnus patch) and 0\secspt000016 (for the Aquila stream), and for their distances 40,000 and 60,000 parsecs.}
\endmarg
The parameter $\rho$ requires some explanation and a warning. It was in this paper defined as $5 \log r$, so some sort of distance modulus. The distance $r$ is in units of 10 pc, following Kapteyn’s convention that would eventually lead to our current definition of absolute magnitudes. Pannekoek apparently regretted this and in later papers the definition was retained by him, {\it except that the unit of $r$ was changed to 1 pc}.

For Cygnus m = 11.5 then corresponds to M = -6.5 and for Aquila m =12 (and a bit) to M = -6.9 (and a bit). The peak in the luminosity functions would then occur at m = 27.0 and m = 27.9 respectively. The luminosity function used has a standard deviation of 6.10 magnitudes, so the fitting is done at about 2.5 standard deviations.

\begin{figure}[t]
\sidecaption[t]
\includegraphics[width=0.64\textwidth]{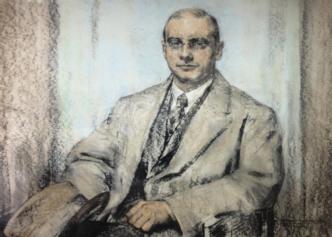}
\caption{\normalsize Pieter Johannes van Rhijn (1886--1960). This crayon drawing has been produced by Eduard Gerdes (1887--1945), a Dutch painter and art teacher and dates from 1926. Later he sympathized with the National Socialist Party, and held important positions relating to art policy during WWII. Gerdes mysteriously died soon after the War ended. He was posthumously found guilty of collaborating with the German occupation. This drawing resides in the Kapteyn Room of the Kapteyn Astronomical Institute in Groningen. See also van der Kruit (1922). Courtesy Kapteyn Astronomical Institute.}
\label{fig:vRhijn}
\end{figure}

The final check Pannekoek performed was to calculate the surface brightnesses of the patches that followed from his models. These are all around the equivalent of  0.20 to 0.26 stars of m = 1 per square degree. The observed values to compare this to are from the PhD theses by two students of Kapteyn, Lambertus Yntema (1979--1932) and Pieter Johannes van Rhijn (1886--1960). The PhD thesis of Yntema (1909; see Appendix B for a brief c.v.) is published in full, but the work  of van Rhijn (see Fig.~\ref{fig:vRhijn}) was published only in preliminary form in van Rhijn (1919); the full publication (van Rhijn, 1921b) appeared after Pannekoek’s paper. The methods  used by Yntema and van Rhijn and the results have been discussed extensively in van der Kruit (2015, 2021a) and the results are surprisingly good compared to modern values. Pannekoek’s model values are  roughly three times brighter than typical values of Yntema and van Rhijn, which Pannekoek regarded as not unreasonable for bright patches in the Milky Way.

I will leave a detailed consideration of the actual counts used and their reliability to the
discussion section below.

\section{Further work on the distribution of stars}

The 1919-paper on the distance to the Cygnus and Aquila clouds was not at all the end of Pannekoek’s work in the area of statistical astronomy and Galactic structure. He had a close look at possible systematic errors in the {\it `Durchmusterung of Selected Areas'} (Pannekoek, 1922a), in which he derived corrections to the magnitudes and therefore counts using information provided by Pieter van Rhijn.  Already three years after his 1919-paper, he returned to the Cygnus/Aquila region, in particular studying the distribution of foreground stars (Pannekoek, 1922b), -- this paper was not mentioned by both van den Heuvel and Smith. This gave a complicated picture of the distribution of foreground stars compared to the Cygnus cloud (Pannekoek, 1922b; p.56):
\begmarg
So it appears that the region of Cygnus has a more complicated structure than would be expected at first sight. Different formations are situated in this direction at different distances, which are seen partly projected upon each other, and combine their effects in the Galactic light. 
\endmarg
This changed the apparent magnitude where the stars of the Cygnus Cloud were seen in the counts and as a result of this the distance to it had the be substantially revised by more than a factor two, bringing it in from 40 kpc to 18 kpc. Although this result is not mentioned in Tai, van der Steen \&\ van Dongen (2019), it has not escaped Tai (2021), who does treat it in some detail. The revision followed from a revised, higher foreground density of stars, but also because of van Rhijn’s (1922) revised surface brightness value, which triggered a revised normalization of the model density in the cloud.  In the paper at the end Pannekoek remarked on dark, `absorbing material' (p.56):
\begmarg
As a third feature determining the distribution of the stars and the galactic light we must add absorbing nebulous masses. 
\endmarg
But no consequences for the distance of the Cygnus cloud were attached to this: dust extinction was believed  to be strictly limited to the dark clouds seen projected onto the Milky Way on the sky.
\bigskip

Not only did Pannekoek have to reduce the distances of the Cygnus Cloud by a factor two in 1922, but a year later he had to retract his result altogether. In a paper, not referenced in either the Amsterdam symposium proceedings and Tai’s thesis, he studied the effects of his assumption of a uniform luminosity function everywhere in the stellar system. This paper, entitled {\it Luminosity function and brightness for clusters and galactic clouds} (Pannekoek, 1923a), is not written in a very lucid manner. It does not have any statement at the beginning about what questions he will address; it immediately delves into the mathematical background with a section `Derivation of formulae’. The sections do not close with definite conclusions and the paper itself ends abruptly without a summary or presentation of the conclusions.

Pannekoek started with a Gaussian luminosity function and calculated the total light from a star cloud by integrating over all absolute magnitudes. He then derived that the `most contributing absolute magnitude’ in this case is $M = -3.1$ (Pannekoek, 1923a, p5.):

\begmarg
[...] the total light of a mass of stars, following the Kapteyn luminosity curve, is 6.3 times the light of all stars of the most contributing magnitude.
\endmarg
So, if you have star counts and a total luminosity you can deduce what the most contributing magnitude is (without Pannekoek mentioning it, it is obvious that this must be in an interval of one magnitude) and determine what the fraction of the total luminosity is contributed by the stars of this magnitude. This then can be compared to what is expected from Kapteyn’s luminosity function and in this way the universality of that function can be studied. Pannekoek noted this can be applied to an individual star cloud, but also in terms of surface brightness, and proceeded to apply it to various cases.

As a first case he turned to globular cluster M3. On average the data were consistent with, but as Pannekoek stressed did not prove, a Kapteyn function. Analyzing available data in rings, Pannekoek found that the luminosity function changes with radius, the brighter stars being more concentrated toward the center. This conclusion was judged to be tentative due to incompleteness and the fact that crowding on photographs in the inner parts tends to create bright stars from superposing of two fainter ones. Now, brighter stars in globular clusters are giants which evolve from the brightest and most massive stars on the Main Sequence. Mass segregation of this kind is known now to be quite common. The Galactic clusters  M37 and M11 were found to have much narrower luminosity functions than Kapteyn’s, and M35 intermediate. So Kapteyn’s curve does not look like being universal! Data on the Small Magellanic Cloud were insufficient (no accurate surface brightness), but for the bright part of the Scutum Cloud the provisional conclusion seemed to be that the small differences with Kapteyn are not significant.

The luminosity functions of the Cygnus and Aquila Clouds, for which Pannekoek had first found such large distances and then subsequently had to reduce these by a factor two, came out problematic. First he noted (Pannekoek, 1923a; p.11):
\begmarg
The number of stars below 12$^{\rm m}$ in these regions show a sudden and strong increase, indicating the appearance of the brightest cloud stars among the nearer system stars, which are soon wholly outweighed by them. Assuming the validity of Kapteyn’s luminosity curve for the remote clouds I deduced a distance beyond $\rho$ =20 from the point of inflexion [his spelling] in the apparent luminosity curve (M.N. 79, 500; B.A.N. 11). It has been shown, however, by Dr. A. Kopff at Heidelberg that in this case the surface brightness of the Cygnus cloud should be far greater than is given by observation; the Kapteyn luminosity curve does not hold for this cloud, so that also the distance deduced by it looses its foundation.
\endmarg
The parameter $\rho = 5 \log r $ was now the redefined version with $r$ in pc, so that $\rho = 20$ corresponds to 10 kpc.

And to continue (p.12):
\begmarg
The stars constituting these Galactic clouds [Cygnus and Aquila] appear to be concentrated within an extremely narrow range of magnitudes, in the same way or still more so than in open clusters. The data are not sufficiently accurate to allow more exact deductions on the luminosity curve. If we might suppose that these stars are giant stars of the same absolute magnitude as in open clusters, we could derive some value for the distance; [...]

Still more remarkable than this great Cygnus cloud seems the region of Sagitta and the adjacent parts of Aquila. While the moderate brightness of the galactic light shows nothing particular, the number of small stars is as excessive as in the brightest parts. […]  This abundançe of stars extends only over very few magnitudes, indicating some kind of giant stars of one definite brightness. 
\endmarg 
The evidence was that the actual luminosity distribution of the stars was far from universal and his method would not work. His attempt to measure the distances to the Cygnus and Aquila Clouds in the end had failed. This paper (Pannekoek, 1923a) is not referred to in either Tai, van der Steen \&\ van Dongen (2019), nor Tai (2021).
\bigskip

The relation of Pannekoek’s studies with ongoing studies in the Netherlands by Kapteyn (who, however, retired in 1921 and died in 1922) and his successor Pieter van Rhijn, amateur astronomer Cornelius Easton and Utrecht student  Isidore Nort has also been covered in much detail in Tai’s (2021) thesis, particularly chapter 2. I will make a few remarks, and recommend the reader of this article interested in more details to examine this thesis.

\begin{figure}[t]
\sidecaption[t]
\includegraphics[width=0.545\textwidth]{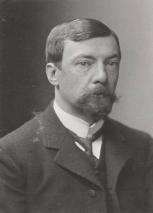}
\caption{\normalsize Cornelis Easton (1864--1929) in 1906. He was a Dutch reporter and newspaper editor, who also was very successful as an amateur astronomer, publishing in professional astronomical journals. He received for his work in 1903 an honorary doctorate from the University of Groningen (see van der Kruit, 2021c). This photograph appears in the \textit{Album Amicorum}, presented to H.G. van de Sande Bakhuyzen on the occasion of his retirement as Professor of Astronomy and Director of Leiden Observatory (after Leiden Observatory, 1908). In the public domain under Creative Commons CC. }
\label{fig:Easton}
\end{figure}

I will first turn to Easton (see Fig.~\ref{fig:Easton} and Appendix B for a short c.v.). Cornelis Easton (1864--1929) argued that there was a strong correlation between the distribution of bright stars and the Milky Way background. Bright areas were accompanied by an excess of bright foreground stars, small proper motions and stars of type A and B (that were by that time known to be bright intrinsically). So, the brightest stars of Pannekoek’s star clouds were affecting star counts at much brighter apparent magnitudes than Pannekoek had deduced and these clouds consequently were much closer. As he wrote (Easton, 1921; p.255) 
\begmarg 
[…] the cumulation of evidence seems to point rather convincingly to the conclusion, that in several parts of the zone the Galactic aggregations affect the density, the proper motions, and the spectral type of the relatively bright stars nearer to our Sun. Thus, unless we are prepared to assign to a considerable number of `cloud stars’ intrinsic brilliances of a million or ten million times that of the Sun, it is difficult to conceive how those Galactic clouds can be situated at distances of the order of 50,000, or even 10,000 parsecs.
\endmarg

Details of the arguments between Pannekoek (1922b) and Easton (1922) have  been documented very well in Tai’s (2021) thesis. The matter was discussed extensively at the May 28, 1922  ‘conference’ during Shapley’s visit to Leiden after the first Rome IAU General Assembly with both Easton and Pannekoek present (Tai, 2021), but the case was not concluded definitely in favor of either of the two.

Next I mention Utrecht student Henri Nort (again for a brief bio see Appendix B). Isidore Henri Nort (1872--1943) in his PhD thesis (Nort, 1917) used Pickering’s {\it Harvard Map of the Sky}. A set of 55 double contact prints on glass plates covering the whole sky had been put at astronomer's disposal, counts from these maps had been performed by Hans Henie (1887--1950) at Lund Observatory in Sweden. Nort redetermined for each plate the limiting magnitude and produced star counts in 5000 small areas down to magnitude 11. From this he derived some statistical properties of the distribution of the brighter stars. In his thesis Nort had determined the shape of the system of stars down to magnitude 11.0 and found this to be roughly an ellipsoid with three unequal axes  with ratios 0.53 : 0.86 : 1.00, the shortest perpendicular to the Galactic plane and the longest pointing towards (new $l^{\rm II}$) longitudes 80\degs \ and 260\degs. The center was in the direction in $l^{\rm II}$ = -1\degs, $b^{\rm II}$ = -19\degs. The analysis did not involve absolute distances.

\begin{figure}[t]
\sidecaption[t]
\includegraphics[width=0.64\textwidth]{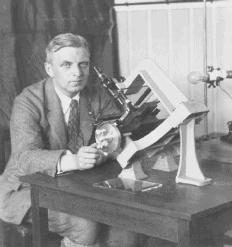}
\caption{\normalsize Egbert Adriaan Kreiken (1896–1964) at the measuring machine of the Astronomical Institute of the University of Amsterdam. Kreiken studied with Kapteyn, but only completed his PhD thesis on stars in the Scutum Cloud in 1923 under van Rhijn after Kapteyn’s retirement and death. After that until to 1928, he taught physics and astronomy at a secondary school in Amsterdam, from 1926 also as Privaat Docent in stellar astronomy at the University of Amsterdam. This photograph must have been taken in this period. He subsequently worked at a number of different places, notably the Bosscha Observatory in Lembang, Dutch East-Indies, and finally settled in Turkey, where he founded in Ankara the Observatory that is now named after him. Courtesy Anton Pannekoek Institute, Amsterdam.}
\label{fig:Kreiken}
\end{figure}

The final person involving himself with this was Egbert Kreiken (see Fig.~\ref{fig:Kreiken} and for a short c.v. see Appendix B). Egbert Adriaan Kreiken (1896--1964) had studied under Kapteyn and obtained his PhD degree with van Rhijn in 1923. While working in Amsterdam at a high school, he published a paper on the local star-system (Kreiken, 1926).  The local irregularities are due, according to Kreiken,  to the influence of the dark nebulosities in Taurus, Ophiuchus, etc. The star density in the local system is at maximum at a distance of 2270 pc,  more or less in the direction of the Galactic center as we know it now.  The density distribution is elongated with the longest axis pointing in that direction.
\bigskip

Pannekoek decided to produced his own version of the determination of the local  structure (Pannekoek, 1922c), followed by a set of three papers on the subject (Pannekoek, 1924, 1929a, 1929b) and a summary in (Pannekoek, 1929c). The 1924-paper is a long, comprehensive and very  profound study. It starts with writing down the fundamental equations of statistical astronomy and then these are adapted for the three cases of condensations (clouds or local enhanced star densities), voids (local deficiencies of the same) and ‘absorbing matter’ (localized dark clouds causing extinction). Remarkable is that he still used the Kapteyn luminosity function, represented by the canonical  Gaussian (‘quadratic-exponential’) function, which — in spite of his conclusion a year before  (Pannekoek, 1923a) — he still assumed universal (Pannekoek, 1924; p.4):
\begmarg
The luminosity function deduced by Kapteyn follows this function to the utmost exactness,
\endmarg 
\bigskip

The extensive studies by Pannekoek on local structure in the stellar distribution using these backgrounds has been the subject of a very detailed presentation and discussion in  Tai’s (2021) thesis,  especially in his chapter 2.1, and I will not treat it in the same detail here.

\begin{figure}[t]
\begin{center}
\includegraphics[width=0.98\textwidth]{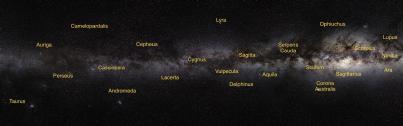}
  
\includegraphics[width=0.98\textwidth]{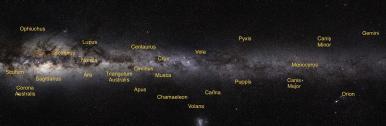}
\end{center}
\caption{\normalsize  Map of constellations in on near to the Milky Way on the the ESO {\it Milky Way Panorama} (ESO, 2009) in two parts with the area of the Galactic center repeated for clarity. The clusters h and $\chi$ Persei are the two dots to the lower left of the C in Cassiopeia. Credit: ESO/S. Brunier.}
\label{fig:MWcons}
\end{figure}

In brief terms, he analyzed star counts using Kapteyn’s luminosity function to deduce the detailed distribution of stars near the Sun from irregularities in the distribution of star counts across the sky. For example in Pannekoek (1924), he derived from van Rhijn’s (1921a) counts as a function of latitude a smoothed-out `Schematical Universe’, which was a relatively flat, oblate spheroid. He compared this to the {\it Bonner Durchmusterung} in  the north and the {\it C\'ordova Durchmusterung} in the south, after carefully discussing the magnitude scales, to  produce maps of deviations in those counts relative to those of the Schematic Universe at different apparent magnitudes. This showed a large-scale pattern that displayed an irregular structure. He next identified bright condensations in the Milky Way and studied the {\it Bonner} and {\it C\'ordova Durchmusterung} stars in front of these. The surpluses were then used to derive distances, which resulted in a schematic drawing with the positions in space of condensations or local enhancements.  Finally he located deficiencies of stars to indicate positions of `absorbing clouds’.  He was aware that dust extinction played a role, but the belief at the time still was that this was restricted to the dark clouds and rifts that are easily observable by examining the structure in the Milky Way. As a guide to appreciate the following quotations of Pannekoek for those unfamiliar with constellations I provide in Fig.~\ref{fig:MWcons} a map indicating their positions on the {\it ESO Milky Way Panorama} near the plane of the Galaxy. He wrote (Pannekoek, 1924; p.119):
\begmarg
It must be borne in mind that the number of stars observed and counted in these catalogues, is smaller than the real number. For part of them is obscured by the large clouds of absorbing matter, situated on one side of us, towards Taurus, at nearly 140 parsecs, extending with less density through Perseus and Camelopardus, and at the other side, at perhaps somewhat smaller distance, towards Ophiuchus and Scorpius. Probably the dark rift between the branches of the Milky Way through Aquila as far as $\beta$ Cygni is due to an appendix of these nebulae, and the absorbing nebulae in the northern part of Cygnus, in Cepheus, Lacerta, Cassiopeia are perhaps a continuation of them.
\endmarg

But there were a few significant concentrations he identified  (p.120):
\begmarg
The part of space considered here shows several larger and smaller condensations of stars. The most remote that can be detected in the DM [Durchmusterung] stars is situated at 800 parsecs towards Monoceros; the largest is the Cygnus condensation at 600 parsecs, extending with a broad tail over 90\degs\ in longitude, as far as the clusters h [and] $\chi$  Persei and filling the space down to the surroundings of our Sun with surplus density. […] The third condensation, situated towards Carina at 400 parsecs, is smaller, but much more concentrated than these. 
\endmarg
In the sequels (Pannekoek, 1929a,b) he made the very valid point that distances of individual stars are needed to accurately map the stellar density around the Sun. Since this is not possible with only  trigonometric or secular parallaxes, also spectroscopic parallaxes are required and hence spectral types. This limited the work to stars with small dispersions in their intrinsic luminosities, in particular K-giants and A and B-type Main Sequence stars; in his own words (Pannekoek, 1929a, p.1):
\begmarg
	The distribution of stars in the {\it Durchmusterung Catalogues}, discussed in Nr. 1 of these {\it Publications} by means of the general luminosity function, cannot give information about the space density in the immediate surroundings of our sun. Here the space distribution of the stars must be studied by methods, which are able to give approximate distances for the separate stars, by measured parallaxes, by proper motions, or by spectra. The spectrum may be used for such stars, where the dispersion in absolute magnitude is small. This is the case with the A-type stars, which therefore have often been used for finding the distance of some stellar agglomeration; also for the giant K stars, which are more numerous than the G and M giants, the dispersion is small. For the B-type stars the absolute magnitude varies considerably with the spectral division, but for each division it is generally assumed to be constant.

	The completion of the Draper Catalogue containing magnitude and spectrum for all stars down to the ninth magnitude gives the possibility of finding the space distribution of the stars of these spectral classes. Of course this distribution will not always be identical with that of the bulk of the stars ; differences are already known, e.g. the scarcity of the stars of class B relative to class A in the Cygnus condensation. The other spectral classes seem to be more equally intermingled; [...]
\endmarg
The point is that the completed {\it Henri Draper Catalogue} (Cannon and Pickering, 1918-1924) with almost a quarter of a million stars allowed a new study as outlined.

\begin{figure}[t]
\sidecaption[t]
\includegraphics[width=0.64\textwidth]{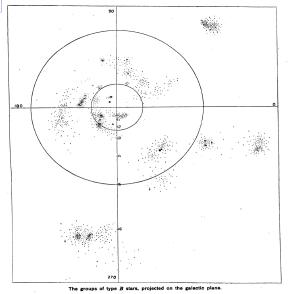}
\caption{\normalsize Schematic representation of the distribution of B-stars around the Sun projected on the Galactic plane. The longitudes are of course the old system with the direction of the Galactic center at 327\degs; the circles denote distances of 0.3 and  1.0 kpc from the Sun. From Pannekoek (1929a)}
\label{fig:Ostars}
\end{figure}

The assumption was that in particular the K-stars, but to some extent also the A-stars,  were distributed like the common stars. For the B-stars, Kapteyn, who designated them  `helium-stars’, had shown this to be not the case. Kapteyn (1914, 1918a,b,c) had extensively studied their spatial distribution in two very long {\it Astrophysical Journal} papers (the second published in two parts because of its extreme length). These studies went only to the sixth magnitude, while Pannekoek’s limit was three magnitudes fainter. Pannekoek does not refer to Kapteyn’s work; but on the other hand in his presentation for the Royal Academy (Pannekoek, 1929c) he did credit Kapteyn for pointing out that the small dispersion in absolute magnitude for B- and A-stars, which  did make such a study possible. Pannekoek (1929a) derived a schematic drawing of the clumpy distribution of B-stars around the Sun (Fig.~\ref{fig:Ostars}). This is a precursor of the famous map of ‘aggregates’ of O-, B- and A-stars of Morgan, Whitford \&\ Code (1953), often hailed as the discovery of spiral structure in our Galaxy. But it also is an improvement and an extension if Kapteyn's work. Distances go out to 3 kpc or so; the local spiral arms running in the directions 210\degs -30\degs\  (also old coordinates).  Not surprisingly with these small distances spiral structure is not clearly delineated. Pannekoek probably was not looking for spiral structure, but this study stands as a prototype for those that led to the discovery of spiral structure in the Galaxy, as outlined in  the distribution of (blue) stars,  just before it was confirmed using the hydrogen 21-cm spectral line.
\bigskip

As a general conclusion, Pannekoek found that the general stellar distribution was more complicated than an ellipsoid, and that stars of a particular spectral type (such as B, A or K) were concentrated in condensations  with, however, the ones of B-stars not always coinciding with those of A-stars and ones of K-stars. The structure of the stellar distribution in space near the Sun was extremely complicated. In Pannekoek’s words (1929c; p.14):
\begmarg
Our results show that a local system in the sense of a separate large condensation, does not exist. We may use the word in a technical sense, comprising in it the whole of the near stars contained in and studied by means of our catalogues. But in reality it consists of a number of separate smaller or larger condensations and clusters which probably have no more connection with another than with the more remote groups.
\endmarg

\section{Dynamics}
When he set out to work on the stellar distribution, Pannekoek found large distances, which held promise to study the structure of the Galaxy on a large scale. In the end his analyses revealed only details of the local structure, while the major advances on the nature of the Galaxy came from studies of the large-scale structure and kinematics. Tai, van der Steen \&\ van Dongen (2019) and Tai (2021) and others in reviews of Pannekoek’s work on the structure of the Galaxy, did not comment on the studies in the areas of the kinematics and dynamics of the Stellar System, and that of distances of dark, rather than stellar, clouds. The latter will be treated in the next section, and here I turn to dynamics. But we first need a bit of background.

Probably the most important aspect of Kapteyn’s work was his insight that any model for the structure of the Sidereal System would only be complete when it was not just restricted to the distribution of stars, but also included their kinematics and an explanation of the equilibrium (or near-equilibrium) between the gravitational field and the kinematics in terms of dynamics. The foundations of stellar dynamics had been laid in the U.K. by Arthur Stanley Eddington (1882--1944) and James Hopwood Jeans (1877--1946).  Oort (1981) told how he had been inspired by Eddington’s  (1914) {\it Stellar movements and the structure of the universe} and Jeans` (1919) {\it Problems of cosmogony and stellar dynamics}, books that Kapteyn would have introduced him to. Kapteyn was the first to include dynamics in his attempt to understand the `construction of the heavens’, as Herschel described the field when he started it. It induced Jeans (1922) to apply his hydrodynamic equations (Jeans, 1919) to what he coined `a Kapteyn-Universe’.

Kapteyn’s modeling of the Sidereal System consisted of two parts, the first being the determination of the distribution of stars in space  (Kapteyn \&\ van Rhijn, 1920) and the second the dynamics (Kapteyn, 1922). He calculated the gravitational field and assumed dynamical equilibrium and the velocity distribution to be Maxwellian (that is having a Gaussian shape). His dynamical model consisted of two components, the first being the notion that the random motions of the stars in the vertical direction maintained the distribution perpendicular to the plane of the Milky Way. Kapteyn's in this way determined value of the space density in the solar neighborhood still holds up to the present. His estimate for the total density near the Sun in the Galactic System came out as 0.099 solar masses $M_{\odot}$ pc$^{-3}$ -- according to Oort (1932), who translated Kapteyn’s result into these units. The recent analysis by Kuijken \&\ Gilmore (1989) concluded that it is 0.10 $M_{\odot}$ pc$^{-3}$ and that there is no conclusive evidence for additional unseen, dark matter. Jeans (1922) with his analysis using the hydrodynamic equations found 0.143 and Oort (1932) 0.092 again in the same units. Jeans (1922) had also assumed dynamical equilibrium and noted that this could neither be proven nor disproved, but seemed reasonable for the moment. That this was not trivial is shown by the fact that Eddington (1918) had argued for the case of the Galaxy not being in dynamical equilibrium,

Now Pannekoek (1924) realized that no revision of Kapteyn's `admirable explanation’ of the vertical dynamics was required by the new view of a much larger system.  For the dynamics {\it in} the plane of the Milky Way this was a different matter. Kapteyn included there in addition to the effect of random motions the centrifugal force resulting from rotation in the stellar system, which he identified as his famous Star Streams, presented first at the 1904 Saint Louis World Exhibition, celebrating the centennial of the Louisiana Purchase the year before. He interpreted these as two rotations  in opposite directions around the center of his system of about 20 km sec$^{-1}$ each (Kapteyn, 1904). Pannekoek commented on it in this 1924-paper. This center of rotation, according to Kapteyn (1922), had to be at 0.65 kpc in a direction perpendicular to his two streams. The stream vertices were in directions  $l^{\rm I}$  = 167\degs\  and 347\degs, $b^{\rm I}$ = 0\degs\ ($l^{\rm II}$ = 200\degs\ and 20\degs) in the constellations Gemini and Scutum, so perpendicular to these was either in the constellation Cassiopeia or in Carina. Kapteyn has chosen for Cassiopeia because  it `is in good accordance with the investigations of Herschel, Struve, and myself’. According to  Pannekoek, Shapley disagreed and would have chosen Carina, because the Milky Way is brighter there. This seems a much more logical choice than Kapteyn’s, since the stellar density would be highest towards the center and Pannekoek concurred with Shapley. Actually, using the IAU conventions of the constellations, the direction ($l^{\rm II}$  = 110\degs) is just in Cepheus close to the border with Cassiopeia, and ($l^{\rm II}$  = 290\degs) only just in Carina, close to the borders with Vela and Centaurus.

Our understanding of the nature of Kapteyn’s Star Streams would soon change when Oort had detected the rotation of the Galaxy. The streams in fact were only apparent streams, but are in reality the result of anisotropies in the velocity distributions, as Karl Schwarzschild had pointed out as an alternative interpretation already very soon after they were announced. These velocity distributions take the form of a ‘velocity ellipsoid’ whose longest axis manifests itself as the streams and according to stellar dynamics was later found to have to point towards the Galactic center and anticenter. When Oort had determined the center of rotation it was found that instead the vertices of the Kapteyn Streams did point 20\degs\ away from the center-anticenter line. This has been named Deviation of the Vertex. The deviation turned out to be real and accurate and Kapteyn’s determination in 1904 was correct and of excellent quality; it was explained by Oort (1940) as due to the gravitational effect of spiral arms. But all of this was unknown to Pannekoek in 1924. So he asked himself, how these Kapteyn Streams would fit into a kinematical and dynamical picture of the larger stellar system. This somewhat long expos\'e serves to establish that the Streams are real and well understood, and have the property of being highly regular and symmetric.

Pannekoek (1924) had used the {\it Bonner Durchmusterung} in the north and the {\it C\'ordova Durchmusterung} in the south to study the distribution of stars in the solar neighborhood and had found this to be complicated and clumpy. The space near the Sun contained a number of concentrations of stars. As the first option to explain Kapteyn's Star Streams Pannekoek then examined whether these condensations could give rise to a local phenomenon in the form of motions induced by their gravitational pull.  But he found that to be impossible for 20 km sec$^{-1}$. With all the detailed structure the motions would be expected to be more irregular anyway than Kapteyn’s simple highly-symmetric two-stream phenomenon, and this effort was doomed to fail. He next considered the Galaxy to be not in equilibrium, the clumpy, irregular structure not looking at all like  a steady state. Earlier Jeans (1916) had also argued that the distributions were more chaotic than expected of a system in equilibrium. In order to make progress in understanding the dynamics the motions of all these  clouds and concentrations needed to be determined first, which would be an enormous task. Pannekoek did not come up with an explanation for Kapteyn’s Star Streams. 

\begin{figure}[t]
\sidecaption[t]
\includegraphics[width=0.64\textwidth]{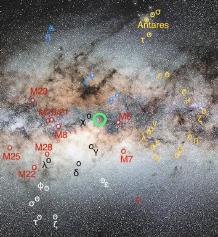}
\caption{\normalsize  The central part  of the Milky Way from the \textit{ESO Milky Way Panorama} (ESO, 2009).  The longitude range is approximately from longitudes 18\degs\ to 340\degs, latitudes run from approximately -18\degs\ to +19\degs. The four stars used by Pannekoek to identify the Sagittarius Cloud,  $\lambda$, $\delta$, $\gamma$ and X Sagittarii are indicated in black. The first three, together with the white stars, $\epsilon$, $\phi$, $\sigma$, $\tau$ and $\zeta$ Sagittarii, form the well-known `Teapot' pattern. The position of the Galactic Center has been indicated by the green circle.  Oort's rotation center is about 4\degs\ to the right (near Messier 6), and Shapley's center of the globular cluster system halfway in between. Antares ($\alpha$ Scorpii) and other stars in Scorpius are shown in yellow, in Ophiuchus in blue. Messier objects are in red. Credit: ESO/S. Brunier.}
\label{fig:Sagcloud}
\end{figure}

 Pannekoek's follow-up studies (1929a, 1929b) based on the {\it Henry Draper Catalogue} and the {\it Cape Photographic Durchmusterung} contain no comments on the dynamics of the Stellar System.
\bigskip

When Oort (1927) had discovered Galactic rotation with the center near Shapley’s center of the system of globular clusters he surmised that there had to be an attracting mass in the center of  $ 8 \times 10^{10}$ M$_{\odot}$ at a distance of 6 kpc. Pannekoek (1927)  asked himself: could this be the Sagittarius Cloud? This cloud of stars is illustrated in Fig.~\ref{fig:Sagcloud}.  The figure includes the Galactic center. Many objects in the field look like stars but in fact are some other objects. These are in red. M6 (the Butterfly Cluster), 7 (Ptolemy's cluster), 21, 23 and 25 are open star clusters, M22 and 28 are globular clusters and M8 (Lagoon Nebula) and M20 (Trifid Nebula) diffuse, HII (ionized hydrogen) regions of star formation

Using again Kapteyn’s Luminosity Function and Eddington’s Mass-Luminosity Relation he found that the mass of the Cloud fell short by a factor one thousand for this. He also rejected the possibility that there is interstellar extinction, because than the central star cluster will shine through holes in the dust distribution and would give rise to very bright spots, a hundred times more brilliant than observed. Actually Baade’s Window of exceptionally low extinction is the bright spot just above $\gamma$ Sagittarii and is part of the Sagittarius Cloud; it is only about 4 degrees from the Galactic Center and because of the low extinction is very bright.  But indeed not that bright. The central star cluster inferred by the rotation is much more concentrated than the Sagittarius Cloud and hidden by  many magnitudes of extinction. Pannekoek also rejected the possibility that it would be in the form of dust, as  this could also not provide sufficient mass. According to Pannekoek the central attracting mass could be in the form of concentrated interstellar gas, the existence of which had been hypothesized by Eddington (1926) in his Bakerian Lecture. But no proof of  this could be found.
\bigskip

\section{Distances of dark clouds}

Pannekoek was among the first to research dark clouds and he should be counted as one of the pioneers in this area. Edward Emerson Barnard (1857--1923) at Yerkes and Maximilian Franz Joseph Cornelius Wolf (1863--1932) at Heidelberg photographed large numbers of dark nebulae that the Herschels had interpreted as holes in the stellar distribution (Steinicke, 2016). Barnard  (1919) had provided a catalog (and later an atlas; Barnard, 1927) and concluded from visual inspection that most — but undoubtedly not all, as he was careful to cover his back — of these are ‘obscuring masses of matter in space’ in front of the background of stars in the Galaxy. Frank Watson Dyson (1868--1939) and Philibert Jacques Melotte (1880--1961) at the Royal Greenwich Observatory were the first to estimate the distance of a number of such a dark nebula by noting that from apparent magnitude 9 to at least 14 the number of stars projected onto them was quite uniformly one-fifth of what it was in adjacent regions and arrived at 200 to 300 parsecs (Dyson \&\ Melotte, 1919). Seeley \&\ Berendsen (1972) described the history of interstellar extinction research, concentrating mainly on the diffuse extinction; their treatment of discrete dark nebulae is weighted heavily towards Barnard’s contribution  and ignores the work of Max Wolf and that of Pannekoek.

\begin{figure}[t]
\sidecaption[t]
\includegraphics[width=0.54\textwidth]{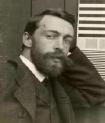}
\caption{\normalsize   Ejnar Hertzsprung (1873--1967) as a young man. This photograph in the Hertzsprung archives in Aarhus, Denmark, is undated. The pattern in the background suggests a relation to photography, so a possibility is that this has been taken, when he studied photochemistry at Leipzig University. This puts it at 1901, when he was around 28 years of age. Courtesy AU Library, Institut for Fysik og Astronomi, Aarhus Universitet, Denmark.}
\label{fig:Hertzsprung}
\end{figure}

Pannekoek (1919a) took up the approach of Dyson and Melotte and studied a small region in the Aquila Cloud using Franklin-Adams plates provided by Ejnar Hertzsprung (1873--1967; see Fig.~\ref{fig:Hertzsprung}). The area measured about 2\degspt5 square roughly centered on the star {$\gamma$} Aquilae (see Fig.~\ref{fig:CygAqu}). There is a dark spot in this area. He wrote (p. 1333):
\begmarg
As a first explanation we may admit that this cause consists in a local diminished space-density of the stars, so that there is an actual hole between  and in the dense star-clouds that constitute the galaxy. In this case the nearer stars are not influenced thereby, so they must show no thinning, the brightest stars will be relatively more numerous than the faint ones, and the gradient must be smaller than in the denser regions. Of this the numbers show nothing; the stars of  the 10th to the 14th magnitude are all diminished to an equal rate.
\endmarg

This would  rule out convincingly sparsety of stars in space. Pannekoek noted (same paper, p.1334; his italics):
\begmarg 
If the nebulous matter should exist in the regions of the galactic condensations, only the more distant stars would be dimmed, and the phenomena would be [...], a relative excess of brilliant stars. From the numbers found, it therefore becomes evident, {\it that the absorbing dark nebulous mass causing the tripartite hole, is so near as to dim also the majority of the stars of the 10th and 11th magnitude. It stands in no organic connection to the galactic clouds, being only accidentally projected against that clear background.}
\endmarg

However, since this work averaged over extended regions, he was careful not to definitely rule out effects of the stellar distribution in space, but it certainly opened the strong possibility of an effect of extinction. Note that this was in 1919, the same year in which Barnard had published his catalog of dark nebulae and Dyson and Melotte had made the first, educated estimate of a distance based on comparison of star numbers on and next to a nebula.

Pannekoek (1921a) made the first serious attempt to derive a distance of a cloud of absorbing or scattering material using quantitative procedures. This did not concern the relatively small dark clouds in Barnard’s catalogue,  but more extended areas in deficiencies in star density. In Pannekoek (1921a; p.708) he wrote: 
\begmarg
 The wide extension of this absorbing substance became evident in yet another way, by an investigation of the general distribution of the stars up to the 11th magnitude [Pannekoek, 1919c]. It was found here that around two places with a considerable deficiency of stars, in Taurus and Ophiuchus, as around two centres of obscuration, there are wide regions where the number of stars is below the normal.
\endmarg

Pannekoek then proceeded to present his way of comparing star counts towards the dark nebula to those in a comparison area. This was similar to what Dyson and Melotte had done but now in a more rigorous manner. He assumed that everywhere stars were distributed according to Kapteyn’s luminosity function. The counts as a function of apparent magnitude then followed from integrating the contribution for each distance along the line of sight, which is the local total density of stars multiplied by the value of the luminosity function at the absolute magnitude corresponding to that distance. As explained elsewhere (e.g. van der Kruit, 2021c, 2022) the problem is that the counts are what is known and the density as a function of distance is to be solved for,  so that the integral equation has to be inverted. Karl Schwarzschild had developed a numerical procedure in case the luminosity function is a Gaussian (as indeed Kapteyn’s version was) and this very likely played a role in Schwarzschild’s honorary doctorate at the University of Groningen on the occasion of its tricentennial in 1914 (van der Kruit, 2021c). Pannekoek used the same numerical values for the luminosity function as mentioned in section 3 in connection with Pannekoek (1923a).

Using an adopted luminosity function, the method would be to solve for the run of stellar density along the line of sight in the direction of the dark nebula and that next to it and derive a distance for the cloud and the total amount of extinction by comparing the two. Pannekoek (1921a) used a more direct, analytical approach. He used Kapteyn’s Gaussian luminosity function and assumed the density could be approximated also by a Gaussian, and compared  to van Rhijn’s counts in the general area of sky  of latitudes between 20\degs\ and 40\degs, the latitude range that included the Taurus area. He then introduced an infinitely thin screen producing a certain number of magnitudes of extinction.

\begin{figure}[t]
\begin{center}
\includegraphics[width=0.98\textwidth]{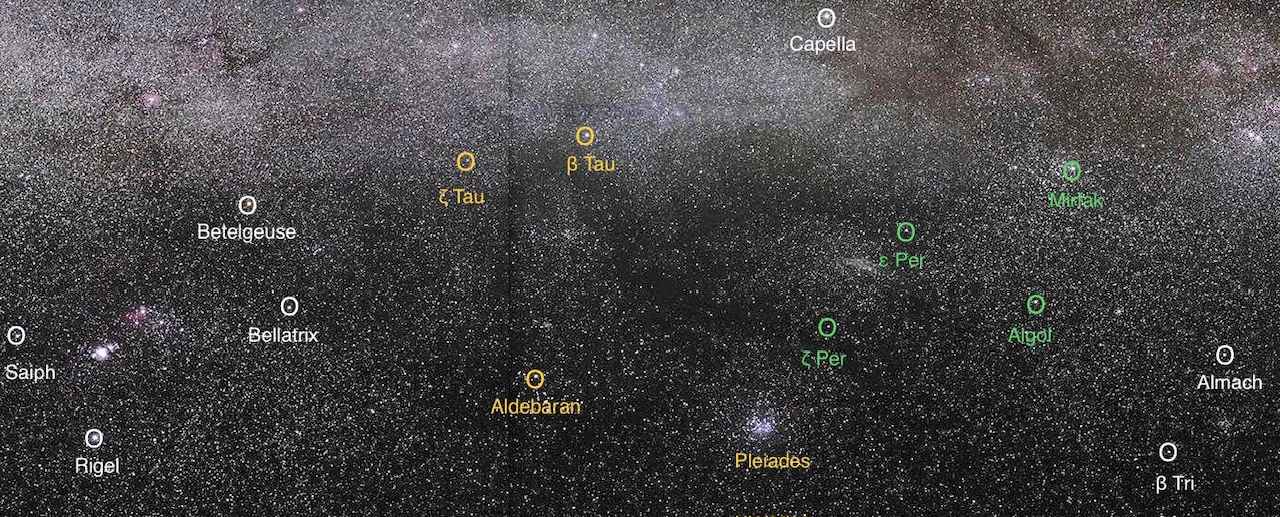}
\end{center}
\caption{\normalsize  The constellations Orion (left), Taurus, Auriga and Perseus with the dark clouds studied by Pannekoek (1921) from the ESO {\it Milky Way Panorama} (ESO, 2009). The picture runs from roughly Galactic longitude 218\degs to 124\degs, latitude -28\degs\ to +12\degs. The stars in Taurus are indicated in yellow and in Perseus in green; in the other constellations in white. The following named stars have been  indicated: Saiph ($\kappa$ Orionis), Rigel ($\beta$ Orionis), Betelgeuse ($\alpha$ Orionis), Bellatrix ($\gamma$ Orionis), Aldebaran ($\alpha$ Tauri), Capella ($\alpha$ Aurigae),  Algol ($\beta$ Persei), Mirfak ($\alpha$ Persei), and Almach ($\gamma$ Andromedae).   The stars Pannekoek (1921a) used to describe the location of his dark clouds are $\beta$ Tauri, the Pleiades and $\zeta$ Persei. The photograph also shows at the left the Orion Nebula, and the two dots on the far right above the middle are the clusters h and $\chi$ Persei. A line from about halfway between $\zeta$ and $\beta$ Tau to the Pleiades is roughly east-west, so north is towards the top-right. The dark, vertical line near Aldebaran shows where the two ends of the panorama have been joined. Credit: ESO/S. Brunier.}
\label{fig:OriPer}
\end{figure}

It again is instructive to see what the dark clouds Pannekoek studied look like on the sky in order to appreciate what his approach was. The areas Pannekoek concentrated on can be located in Fig.~\ref{fig:OriPer}, using his description that the areas studied are slightly southwest of $\zeta$ Persei, between the Pleiades and $\beta$ Tauri and southwest of $\beta$  Tauri, and can be recognized as somewhat extended areas of diminished numbers of stars. Pannekoek gave a long discussion concerning non-uniform plate material (especially {\it Carte du Ciel}) and the non-uniform distribution of the apparent extinction. The  method did not work well in small areas as they contained too few bright (that is to say {\it Bonner Durchmusterung}) stars for reliable statistics. There were two Kapteyn {\it Selected Areas} he could use, counts from the Paris Observatory as part of the {\it Carte du Ciel} and the Franklin-Adams results of Dyson and Melotte (1919). Pannekoek noted that although in general {\it Carte du Ciel} counts are problematic because of great variation of the limiting magnitude from plate to plate, in this particular case all material was published so that he could make the material uniform using the property that every plate corner is the center of another plate. In the end the conclusion did depend on all of this material but to a very large extent on the {\it Bonner Durchmusterung} stars. 

The analysis was performed in eight smaller areas (20 - 45 square degrees) defined in the text with reference to a map in the Dyson and Melotte paper. The two ‘darkest regions’ (designated ‘A’ and ‘B’) were located respectively at the borders between Taurus, Perseus and Aries almost halfway from the Pleiades to Algol near $\zeta$ Persei, and in Taurus about halfway between $\beta$ Tauri and the Pleiades (indicating and labeling these in Fig.~\ref{fig:OriPer} would have interfered too much with illustrating the dark areas). When he applied his method of analysis, he found for the distance of the Taurus Cloud 140 pc with a reliability range of 100 to 200. As Pannekoek noted this would  mean behind the Hyades, actually three to five times as far away from us. The total extent on the sky is of order 30\degs\ or about 70 pc. Region A of about 9\degs\ by 3\degs\ corresponds to 20 by 7 pc. In the area there are three objects in Barnard’s Catalogue which then would have diameters of order 40,000 astronomical units (0.2 pc). The extinction values did vary considerably over the field, but were mostly of order between 1 and 2 magnitudes.
\bigskip

Pannekoek (1921b) then continued to do something that -- as far as I know --  had not been attempted before and that is to estimate the masses in the dark clouds. This required knowledge of the physical process of the extinction. He considered the mechanism to be Rayleigh scattering (as Kapteyn had supposed and which is strongly wavelength dependent and would give rise to reddening). This happens in the Earth’s atmosphere, resulting in blue skies, and involves small particles, the molecules of air. Measurements of the daytime sky brightness at Mount Wilson  relative to the Sun were performed by Abbot  (1914) at the request of Kapteyn to determine what percentage of starlight at Mount Wilson is scattered in the atmosphere (he found $\sim$ 5\%\ of the light arriving from the zenith). Assuming a thickness along the line of sight of 10 pc for the $7 \times 20$ pc region A then gives a mass of $10^{10} M_{\odot}$. This is for air molecules, but using Rayleigh’s equation for hydrogen gas he found a similar value,  $4 \times 10^9 M_{\odot}$. Pannekoek noted that this is as much `mass as all the stars within a globe extending 20 times further' (than 140 pc). This clearly is unrealistic. He showed that it would give rise to unrealistically high velocities and proper motions. And the Sun would in fact have to be in a very elliptical orbit around the nebula with a period of two to three million years.

If indeed the low star numbers indicated voids of stars the counts required that these extended at least ten times as much along the line of sight as they did on the sky, so
\begmarg
we come to the hardly acceptable assumption of protracted, tubular cavities, all running in the direction of the line of sight.
\endmarg
The only way out would be (Rayleigh) scattering on electrons since these are so much lighter than hydrogen molecules. With the strong dependence then on wavelength this would clearly turn up in reddening, but only for stars of magnitude twelve or fainter and obtaining spectra to derive expected colors to compare with observed ones was at the time not feasible.

There appeared a fundamental problem of mass with Rayleigh scattering. Note that  the unrealistically strong wavelength dependence of $\lambda ^{-4}$ did not contradict observations yet, since the actual approximate $\lambda ^{-1}$ was only found in the 1940s (see discussion in van der Kruit, 2022). In a postscript to the paper Pannekoek reported that Willem de Sitter (1872–1934), the director of Leiden Observatory, proposed `opaque articles’. This would have the property that there would be no wavelength dependence and reddening. That the scattering particles could have dimensions comparable with the wavelength of optical light, so that scattering would be wavelength dependent and  produce reddening, had apparently not occurred to either de Sitter or Pannekoek.
\bigskip

In his autobiographical notes  he described this episode including work on dark nebulae as follows (Pannekoek,1982; p.248; my translation):
\begmarg

[...]  in November 1919 a map of the Taurus Nebula was published in the MN by Dyson and Melotte, with a rough estimate of the distance. I then, making use of Kapteyn's luminosity curve, developed a numerical method for distance of such dark nebulae, and from a collection of all available star counts calculated the distance of the Taurus Nebula. This piece was published in 1920 in the Reports of the Academy, with another somewhat fanciful afterthought on mass.  
\endmarg

\begin{figure}[t]
\sidecaption[t]
\includegraphics[width=0.54\textwidth]{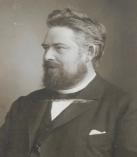}
\caption{\normalsize Maximilian Franz Joseph Cornelius Wolf (1863--1932).   This photograph comes from the {\it Album Amicorum}, presented to H.G. van de Sande Bakhuyzen on the occasion of his retirement as Professor of Astronomy and Director of Leiden Observatory in 1908 (after Leiden Observatory, 1908). In the public domain under Creative Commons CC.}
\label{fig:Wolf}
\end{figure}

This is a rather unpretentious description. In fact, it was significant, new and fundamental work. However, two years later a paper appeared by Max Wolf, that became the standard reference (Wolf, 1923). Wolf (see Fig.~\ref{fig:Wolf}) plotted star counts on and next to a dark nebula.  The nebula he used, NGC 6960, is the Veil Nebula (part of the Cygnus Loop). This may look like a strange choice, but Wolf justified it as follows (NGC 6992 is that part of the Cygnus Loop opposite to NGC 6960; p.110 my translation):
\begmarg
The nebula NGC 6960, which passes in a long band almost exactly through the star 52 Cygni [...]  approximately north-south […] forms a bright spot seen against a large dark cloud, so to speak. It is actually the western part of a very extended nebular mass, which, however, is only detectable in the longest exposures and continues into the also elongated bright nebula NGC 6992 [...]. Therefore NGC 6960 is to be regarded as one edge of this very large Milky Way nebula. While the edge formed by NGC 6992 is not bordering on any or only difficult to detect dark masses, the western edge, i.e. NGC 6960, is bounded by a conspicuous star void. 
\endmarg
While the two curves (the archetypical case is in Wolf, 1923; p.113) corresponding to the two star counts, on and off the dark nebula, coincide at bright levels,  at some apparent magnitude they start to deviate but then become parallel again. The magnitude where this starts is an indication of the cloud’s distance (taken to be that of the average star of that apparent magnitude) and the distance between the parallel lines the total amount of extinction. These Wolf diagrams or Wolf curves have become the standard and appear in many textbooks on elementary astronomy. But Pannekoek’s work preceded it!

Wolf does acknowledge that Pannekoek’s method is in fact more accurate (p.113, my translation):
\begmarg
One could perhaps try to calculate more exact values according to the method of Pannekoek [footnote to reference]. With the uncertainty of my magnitudes and the inaccuracy of the figures for the average parallaxes, it serves little purpose for the time being.
\endmarg

Pannekoek's method may be more accurate, it is also more cumbersome, and therefore many researchers might have felt Wolf diagrams were good enough and the use of these would  therefore have become popular. Bartholomeus Jan (Bart) Bok (1906--1983) recommended the use of Wolf diagrams to obtain the total amount of extinction in a dark cloud, but preferred for procedures solving for the distance `the method developed by Pannekoek in his well known paper on the dark nebula in Taurus’ (Bok, 1931).  He advocated the use of so-called ($m$ - log $\pi$) tables, developed by Kapteyn, which Bok as a student of van Rhijn was very familiar with. Then no assumptions on analytical functions are required. The tables, also called ($m$ - log $r$) tables with $\pi$  parallax and $r$ distance, lost their usefulness when electronic computers arrived, but survived as the principle of which the $V/V_{\rm max}$ method for the study of quasars is an extension (Schmidt, 2000). A detailed investigation of the effects of the width of the luminosity function, fluctuations in the star counts and the extent of the dark cloud along the line of sight led Freeman Devold Miller (1909--2000) of Swasey Observatory of Denison University at Granville, Ohio, to warn against uncritical use of the Wolf diagrams (Miller, 1937).
\bigskip

In the mean time Pannekoek had been pointed to the work of Meghnad Saha. The way this had happened was through his successor as teacher at a secondary school in Bussum. This was Herko Groot (1890--1974), who had defended a PhD thesis in Utrecht in 1920, entitled {\it Stralingsdruk beschouwd in verband met zijn beteekenis voor de physica der zon} (Radiation pressure considered in the context of its significance for solar physics) with Willem Henri Julius (1860--1925), founder of solar physics research in Utrecht, as supervisor. Groot had sent this thesis to Saha, who then had sent reprints of some of his papers in return, which he then had passed on to Pannekoek. The latter had picked this up and this, as described more competently than I can, led to him becoming (van den Heuvel, 2019; p.45)
\begmarg
the pioneer in numerically calculating the structure of stellar atmospheres, and the spectra produced by these atmospheres.
\endmarg
This must have kept him from continuing further work on the structure of the Galaxy and related matters and he returned to dark nebulae only during the Second World War (Pannekoek, 1942). 

This 1942-paper, published as number 7 of the {\it Publications of the Astronomical Institute of the University of Amsterdam}, is a major piece of work, running 74 pages. In the first section of the paper he summarized the theory behind his method, including the formal calculation of Wolf diagrams, treatment of errors and distribution of parallaxes and proper motions. The latter, as Pannekoek noted, had been the subject of a PhD thesis in Groningen by Broer Hiemstra, to which I will return below. At the start of the second section Pannekoek provided the justification for the undertaking:
\begmarg
The amount of labour spent by different astronomers on star counting, using well standardized plates taken especially for this purpose and often based on accurate measurements of the magnitude of a great number of stars, makes it worth while to submit their results to a more exact discussion. Especially the vague and inaccurate indication that the absorption sets in at the mean distance of one and extends to the mean distance of another magnitude has to be replaced by a computation of the distance at which absorbing masses have to be placed to produce the observed deficiency stars. 
\endmarg
This is in fact criticism of the use of Wolf diagrams without mention Wolf by name.

 The second section includes the application to a large number of dark patches, clouds or extended regions, including the Cygnus Rift that can be seen in Fig.~\ref{fig:CygAqu} as the extended dark area below the Cygnus Cloud. Although he arrived at an indicative, but far from definitive distance of 2000 pc for this rift, he remained uncertain as to which counts to compare the ones in front of it to compare. For the Taurus cloud, for which he previously had found a distance of 140 pc, he derived a new value 290 pc, but there is no discussion on what caused this large correction by a factor two. There is no summary of results.

In the third section an analysis of the star counts to magnitude 9.3  the zone between galactic latitudes +25\degs\ and -25\degs\ was performed. For this the counts in the {\it Bonner} and {\it C\'ordoba Durchmusterungen} were used to calculate the total numbers in 1\degs\ squares and compared to the averages at 5\degs\ intervals. This resulted in colored maps with deviations (positive or negative) and a list of 130 well-defined and strong deficiencies. Deficiencies were measured in the logarithm of the number of stars in these areas. He argued that the natural uncertainty of these numbers (the square-root of the number or $\sqrt{N}$) is only 1 to 10\%\ in the log and therefore the observed variations are significant compared to random fluctuations. The (maximum) distances range from 110 to 420 pc. By way of summarizing (p.68):
\begmarg
Nearby absorbing matter is recognized as void regions, poor in visible stars; sometimes (as in Taurus and the Scorpion) there is a faint continuous light due to the nebulous matter reflecting the light of the imbedded stars. Whereas the Milky Way drawings show little concordance with our colored maps of the [deficiencies] because the visual Milky Way depends to a very small extent on the nearby stars and nebulae — the photographic Atlas may serve as a bridge, explaining them both as simplified expressions of a different characteristic of the spatial structure.
\endmarg

There is little in terms of developing the large picture of ‘absorbing matter’ in the Galaxy. Unfortunately Pannekoek does not go beyond merely  mentioning work by others. It would have been interesting had he used his approach on the dark clouds studied by Broer Hiemstra (1911--1994; see Appendix B for a short c.v.), who wrote a clever thesis under van Rhijn in  1936 developing a method that is a variation on Pannekoek’s (for more on this see van der Kruit, 2022). His idea (or that of his supervisor Pieter van Rhijn) was to use counts as a function of proper motion rather than apparent magnitude. The equations have the same structure in that case (the luminosity function is replaced by the distribution of
linear velocities, which could be described as a Schwarzschild velocity ellipsoid). Hiemstra measured proper motions in four Kapteyn {\it Selected Areas} of the {\it Special Plan}. Briefly for those readers not familiar with this: the {\it Plan of Selected Areas} consisted of two parts. In addition to the {\it Systematic Plan} of 206 areas distributed evenly across the sky, on the insistence of Edward Charles Pickering (1846--1919), director of Harvard College Observator, an additional 46 were chosen for a {\it Special Plan} to be particularly suitable for more detailed studies of structure within the Stellar System. Some of these were selected to contain dark nebulae and Hiemstra studied four of these. The four dark clouds were found to be at distances ranging from 300 to 1000 parsecs and their total extinctions at least 0.5 to 2 magnitudes. Hiemstra presented a `comparison of results with those of other authorities’. These comparisons come out reasonably well (particularly Miller, 1937), except for those of Wolf, but Hiemstra noted that the latter’s method is `not unobjectionable’. Pannnekoek mentioned Hiemstra's study in passing, but it would have deserved more space.

\section{Visual maps of the Milky Way}
One remarkable and admirable set of products of Pannekoek’s work and research were isophotal maps of the Milky Way. These were accompanied by extensive descriptions of features and areas in it. He was interested to see if there were changes with time, but he also was sensitive to the phenomenon that the visual appearance varied from observer to observer. Part of his effort was to arrive at an ‘average’ picture that would be a representation acceptable to all observers. There is much more to be said about these studies of the brightness distribution across the Milky Way, but this has been addressed extensively in Tai’s (2021) PhD thesis, the aims of which he summarized as (p.170):
\begmarg
[...]  how did astrophotography impact Pannekoek’s astronomical research; what epistemic virtues did he pursue and how did these impact his research; and what connections can be found between his astronomy and Marxism.
\endmarg 
I will not enter into these lines  but concentrate on how accurate his isophotal maps are compared to modern data. Before doing this I will have to look in some detail into his observational techniques and his reduction and calibration procedures. But first I will address the question of how he came to do all this mapping. For more on this see section 1.2 in Tai (2021).

\begin{figure}[t]
\sidecaption[t]
\includegraphics[width=0.54\textwidth]{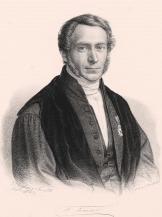}
\caption{\normalsize Frederik Kaiser (1808--1872). Born in Amsterdam in an immigrant family from Germany, he was raised by his uncle, who arranged for him to become `observator’ at Leiden University. He made quite a name in 1835 predicting the time of perihelion passage of Halley’s Comet with an unprecedented accuracy of 1.5 hours. In 1837 he was appointed director of the Observatory and in 1845
full professor of astronomy. He arranged for the new Observatory to be built and
opened in 1861. Kaiser’s main activity concerned positional astronomy or astrometry. Photograph of a drawing made in 1855 by engraver and lithographer Johann Peter Berghaus (1810--1870) in the public domain in  Wikimedia (commons.wikime dia.org/wiki/File:Frederik-Kaiser-2.jpg).}
\label{fig:Kaiser}
\end{figure}

In his autobiographical notes, Pannekoek (1982) described how he became interested in observing the starry sky. His teacher in physics had given him some literature and he became interested in buying books at book auctions. He mentioned  in particular two books (my translation),
\begmarg
Br\"unnow’s {\it Lehrbuch der sph\"arischen Astronomie} and Argelander’s Manual for amateur observing, translated by Kaiser.
\endmarg
Franz Friedrich Ernst Br\"unnow (1821–1892) was director of Berlin Observatory when he wrote the book, later he became the first director of the Detroit Observatory of the University of Michigan in Ann Arbor, but eventually returned to Germany, only to become shortly after that Astronomer Royal of Ireland and director of Dunsink Observatory. The book not only had a large influence on Pannekoek: it was a leading textbook, having in the end been translated into a fair number of languages. The book by Friedrich Wilhelm August Argelander (1799--1875), the driving force behind the {\it Bonner Durchmusterung}, had a major influence as well. Pannekoek described both in his autobiographical notes and his first presentation of the brightness distribution of the northern Milky Way (Pannekoek, 1920). He noted that his first drawings date back to 1889 and 1890, the years in which he became sixteen and seventeen and was still in highschool. The Argelander book, in the form of the translation into Dutch, must have induced him to start making maps of the Milky Way. So let us have a closer look at this book.

The Dutch translation in 1851 of Argelander (1844) had  the title {\it Invitation to friends of astronomy to make observations on several important branches of celestial science that are as interesting and useful as they are easy to perform, translated from the High German by W. F. Kaiser, with a preface, notes and appendices by F. Kaiser} (my translation). It is amusing to quote from the Preface by Frederik Kaiser (1808--1972; see Fig.~\ref{fig:Kaiser}), who founded Leiden Observatory and did provide the groundworks for modern astronomical research in the Netherlands (my translation):
\begmarg
In recent years, Germany has been extraordinarily prolific in writings on astronomical subjects; but this prolificacy was largely a result of ignorance driven by delusion. Some of these writings have been transferred to our soil and it seems that publishers were the more easily found when translators  more clearly betrayed their lunacy. On the other hand, when a famous astronomer published a popular text, in which to a simple mind he proclaimed the definite statements of science, only a few people took notice of it, and no one seemed to feel inclined to bother with a translation, when it might, in fact, have been of real use.

Thus the follies of Sch\"opfer and Schmitz, immediately after they appeared, were published in Dutch, while the beautiful and important work of the famous Argelander, in the Yearbook of Schumacher, could not find a translator with us until now. The well-known Professor Schumacher at Altona, published an astronomical yearbook from 1836 to 1844, to which every year some popular essays by the most famous naturalists and astronomers in Germany were added. Olbers, at the request of Schumacher, was to produce  a treatise on a most important astronomical subject, but was prevented by death from fulfilling his promise, and the difficult task, which Olbers had taken upon himself, passed, after his death, into the best hands it could find, namely those of Argelander. The contribution of Argelander, together with two others, from Steinheil and Moser, were included in the volume of 1844.
\endmarg
Heinrich Christian Schumacher (1780--1850) was the founder of Altona Observatory just outside Hamburg, operated between 1823 and 1871, and of the {\it Astronomische Nachrichten}. I continue with Kaiser:
\begmarg
The desirability of a Dutch translation of Argelander’s contribution has become clear to me, especially in recent times, and since no one else seemed to decide to do it, I have entrusted it to my son Willem Frederik, who was not entirely a stranger to the subjects dealt with by Argelander. The translation of such a piece was a difficult task for a young man of eighteen years, but where his strength was lacking, I came to his aid, and I believe the piece was very well translated, and I believe that the contribution, through his translation, has gained considerably in clarity. 
\endmarg

The seven chapters treat the aurora, zodiacal light, shooting stars, twilight, Milky Way, magnitudes and colors of stars, and variable stars. In the chapter on the Milky Way, Argelander argued that much was to be improved over the drawings of earlier researchers such as Johann  Bayer, John Flamsteed, William and John Herschel. Kaiser added at the end of each chapter a few remarks by himself. For the chapter on the Milky Way he noted that much remained to be added particularly to the drawings of the southern Milky Way by John Herschel, This enticed Pannekoek to start an observing program to improve the northern part; of course the southern Milky Way was outside Pannekoek’s reach. In his description (Pannekoek, 1920; p.1; my translation from the German):
\begmarg
My first attempts at observing the Milky Way, stimulated by the translation of Argelander’s `Invitation’ produced by F. Kaiser, date from 1889 and 1890. The method of observation followed in these and the following years consisted in drawing boundary lines, or more correctly lines of equal brightness, with chalk on large black cards on which the stars were represented by white disks. No artificial illumination was necessary; the eye could remain completely in the dark, since the starlight was sufficient to see the asterisms on the maps. The procedure was to trace some boundary line along the Milky Way, always comparing the brightness back and forth, usually over an area of 30 to 60 degrees longitude; when the position of the line was well impressed on the mind, it was drawn and then either its continuation or the next brighter or fainter one was tackled. In this way a number of drawings of a large part of the Milky Way were made in the years 1890-92. This method of work has been described and recommended in the invitation to observe the Milky Way, published by the author in 1897 in some journals. The results obtained in this way served as a basis for the representation of the Milky Way, which was presented to the annual meeting of the `Association of Friends of Astronomy and Cosmic Physics’ in M\"unster in 1893.
\endmarg

The presentation at the Association in M\"unster was by a Joseph Plassmann (1859--1940), in daily life a Gymnasium teacher in that city. The Association was co-founded together with Wilhelm Julius F\"orster (1832--1921) of the Berlin Observatory. Pannekoek had sent a map he had produced to F\"orster. His 1897-papers, that Pannekoek referred to,  form an article published  in three parts in {\it Popular Astronomy}, Pannekoek (1897, 1898a,b), but appeared partly also in other amateur periodicals. In the 1898a-paper he described new star charts he had prepared and which had been lithographed by Cornelis Easton and made available to whoever was interested. These were star charts in Galactic coordinates which were not otherwise available. He urged others to use these also to map the Milky Way, stressing that the general outline should be done by drawing the maps, but details should be noted in writing and added to these maps later, but not too long after the actual observing. His way of improving the accuracy was to average results by various observers, but the response to the invitation in Pannekoek (1898a) to request copies of these charts from Easton was far from overwhelming. 
\bigskip

The production of such maps of the Milky Way was long, patient and tedious work. Certainly, Pannekoek was not the first to do this. An interesting review
of the history of the subject has been written by Colin Henshaw (2014). Each part was observed by Pannekoek a number of times until a consistent picture emerged from the various individual traces. And it required an observer with excellent eyesight. Now Pannekoek was a very accurate observer with superb eyesight, which usually is illustrated by his visual discovery that $\alpha$ Ursae Minoris, the North Star or Polaris, is a variable. As a teenager he noted that compared to nearby stars it showed a small variation with a period of about four days. When Hertzsprung (1911) confirmed that it in fact is a Cepheid variable,  he credited Pannekoek in the introduction of his paper (Hertzsprung’s German in my translation, the Pannekoek quote was in English) as follows:
\begmarg
The most important, however, is a footnote that Pannekoek (1906) added to his discussion of the ac-stars of Antonia C. Maury: `In this connection may be mentioned that in 1891 the author thought he detected a variability of $\alpha$ Ursae Minoris with a period of a little less than 4 days. The small amplitude and the great influence of biased opinions on estimations of brightness after Argelander's method in cases of short periods of almost a full number of days, made it impossible to obtain certainty in either a positive or a negative sense. Campbell's discovery that it is a spectroscopic binary system with a period of 3$^{\rm d}$ 23$^{\rm h}$ 4$^{\rm m}$ makes me think that it has not been wholly an illusion’. It would be  interesting to see whether Pannekoek’s observations are consistent with the Ephemerides quoted below.
\endmarg

The publication by Hertzsprung prompted Pannekoek to go ahead and publish his observations a few years later (Pannekoek, 1913), when he resided in Bremen. While Hertzsprung quoted the year of Pannekoek’s observations as 1891, in his autobiographical notes (Pannekoek, 1982) he wrote December 1890. In Pannekoek (1913) he explained that in those years 1890 and 1891 as a teenager he observed stars of the second and third magnitude, estimating their brightness using the Argelander step method (explained below a few paragraphs hence). The observations continued up to 1899, the year he joined the Observatory in Leiden. Hertzsprung found the star was a Cepheid and derived in this paper an amplitude of  0.171($\pm$ .012) magnitudes (corresponding to 17\%) and a period of 3.97 days. This small variability then must have been sufficient to be visible to Pannekoek’s eyes. What then about Campbell’s almost-four-day spectroscopic binary, if it is a Chepheid? Polaris is  a triple system at about 110 parsec or 350 lightyears from us, consisting of a narrow binary and a wide third component. The inner binary consists of the bright Cepheid with a mass of  5.5 $M_{\odot}$ and a much fainter 1.3 $M_{\odot}$ companion in an orbit with half major axis  $\simgt$3 AU and orbital period about 30 years. The third component has a mass of about 1.4 $M_{\odot}$ and is seen at a projected distance of 18 arcsec or about 2000 AU. This apparent separation implies an orbital period of order ten thousand years.

The 30 year period of the compact inner binary manifests itself in the form of a wavelength variation in the spectral lines, but the main variation in the spectral lines seen by Campbell is not the sign of a spectroscopic binary, as he inferred, but reflects the radial pulsation of the main Cepheid star. The inner binary could only be resolved with Hubble Space Telescope (see Hubblesite, 2006; Evans {\it et al.}, 2008).
\bigskip

\begin{figure}[t]
\begin{center}
\includegraphics[angle=90,width=0.75\textwidth]{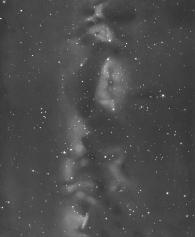}
\includegraphics[width=0.75\textwidth]{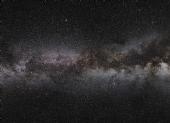}
\end{center}
\end{figure}

\begin{figure}[t]
  \caption{\normalsize This figure shows the area (l$^{\rm I}$, b$^{\rm I}$) = (65\degs\ to 348\degs, -15\degs\ to +15\degs), roughly (l$^{\rm II}$, b$^{\rm
II}$) = (98\degs\ to 21\degs, -15\degs\ to +15\degs), running through the constellations Cepheus, Cygnus, Sagitta, Aquila and Scutum. The upper part is a drawing by Pannekoek (1920, Plate or `TAFEL ’ I). The lower part is the corresponding area from the {\it ESO Milky Way Panorama} (ESO, 2009). Note that apparently there are deviations from linear angular scales in Pannekoek’s drawing, so the bright stars in both panels do not line up perfectly. Credit: ESO/S. Brunier.}
\label{fig:Maps1}
\end{figure}

The estimation of brightnesses was done visually using using the ‘Argelander step estimation method’, in which one  estimates where the  brightness of a particular star is positioned as a step value for the fraction of the magnitude difference between two comparison stars. The fascinating history of magnitude estimations from the Herschels to Argelander and beyond is summarized by John Hearshaw (1996: Chapter 2). Pannekoek first chose five points in Cygnus, Lacerta and Cassiopeia as standard reference and then for the winter sky three points  in the Auriga-Gemini area, and for the summer sky five in the Aquila area. These points he then compared as often as he could so that this set could serve as a reference for further brightness estimates. For these he used four step values. This was done repeatedly until he had a set of 18 consistent relative brightness estimates over the northern Milky Way. 

With this framework of 18 reference positions Pannekoek next listed estimates of the brightness in 128 positions in the Milky Way. These are in principle the four step values, including decimal values, and because some positions were brighter (and darker) than the reference values extended to below unity and up to six or so.

\begin{figure}[t]
\begin{center}
\includegraphics[width=0.75\textwidth]{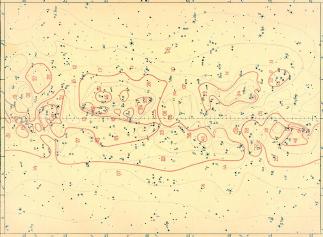}
\includegraphics[width=0.75\textwidth]{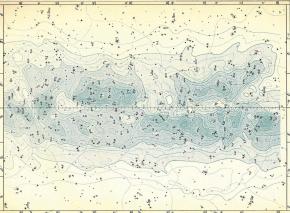}
\end{center}
\end{figure}

\begin{figure}[t]
 \caption{\normalsize The same area as in Fig.~\ref{fig:Maps1} in the form of isophotes obtained by Pannekoek (top) and the average of four observers (bottom), both taken from Pannekoek (1920). These are respectively TAFEL  IV and VII of that publication. In the upper part the red isophotes are at levels 2.0 and 4.0 (numbers in the map have the decimal point deleted); for the odd values isophotes are thin, but full-drawn lines. Intermediate values are drawn with dashed lines, but have sometimes been omitted. In the lower panel more isophotes are drawn; the shading changes at the four step levels. The very thin lines in the background, which are the same in both panels are the borders of constellations adopted by Pannekoek.}
\label{fig:Maps2}
\end{figure}

The basic values were the four steps. The ratio between these was calibrated in a clever way. He took two spots of different step level and observed these when the brightest was visible at low altitude and estimated when they were similar. The atmospheric attenuation, which was calculated geometrically from the ratio in pathlength through the atmosphere, could then be used to determine the step ratio. Another way was observing when particular lines  (we would say isophotes) became visible in twilight and compare that to stars becoming visible and using the magnitude difference between these stars to derive that for difference for the isophotes. He found that his step-size varied between 0.30 and 0.37 magnitudes. Deriving absolute values for his luminosities (surface brightness levels) was not possible. There was no way to go from brightnesses of single star-like images to surface brightnesses.  There were also no reliable, absolute values to derive a zero-point; the surface brightness levels found by Yntema and van Rhijn in their PhD theses (for the latter these were only preliminary results)  were affected by what the former had dubbed `Earth light’ and accurately correcting for this was not possible. Using some very rough guesses for the brightnesses of Earth-light and Galactic light, Pannekoek showed how sensitive the results were to the assumptions. Therefore no reliable calibration of his steps was possible.

The results are in Figs.~\ref{fig:Maps1} and \ref{fig:Maps2}. The upper panel of Fig.~\ref{fig:Maps1}, taken from Pannekoek (1920), shows a drawing representing the impression to the eye of a part of the Milky Way and it is compared in the lower panel to a photographic image taken from the {\it ESO Milky Way Panorama} of the same area. The comparison shows excellent agreement. Fig.~\ref{fig:Maps2} shows the same area after drawing isophotes on the basis of observations by eye. The top panel gives the results obtained by Pannekoek himself and the lower panel after averaging drawings of isophotes  by the four individuals as described below. The contours have been drawn based on the sketches made during visual observing; the numbers are independent and have been used to calibrate the isophotes on the scale of the relative levels using the step method. The numbers in the figures are the brightnesses in the 128 positions with the decimal point deleted. The thick, red isophotes in the top panel correspond to even values of the steps, the odd values are represented by thin, full drawn lines. In the lower panel the shading changes at these four step levels. 

In addition to contour (isophotal) maps Pannekoek produced listings at one degree resolution, the production of which must have been a time consuming business. It would be instructive to turn the numerical listings into a gray-scale representation, but that requires significant effort to type in all the numbers. An approximate and partial way of doing this is in Fig.~\ref{fig:blur}. This gives a gray-scale impression of a quite similar part of the sky from the tabular data as in the previous figures (see the caption for how this was produced). This is a different representation of the same numerical data as the contour diagram in Fig.~15, upper panel, at least for the fainter isophotes, and is more objective then the artist impression of Fig.~\ref{fig:Maps2}, upper panel. Again the agreement with the {\it ESO Panorama} is very good, testifying to the excellent observational capabilities of Pannekoek.
\bigskip

\begin{figure}[t]
\begin{center}
\includegraphics[width=0.94\textwidth]{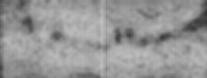}
\end{center}
\caption{\normalsize This gray-scale impression illustrates the numerical measurements of the surface brightness of the Milky Way from visual observations  (Pannekoek, 1920), based on the observations of Pannekoek and Easton. This image was obtained from taking scans of the part of his printed, tabular data (l$^{\rm I}$, b$^{\rm I}$) = (70\degs\ to 350\degs, -15\degs\ to +15\degs), roughly (l$^{\rm II}$, b$^{\rm II}$) = (103\degs\ to 23\degs, -15\degs\ to +15\degs). These scans were then severely smoothed; a gray-scale impression resulted, since  numbers with only one figure (1-9) give light pixels, those with two figures and the first a one (10-19) a somewhat lighter pixel, and numbers with two figures (20 and larger) a darker pixel still. The scans were first joined, oriented properly, the aspect ratio corrected and the display put in inverse video. The tables had a resolution of one degree. Compare to Fig.~\ref{fig:Maps1}, upper panel.}
\label{fig:blur}
\end{figure}

There were a few other studies available that could be used for Pannekoek’s approach of averaging results from various observers. The first possibilities would be Star Atlases which have outlines of the Milky Way in the maps. There are two Pannekoek considered, the {\it  Atlas Coelestis Novus} by Heis (1872), and the {\it Uranometrie Generale} by Houzeau (1878). Eduard Heis (1806—1877) was a German astronomer (and mathematician), working at the K\"onigliche Theologische und Philosophische Akademie, predecessor of the Westf\"alische Wilhelms-Universit\"at M\"unster. He was interested in observing the Milky Way and zodiacal light. Belgian astronomer, meteorologist and journalist Jean-Charles Hippolyte Joseph Houzeau de Lehai (1820–1888), known as bibliographer of astronomical literature and leader of solar eclipse and Venus transition expeditions, was director of the Ko\-nink\-lijke Sterrenwacht van Belgi\"e (Brussels Observatory). The Heis atlas turned out available in the Groningen University Library, but  the  {\it Annales de l'Observatoire Royal de Bruxelles} start only with Vol. 2, so the Houzeau atlas is missing. The Heis Atlas has been digitized by the University of Latvia (see reference list) and the Houzeau catalogue and atlas are available through the {\it the SAO/NASA Astrophysics Data System (ADS)} (see list of references). The Heis atlas certainly has beautiful maps (in my opinion at least), but Pannekoek considered them unusable, because (Pannekoek, 1920, p.11; my translation)
\begmarg
the large uniform areas, the straight or angular lines and the lack of details make the impression as if not too much time was spent on the study of the Milky Way.
\endmarg
Much better were the lithographed maps of Cornelis Easton (1893, 1913, 1928), used to outline Galactic spiral structure. Another set of drawings was published by Boeddicker (1892), of which Groningen University Library owns a copy. Otto Boeddicker (1853–1937) was a German astronomer, who worked most of his career as assistant of Lawrence Parsons, 4th Earl of Rosse at Birr Castle in Ireland. It contains drawings much like in Fig.~\ref{fig:Maps1} (but in photographic negative) from which Pannekoek then must have determined isophotes. There were degradations due to the reproduction process, however, but Pannekoek had the secretary of the Royal Astronomical Society send him the original drawings on loan from Birr-Castle, where they were kept. Then there were drawings in the archives of Johann Friedrich Julius Schmidt (1825--1884) at Potsdam, who spent a significant part of his life in Athens. Schmidt never published them, but Pannekoek (1923b) had arranged with Willem de Sitter for them to be published in the {\it Annalen van de Sterrewacht te Leiden}. Finally a few drawings on small, special areas were available from others. The combined map is shown in the lower panel of Fig.~\ref{fig:Maps2}. Note the level of detail, which corresponds well between both panels. This is not entirely trivial because Pannekoek’s contours in the upper panel contain only half or less of the information on which the lower panel is based.
\bigskip

\begin{figure}[t]
\sidecaption[t]
\includegraphics[width=0.54\textwidth]{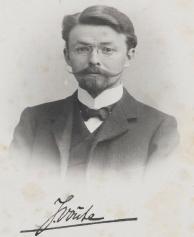}
\caption{\normalsize  Joan George Erardus Gijsbertus Vo\^ute (1879–1963). This photograph comes from the {\it Album Amicorum}, presented to H.G. van de Sande Bakhuyzen on the occasion of his retirement as Professor of Astronomy and Director of Leiden Observatory in 1908 (after Leiden Observatory, 1908). In the public domain under Creative Commons CC.}
\label{fig:Voute}
\end{figure}

This work concerned the northern Milky Way only, and understandably Pannekoek would want to have this extended to the southern part. He had hoped to have the opportunity to visit Johannesburg for some time, where Leiden Observatory had established a collaboration and was setting up a southern station. This did not appear feasible on any short timescale. As described by Tai (2021, sect. 2.1) an extension to the south  was attempted by Josef Hopmann (1890–1975), who worked as an assistant at Bonn Observatory, when in 1922 he went on  a solar eclipse expedition to Christmas Island. He produced a southern map, but Pannekoek found inconsistencies when joining it to his northern data (Pannekoek, 1925) and became skeptical about the large amount of detail in the Milky Way, not realizing that the southern Milky Way  indeed did show much more structure than the northern part. Now Pannekoek had been appointed to the ‘Eclipse Committee’ of the Netherlands Royal Academy of Arts and Sciences (KNAW), which organized expeditions to observe total solar eclipses. This committee was chaired by Willem Julius. When  the latter died in 1925, Pannekoek was appointed to succeed him. At that time the next eclipse expedition went to Sumatra in the Dutch East Indies, where a  complete solar eclipse would be visible on January 14,  1926. This opportunity was gratefully embraced by Pannekoek.

The expedition was a failure, since clouds made observing the eclipse impossible, but Pannekoek stayed until May at the Bosscha Observatory to observe the southern Milky Way. This observatory at Lembang near Bandung on Java had been founded (in fact at the time of Pannekoek's visit  was still being built) by tea-planter, and amateur astronomer Karel Albert Rudolf Bosscha (1865–1928) and  Joan George Erardus Gijsbertus Vo\^ute (1879--1963)  had been appointed director. Vo\^ute (see Fig.~\ref{fig:Voute}) had studied at what is now the Technical University at Delft, but had turned to astronomy afterwards and had worked at the Leiden and Cape Observatories before returning to Java where he was born and raised. Vo\^ute was setting up collaborations with Dutch astronomers and, according to Pannekoek's (1982)  autobiographical notes, had invited him earlier to visit.

\begin{figure}[t]
\begin{center}
\includegraphics[width=0.68\textwidth]{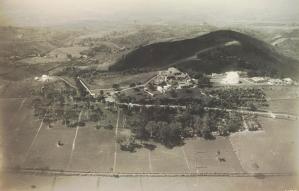}
\includegraphics[width=0.30\textwidth]{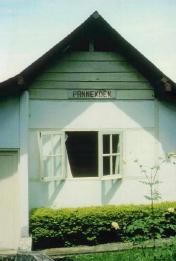}
\end{center}
\caption{\normalsize Left: The Bosscha Observatory at Lembang, Java, Indonesia around 1930. Thisa aerial view is from the northeast. The large dome contains the Zeiss double 60-cm refractor, that has been installed in 1926. Right: The residence of the Pannekoeks during their stay at the Bosscha Observatory in 1926, photographed by Edward P.J. van den Heuvel in 2013. From Wikimedia (commons.wikimedia.org/wiki/File:Bosscha\_sterrenwacht\_te\_Lembang,\_KITLV\_141951) and courtecy E.P.J. van den Heuvel.}
\label{fig:Bosscha}
\end{figure}

It must have taken a while to get acquainted with the southern sky and its constellations.  Observing from the tropics was a new challenge also. I quote from the publication in which the visual mapping of the southern Milky Way was presented (Pannekoek, 1928; p.A6; my translation from the German):
\begmarg
The observing site in Lembang was on the southern edge of the observatory's grounds, right next to our apartment, where there was an unobstructed view of most of the sky. The city lights of Bandung, which was 600 meters lower in the plain at a distance of 10 kilometers, was so disturbing that mats were always hung on the south side to make the light harmless. The summer lightning, never absent on tropical nights, was sometimes so strong that it made observation difficult by the incessant glare; a few times it also offered the advantage that the lightning  betrayed the presence of thin cirrocumuli in apparently clear skies. The southern horizon over the Bandung plain was almost always covered with veils of fog to an altitude of 20\degs\ or more; the few nights when it was clear in that area were a rare exception.

\begin{figure}[t]
\begin{center}
\includegraphics[width=0.88\textwidth]{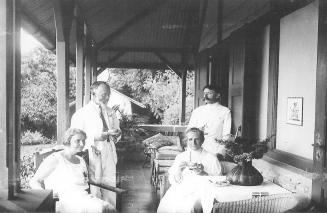}
\end{center}
\caption{\normalsize  Tea at Bosscha Observatory. Standing in the back Pannekoek on the left and Vo\^ute on the right, while the ladies are sitting, Mrs. Vo\^ute on the left and Mrs. Pannekoek on the right. Courtesy Anton Pannekoek Institute.}
\label{fig:panvout}
\end{figure}

In view of the unfavorable circumstances of these observations, the question could be asked whether the value of the result does not lag significantly behind that of the previously explored northern Milky Way. I have come to the conclusion that these unfavorable conditions were more than compensated for by the extensive experience and certainty of the observational method.
\endmarg
Fig.~\ref{fig:Bosscha} shows an aerial view of Bosscha Observatory around 1930 and the house where the Pannekoeks stayed in 1926. The description above that it was on the southern edge implies it probably is seen here somewhere on the ridge behind the large dome. Fig.~\ref{fig:panvout} shows Pannekoek and Vo\^ute and their wives having tea at the Bosscha Observatory.
\bigskip

\begin{figure}[t]
\begin{center}
\includegraphics[width=0.75\textwidth]{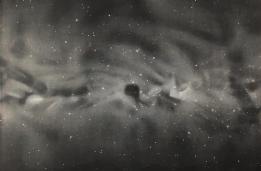}
\includegraphics[width=0.75\textwidth]{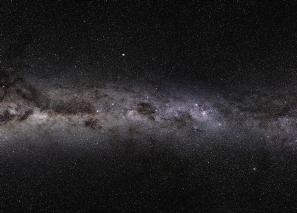}
\end{center}
\end{figure}

\begin{figure}[t]
  \caption{\normalsize    Top panel: The impression of  part of the visual southern Milky Way survey. This is ‘TAFEL II’ in Pannekoek (1928), but after applying inverse video for easy comparison with Fig.~\ref{fig:Maps1}, upper panel. The range is (l$^{\rm I}$, b$^{\rm I}$) = (305\degs\ to 280\degs, -30\degs\ to +30\degs), roughly (l$^{\rm II}$, b$^{\rm II}$) = (338\degs\ to 263\degs, -29\degs\ to +31\degs). Bottom panel: The same part taken from the ESO \textit{Milky Way Panorama} (ESO, 2009).  In order not to overload this caption, some notable  objects have been pointed out in the text. The south celestial pole is near the middle at the bottom edge of both figures.  Credit: ESO/S.Brunier.}
\label{fig:VisS1}
\end{figure}

\begin{figure}[t]
\begin{center}
\includegraphics[width=0.88\textwidth]{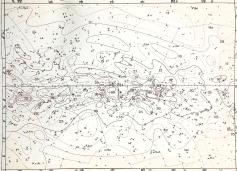}
\end{center}
\caption{\normalsize   The isophote diagram of the area in Fig.~\ref{fig:Maps2} of the southern visual Milky Way survey. This is  `TAFEL V’ in Pannekoek (1928). Thicker contours have been drawn for steps 2.0, 4.0 and 6.0.}
\label{fig:VisS3}
\end{figure}

The results were published in 1928 in the {\it Annalen van de Bosscha Sterrenwacht}. The article was were printed in Amsterdam; in spite of the Dutch title of the series, the paper was not in Dutch, but in German. The presentation was the same as for the one for the northern Milky Way with detailed descriptions of small areas, drawings and isophotal maps, but no one-degree listing of step values. This time Pannekoek divided the survey into three longitude ranges. In order to later make comparisons I take also an area from Pannekoek’s southern visual Milky Way mapping for illustration. Fig.~\ref{fig:VisS1} shows the middle of his three areas. The top panel shows Pannekoek’s visual sketch and the lower panel the same part as it appears in the ESO {\it Milky Way Panorama} (ESO, 2009). It runs from the constellations Ara and Lupus on the left via Centaurus and Crux to Carina and Vela on the right. In the following I mention some noteworthy objects, but in order not to hamper the view of the Milky Way I have not indicated these in the figure itself. The two prominent stars about one third of the extent from the left are $\alpha$ and $\beta$ Centauri (the ‘pointers’) and in the middle one can see the cross-shaped  asterism  -- standing upright -- of the Southern Cross or constellation Crux. The dark patch immediately to the left of Crux is the Coalsack Nebula (sometimes Southern Coalsack), which is at somewhat less than 200 pc from us. On the right of Crux two Galactic clusters are seen, NGC 3532 (sometimes called the ‘Wishing Well cluster’) and slightly to the right and bottom IC 2602 (the ‘Southern Pleiades’). The bright object left of  the middle, well above the Milky Way is the bright galaxy NGC 5128 (the prominent radio source Centaurus A). It was discovered by naked eye in 1826. The bright star at the right well bellow the Milky Way is Canopus ($\alpha$ Carinae).  Fig.~\ref{fig:VisS3} shows the isophotal map of the same area. The units are similar  as before in Fig.~\ref{fig:Maps2}. Without a one-degree listing of the step values I cannot produce a gray-scale representation.

Pannekoek added some more poetic parts, the first one on his experience watching the southern skies, which I will quote in part, while for a guide to constellations I refer back to Fig.~\ref{fig:MWcons} (p.A6; my translation):
\begmarg
Who only knows the parts of the starry sky which are visible in the middle latitudes of Europe, cannot get an idea of the wonderful beauty of the southern Milky Way. Certainly, a dark August night in Europe with the big bright light clouds in Cygnus and the following bay-like band of spots in Aquila and Cassiopeia belongs to the most beautiful impressions of nature in our part of the world. But the splendor of the southern sky is of a completely different order; one would like to associate its higher potency after an incorrect but understandable thought association with the lavish richness of all tropical nature. It is first the much greater brightness which the Milky Way reaches in its southern half. Its light grows gradually from the faint, monotonous glow of the January sky near Orion and Sirius, and in Carina, at 250\degs\ longitude, reaches a brightness that already surpasses the brightest parts of the northern sky. Here the clustering of the most beautiful stars and constellations, the densely scattered groups in the Nave [or Vessel, the constellation Argo Navis, later broken up into Carina, Vela and Puppis], the Southern Cross and the bright Centaurus stars, work together with the irregular bays of the sharply defined light portions of the Milky Way band to create an impressive whole. A little further on a richly articulated series of streams and clouds begins, getting brighter and brighter until they reach a peak in Sagittarius; of this series we still see the last link in Europe in the light patch in Scutum. The brilliance of these bright Sagittarius spots exceeds by far all other formations of the Milky Way; the peculiarity of the impression is still increased by the poverty of these regions at stars visible for the naked eye. In every comparison the contrast is striking, how the Milky Way in Cygnus and in Lacerta is densely covered with scintillating star dust, while the bright Sagittarius clouds appear as if painted with calm luminous paint on a starless sky background.
\endmarg
\bigskip

The second part I want to pay attention to concerns ‘The southern Milky Way in folklore’, and since heritage and cultural anthropology oriented readers of the journal might find this for them possibly less easily accessible piece interesting,  I will quote  it in full  (p.A7; my translation). Fig.~\ref{fig:Oph} is meant to serve as a guide to the stars mentioned.
\begmarg
Considering the conspicuousness of the structure in the southern Milky Way, it is no wonder that  it has also attracted the attention of the original inhabitants of the southern countries. Among the Javanese (as well as several other peoples of the Indies) the Milky Way is called a celestial river, the coal sack a fish basket or cage in this river. Among the constellations and asterisms, which are given special names, we find several sources describing the stars known among different peoples, among others, and Scorpius is called ‘the fighting quails’. There is nothing to be seen at the relatively faint stars of 4th magnitude, which could explain this name; but if one sees, how they both lie as fine double stars, just visible to the naked eye, embedded at the edge of two small Milky Way clouds like two little eyes in a ball of down, the name becomes immediately understandable.

	An even stranger constellation is found in the Milky Way. In the lists of identifications of Malayan constellations with European ones by Alfred Maass [a German ethnologist] Bima-Sakti is translated as ‘Milky Way’. But this identification is correct only in the sense that this image lies in the Milky Way. During a visit to the observatory in Lembang, the well-known archaeological researcher Dr. van Stein Callenfels explained to  me how the Javanese see in the dark spots in Ophiuchus and Scorpius the figure of the giant Bhima. Bhima (or Werkosari), the helper of the gods (hence called sakti, the holy one), who fights the Korawa in the Pandana camp, is the most popular figure in the Javanese wajang play, and known to every child here. The black spot south of $\theta$ Ophiuchi is his face, the dark wedge-shaped stripe directed W is his nose, $\theta$ Oph may well signify his eye, the dark stripes going to the N and to the E and then bending together and merging with each other form his head of hair. His body is formed by the dark masses extending towards $\lambda$ Scorpii, its arm by the narrow dark stripe directed towards Messier 7. One leg between $\epsilon$ and $\mu$  Scorpii stops there; $\mu_1 \mu_2$  Scorpii  are the eyes of the snake (to which further belong the stars $\zeta \eta \theta$ Scorpii, $\alpha \beta \gamma$ Arae) which it has eaten up so far; the other is to be looked for in the dark places between $\mu$ and $\theta$  Sco, and those between $\eta$ Sco and $\epsilon$  Arae. While otherwise the figures of men and gods, which the peoples of the north formed in the sky, always include groups of stars, here we have the case of a constellation formed from the dark Milky Way nebulae.
\endmarg
        
One of the referees suggested that the ethnographical study referred to by German ethnologist Alfred Maass (1863--1931) is that published as a collection in of articles in the Dutch journal for linguistics, geography and ethnography (Maass, 1926). Maa\ss, as he spelled his name, had made an extensively journey  through Sumatra.
\bigskip

\begin{figure}[t]
\begin{center}
\includegraphics[width=0.86\textwidth]{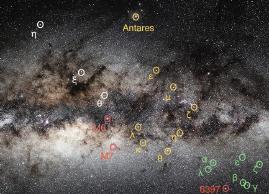}
\end{center}
\caption{\normalsize  The constellations Ophiuchus to Ares from the ESO {\it Milky Way Panorama} (ESO, 2009). Stars in Ophiuchus are in white, in Scorpius in yellow and in Ara in green. Note that $\mu$ and $\zeta$ Scorpii are in fact double, consisting of two naked eye stars, but this is not visible here; $\mu$ is a binary system, $\zeta$ is made up of two unrelated stars.Messier 6 is the (open) ‘Butterfly' cluster, Messier 7 also an open cluster, and NGC 6397 is a globular cluster. This serves to guide the reader towards the stars used by Pannekoek in the text to describe the Javanese constellation Bhima. Credit: ESO/S. Brunier.}
\label{fig:Oph}
\end{figure}

Not long after the publication of the northern Milky Way maps, van Rhijn (1921b) published his results on the sky brightness measurements at Mount Wilson that he had obtained as  part of his PhD thesis. This made it possible to better estimate the absolute surface brightness scale (Pannekoek, 1921c). Although this determination was performed using data in the northern sky, this should apply to the southern data as well, since these had been tied together in the overlap regions. Pannekoek expressed his surface brightnesses as the number of stars of apparent magnitude 0.0 per square degree. It turned out that his step sizes changed from 0.35 magnitudes between step 1 and step 2 to 0.21 between step 5 and step 6, with a mean of 0.27 magnitudes, so somewhat smaller than what he had found in the 1920-paper of 0.30 to 0.37. I will use this below for a comparison to modern observations.

It is worth mentioning that Pannekoek’s unusually sharp eye-sight caused him to notice that there were distinct color differences in the Milky Way. He commented on this only much later in a short publication in {\it Observatory} (Pannekoek, 1953), noting that there had been no time to follow this up.

\section{Photographic surface photometry of the Milky Way}

Going from visual observations to photographic plates was an obvious next approach. This can be seen as a way to both be more quantitative and go fainter. The first, however,  was not among Pannekoek’s objectives. Tai  (1921, p.62) explained:
\begmarg
    […] the eyes and the mind processed the light of many faint stars into a coherent image, which in turn could be used for further scientific research, for example in statistical astronomy. As he [Pannekoek] explained in his Marxist philosophy, usefulness, not truth, was his main criterium for scientific knowledge. The Milky Way image may have been a human construct, but then so were all scientific laws.

    Because the Milky Way was intangible, many different representational methods were needed to capture all of its features. Pannekoek’s depictions of the Milky Way ranged from naturalistic drawings and verbal descriptions to isophotic diagrams and numerical tables of surface brightness. [...]

According to Pannekoek […] photography was inherently incapable of representing the Milky Way without human intervention. Before photography could produce scientific results, meaurement and expert judgement were required from the astronomer. The drawings that resulted from this critical engagement with photography were not the result of nature unveiling itself, but constructed images highlighting the structure of the system. Photography, in this case, replaced visual observation, but not drawing [of isophote maps].
\endmarg

The use of photographic plates had already been the subject of Pannekoek (1912), but this concerned obtaining improved star counts and thus involved photometry of stellar objects. Photometry of stars on photographic plates of course went back long before this, in fact to the 1850s (see Hearnshaw, 1996, chapter 4) and had been extensively used by Kapteyn and Gill in the {\it Cape Photographic Durchmusterung} and in an even more extended application in the {\it Astrographic Catalogue} associated with the {\it Carte du Ciel} effort. Pannekoek had taken up the idea to use out-of-focus plates to spread the light of stars over a larger area, thinking that if the areas taken up by bright stars was limited, enough area would be available to measure the surface brightness of the Milky Way and the photometry could actually be calibrated this way using the surface brightness of the out-of-focus stellar images.

Photographic photometry of stars was usually done by measuring the diameter of the stellar image on the emulsion, calibrating this with stars of known apparent magnitude on the same plate. Surface brightness is another matter, since then one would need to determine in small areas the amount of blackening of the emulsion  as a function of the total incident light, in modern terms the characteristic curve relating photographic density with exposure (intensity times exposure time). `Blackening’, as used by Pannekoek and others before him, is identical to the more recent photographic density, which is minus the logarithm of the fraction of light shone on the emulsion that is not absorbed by the emulsion  and observed on the other side. Much of the technique of using the ‘blackening curve’ is due to Karl Schwarzschild (see Fig.~\ref{fig:Scharzschild}) from the period he worked at at Kuffner Observatory in Vienna (1896--1899), just after having obtained his PhD (see Hearnshaw, 1996, section 4.8), including the use of out-of-focus plates to spread the light of stars over an extended area on the emulsion to avoid over-exposure. 

The problem that arose is what is now referred to as low-intensity reciprocity failure (LIRF), the fact that the photographic density does not scale linearly with the product of the intensity and the exposure time down to faint levels  (the reciprocity also fails at very bright levels, but that is not relevant in astronomy). This effect has later been  particularly important in  photographic surface photometry of galaxies. Hearnshaw (section 4.8)  has documented that it had been noted as far back as 1890 by Edward Pickering and others, but it had only been studied in detail by Schwarzschild (1900) in relation to his development of out-of-focus stellar photometry. According to Habison (1999; abstract only) Schwarzschild claimed an accuracy of 0.05 magnitudes in stellar photometry with out-of-focus plates.

The idea to perform photographic surface photometry of the Milky Way had come to Pannekoek even before his visual survey of the northern part had been published (which was in 1923). In a preliminary investigation to explore the possibilities and practicalities (Pannekoek, 1923c) he mentioned that he had put forward the idea to Max Wolf in 1919, enlisting his help to obtain the photographic plates. In his autobiographical notes (Pannekoek, 1982) he noted that he had discussed it with Kapteyn, who thought it was a good idea. This development has been further documented, including references and discussion of letters between Pannekoek and Wolf, in Tai (2022, section 1.3; Tai in fact described there Pannekoek’s work on Milky Way mapping in great detail). Briefly, the idea is to take photographic plates out of focus so that the light of a star is distributed over a larger area and therefore the emulsion nowhere becomes overexposed. In this way surface brightness could be calibrated from observations of stars, if the brightness distribution of the Milky Way could still be discerned.

\begin{figure}[t]
\sidecaption[t]
\includegraphics[width=0.56\textwidth]{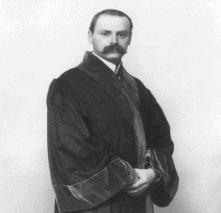}
\caption{\normalsize    Karl Schwarzschild (1873–1916) in academic attire at an unknown date. He is appearing relatively young; this picture might have been taken early on during the period during which he was professor at G\"ottingen University and director of the Observatory, which was between 1901 and 1909. Photograph in the public domain in Wikimedia (commons.wikimedia.org/wiki/ File:Schwarzschild\_in\_academic\_robe.gif).}
\label{fig:Scharzschild}
\end{figure}

The measurement of the plates was performed by Pannekoek with a photometer developed by Johannes Franz Hartmann (1865--1936) at Potsdam (see Hartmann, 1899). This calibrated the photographic density by comparison to a wedge, which is an exposure, `to be made on the same kind of plate as the plate to be measured, if possible’, produced with `a blackening which increases as uniformly as possible’. The optical system employed a double prism with a small part of a side of one prism silvered and made reflecting before cementing it to the other prism, all in such a way that the object on the plate and a part of the wedge were viewed next to each other without any dividing line. This means that a density on the plate could be measured accurately by determining where it was situated along the wedge. This, one should note, calibrated the measurement of photographic density and not the relation between density and exposure (and hence surface brightness). The latter could be done by comparing the (average) density in an out-of-focus stellar images (or the difference with the background) for stars of known apparent magnitude. Pannekoek then calculated for such a star the surface brightness when the light was distributed over the area of the out-of-focus image. 

Because of the similarity of Milky Way photometry to that of external galaxies it is of interest to compare Pannekoek’s procedure to more modern photographic surface photometry, which has been quite an industry in the 1970s and 1980s until CCDs (charge-coupled devices) came into fashion. For a brief history of Galaxy surface photometry see the review by de Vaucouleurs (1979). The first attempt came in 1913 in the central part of the Andromeda Nebula (Reynolds, 1913). He used an ingenious set-up to construct a characteristic curve. In the 1970s and early 1980s (when I myself performed photographic surface photometry of galaxies), this was all done from in-focus plates. One used, for determining the characteristic curve, a wedge which when shining light through it produced a linearly changing intensity along it, or a device that produced a set of sensitometer spots of known intensity ratio. This then was projected and exposed on an unexposed part of each individual plate right after sky-exposure, since characteristic curves may vary significantly  between plates and conditions. This then had to be done for about the same exposure time as on the sky to eliminate effects of LIRF. Pannekoek used a separate plate (‘the same kind of plate as the plate to be measured, if possible’) for determining the characteristic curve, which in galaxy surface photometry would have been considered definitely not sufficiently careful. It should be noted that those more modern emulsions were especially treated such as by `baking’ them for hours in nitrogen or forming gas at 65\degs C. Another difference is that this way only a relative   surface brightness distribution resulted of which a zero-point had to be derived separately from comparison to aperture photometry (or the difference between two diaphragms since galaxy centers were overexposed). There are major differences with Pannekoek's kind of investigation; not determining the characteristic curve on the same plate would now seem imprudent, but was apparently normal practice in Pannekoek’s time.
\bigskip

\begin{figure}[t]
\begin{center}
\includegraphics[width=0.90\textwidth]{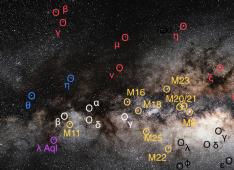}
\end{center}
\caption{\normalsize The Scutum Cloud and the center area of the Galaxy from the \textit{ESO Milky Way Panorama} (ESO, 2009). Stars in Scutum are in white, in Ophiuchus in red, Sagittarius (only part of the `tea-pot’) in black, and Serpens Cauda in blue. Messier objects are in yellow. The star $\lambda$ Aquilae, referred to in the text, is in purple. Most Messier objects (in yellow) are seen also in Fig.~\ref{fig:Sagcloud}; not included there are Galactic clusters M11 and M18 and the star-forming HII region M16 (Eagle Nebula). The figure covers longitudes from about 357\degs\ to 35\degs. Credit: ESO/S. Brunier.}
\label{fig:innerG}
\end{figure}

Now Pannekoek obtained a first set of plates taken for him by Max Wolf in Heidelberg and the results of this preliminary program were published as Pannekoek (1923c).  The position of the plate holder relative to the focus was such that stars produced images with a diameter of half a degree so that naked-eye stars would not overlap. Away from the center these became elliptical. The field studied was the Scutum Cloud, which can be seen in Fig.~\ref{fig:innerG} at the lower left between $\alpha$, $\beta$ and $\delta$ Scuti (see also section 7 and Fig.~\ref{fig:Scutum} in there below). The set of plates were centered on the star $\lambda$ Aquilae, also indicated in that figure. The exposure of the plate used in the end was for over three hours (190 minutes). The measurement of a star involved that of the photographic density with a Hartmann photometer and the Amsterdam Institute had one that was obtained from the Berlin Bamberg Observatory. It included a photographic wedge from Hertzsprung (who by 1923 was in Leiden, but had made it when he was still at Potsdam with Schwarzschild). This wedge had been produced with a linearly increasing exposure time along its length, which Pannekoek had calibrated by putting different `shades’ in front of it and finding positions with the same
difference in blackness. 

For the conversion to magnitudes, Pannekoek adopted a quadratic function in blackness, referring for justification to Hertzsprung (1911d). This paper is a correction of a study of a variable star. In the earlier publication Hertzsprung had interchanged the variables magnitude and blackness. It is {\it A.N.} 4526, which actually consists of two consecutive papers (Hertzsprung, 1911b and 1911c), and Hertzsprung (1911d) corrects the second of these. In these papers Hertzsprung studied the variable star 68 {\it u} Herculi, where the prefix 68 refers to the Flamsteed and {\it u} is to the Bayer designation. It has an apparent mean magnitude of about 5.0 and Hertzsprung worked with plates exposed for 25 minutes or so.  Astronomers did not worry much then about plate-to-plate variations in the emulsion or LIRF.

The question remains how Pannekoek used the out-of-focus images of the stars to go from stellar magnitudes  to surface brightness. This is not straightforward; in Pannekoek (1923c) the light from the star is distributed over an area with a radius found to be 0\degspt325, but it is not uniformly distributed. From mapping -- `measures of a great many points’ -- of the blackness over the disk, produced by the brightest star in the field $\lambda$ Aquilae, corrected for the underlying smoothly distributed light from neighboring points, the relation between the blackness in the center of the disk and the surface brightness, expressed as the number of 0.0 magnitude stars per square degree, could be calibrated. 
\bigskip

Pannekoek did a further thing. He compared his photographic surface brightnesses in the Scutum cloud to his visual map of the northern Milky Way (Pannekoek, 1920). This would provide an estimate of the color index of the Milky Way in the Scutum Cloud. The sensitivity of photographic emulsion and the human eye are broad-band in wavelength, but center around different wavelengths. The response of the photographic emulsions used at the time was mainly in the blue and violet part of the spectrum and that of the eye has a peak somewhere in the yellow. Pannekoek compared his photographic contours with those in the visual using  van Rhijn’s (1921b) Mount Wilson observations as a zero-point. This gave for the Scutum Cloud a color index of 0.43 magnitudes (the second decimal is surely insignificant, but adding more decimals than warranted was common practice), which is that of an F5 star. Discussion of uncertainties resulted in a F0-G0 range.

I will come back to the color index later, but will address the spectral class issue here. Pannekoek noted that his result indicated a somewhat earlier type than determined spectroscopically by Fath (1912). The latter used a spectrograph originally built by him to study zodiacal light (Fath, 1909). It was used from Mount Wilson; an initial exposure on the brightest Galactic area between $\gamma$, $\delta$ and $\lambda$ Sagitarii (the Sagittarius Cloud of Pannekoek, 1927; see Fig.~\ref{fig:innerG}) of some 30 hours proved insufficient, but a second one of about 65 hours showed the stellar absorption lines. A further spectrum of almost 68 hours was obtained on the Scutum Cloud and a final spectrum of about 74 hours was obtained in the area between  $\beta$ and $\eta$ Cygni (see Fig.~\ref{fig:CygAqu}; $\eta$ Cygni is about halfway between $\beta$ and $\gamma$). All these indicated a solar, G-type spectrum, which made one suspect it was severely contaminated by zodiacal light. The spectra illustrated in Fath's paper showed the Fraunhofer lines H \&\ K, G and F, so roughly 4000 \AA\ to  4800 \AA.

Fath had expected a much earlier spectral type because of the more than 30,000 spectral types determined at Harvard about half were A-type. To validate his result he studied star fields with the Mount Wilson 60-inch Telescope to show that fainter stars in these fields are generally redder than the bright ones. This was done following a suggestion by Kapteyn, who had designed it to show that distant, fainter stars are redder as a result of interstellar extinction. The method involved multiple exposures and covering of different part of the photographic plate. Indeed Fath found that fainter stars are on average redder, which he attributed then to an increasing contribution of later spectral types rather than interstellar extinction. Kapteyn had been wary of this effect that would be an alternative interpretation of his discovery (Kapteyn, 1909a, but see 1909b)  of reddening and possible `absorption’ of light in space (for a discussion of this see van der Kruit, 2015, section 11.9).

I note that the integrated light of galactic disks is dominated by late G- and early K-giants, enabling the observation of stellar kinematics from the Mg-b lines around 5180 \AA. For the observations of stellar kinematics in disks of spiral galaxies the spectra of late G- or early K-giants serve as excellent templates to measure the radial velocity and velocity dispersion. For example, even spectra of the disk of face-on galaxies with strong spiral structure (van der Kruit \&\ Freeman, 1986) prominently show the narrow lines of this Mg-b triplet, which means that G- and K-type giants contribute substantially to the integrated light (for a graphical representation, see Fig.~2 and 4 in van der Kruit \&\ Freeman, 1984). In the latter figure also the Fraunhofer E-line of neutral iron (at about 5270 \AA) can be seen on the right, as expected for such giants of similar strength as the b lines. The spectrum of a galactic disk and therefore of the Milky Way may be expected to show an absorption line spectrum of a G- or K-type star. So, Pannekoek’s result does not hold up.
\bigskip

The positive experience resulted  a complete photographic survey of the northern Milky Way, for which Max Wolf  took the plates in Heidelberg (for more details again see Tai, 2022, section 1.3). The result was published as the third volume of the {\it Publications of the Astronomical Institute of the University of Amsterdam} (Pannekoek, 1933). It was not an insignificant commitment of telescope time (it had to be done during the dark of the Moon), eventually resulting in the measurement and reduction of 37 plates with exposure times 2.5 to 6 hours, during 1921 up to 1928. The distance from the focus was not uniformly chosen the same; the scale on the plates was slightly different depending on the out-of-focus positions. Remarkable is that a whole range of emulsions (9 to be precise) and developers were employed. The square plates measured 8 $\times$ 8 cm on a side; Pannekoek used the central  6 cm or a square of about 22\degs\ on the sky on a side. He worked with three types of plates with different distances from the focus, and the diameters of the stellar images were 0.56, 0.66 and 0.78 degrees in diameter (circular in the center, elongated by a factor 1.2 in the tangential direction at the largest distance from the center). The non-uniformity of the observational material is quite significant,  which would raise some doubts as to the confidence in the procedure and its outcome.

Measurements were done as for the Scutum Cloud employing the Hartmann microphotometer and the Hertzsprung wedge. A first problem was that the illumination changed with distance from the center, as Pannekoek explained, because light at the edges reaches the emulsion at an oblique angle and therefore the illumination decreases from the center outwards. Correcting for this effect is important. The plates showed insufficient overlap to use for a determination of this effect (but some were later used for a consistency check), so he used plates that did not have bright Galactic clouds near their centers. This remained problematic because all his plates have Galactic light in them, but he hoped the averaging of 15 of the  plates would still yield the required correction sufficiently accurately. It is a purely geometrical phenomenon, so it would seem that the correction could be determined from considerations of the geometry involved, but Pannekoek did not address this. Pannekoek himself was not very confident, it seems, since he wrote (p.22):
\begmarg
Since the majority of the plates has the center in the brighter areas of the Milky Way, it would be incorrect to take the simple mean of all plates; the calculated course would be then surely too large. Therefore, by excluding a number of plates with bright clouds close to the center, an attempt was made to create a kind of balance between plates with probably too strong and too weak progressions. According to this consideration only 15 plates […] were used for averaging. Of course, it always remains uncertain to what extent this goal was actually achieved.
\endmarg
The illumination came out empirically with the tenth(!) power of the cosine of the angular distance from the center, which seems much more than geometry would predict. Pannekoek made no remark on this. The focal distance is 14.9 cm (the aperture is 33 mm and the focal ratio f/4.5). On the square parts used on plates of 6 by 6 cm, the corner is 4.2 cm from the optical axis, so the obliquety is of order  16\degs, of which the cosine is 0.96 and the tenth power of that 0.67, so the required correction is not minor. 

The area surveyed covered longitudes between 350\degs\ and 185\degs\ (new coordinates 23\degs\ and 218\degs) and latitudes between +16\degs\ and -16\degs. The overlap between the 37 plates actually is considerable (see his Fig.~7).

\begin{figure}[t]
\sidecaption[t]
\includegraphics[width=0.46\textwidth]{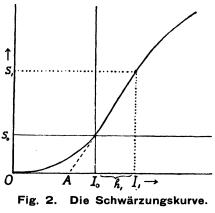}
\caption{\normalsize     Pannekoek’s treatment of the determination of surface brightness from photographic density involved using the `blackening curve' or in modern terms the characteristic curve, which relates blackening or photographic density to the brightness or exposure. For further explanation see the text. From Pannekoek(1933).}
\label{fig:charcur}
\end{figure}

The calibration was done as before. The central photographic density of stellar images minus that of the immediate surroundings was corrected to the average across the image by detailed mapping of some stars and then the curve that related the density to the surface brightness was determined independently for each plate. This requires some explanation. For this I copy Pannekoek’s Fig.~2 in Fig.~\ref{fig:charcur}, which is a schematic drawing of a characteristic curve. Say, the density on the stellar disk is $S_1$ and just next to it $S_0$ and the calculated surface brightness $h_1$ from the stellar magnitude, and it is assumed that the exposure is on the linear part of the characteristic curve. Then this can be used to construct a part of the curve, and repeating that for other stars in the field would yield a good determination. Pannekoek pointed out that the part of the curve  below the point ($I_0$,$S_0$) cannot be determined. The surface brightness of the Milky Way (still to be corrected for other light sources in the background sky) would follow from the distance $I_0$ from the origin $O$, but this position is unknown. The best approximation  Pannekoek could think of is the distance from $I_0$ to the extrapolation point $A$. He noted that (p.8): 
\begmarg
.... to find the general brightness of the sky background itself would require photographs with blackening marks on an unexposed part of the plate, which would allow the scale of blackening and brightnesses to be extended to zero. For our goal, the determination of the relatively small variations of the brightnesses over the sky, it is of no importance whether the brightnesses are counted from O or from an unknown A. It follows at the same time that any uncertainty in the value, subtracted everywhere,  for the unexposed plate is equally unimportant.
\endmarg
Even if it had been possible to obtain a zero-point for the photometric scale, there still is the problem that (p.7)
\begmarg
.... the sky background, which is composed of earth light (aurora), zodiacal light, scattered star light and Milky Way light, and whose amount is unknown.
\endmarg

\begin{figure}[t]
\begin{center}
\includegraphics[width=0.48\textwidth]{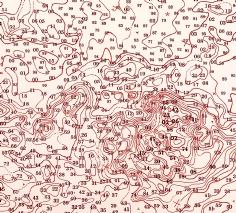}
\includegraphics[width=0.48\textwidth]{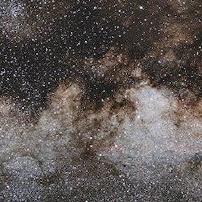}
\end{center}
\caption{\normalsize   Part of the contour maps of Pannekoek (1933) compared to the  ESO {\it Milky Way Panorama} (ESO, 2009). The bright feature is the Scutum Cloud. The region covered is approximately 10\degs\ on a side and centered at ($l^{\rm I}$, $b^{\rm I}$) = (355\degs, -5\degs) or ($l^{\rm II}$, $b^{\rm II}$) = (28\degs, -5\degs). Credit: ESO/S. Brunier.}
\label{fig:Scutum}
\end{figure}

\noindent
So, although he gave surface brightnesses in absolute units (in his words (p.7):
\begmarg
Unit is the light of a $10^m$ star spread over one square degree, thus 0.0001 times the brightness of a $0^m$ star per square degree.’,
\endmarg
\noindent
it has an uncertain zero point and is not corrected for other contributions. But then, Pannekoek stated he is primarily interested in the structure in the Milky Way or the {\it variations} across the sky.
\bigskip

The end product is presented as a set of eight charts with observed surface brightnesses and contours, covering together a strip from new longitude 180\degs\ to 350\degs. These maps contain contours (isophotes) drawn every 10 units, where the lowest values are well below 100 and increase to over 250 in the units mentioned above. As a result the maps are overloaded with contour lines. Fig.~\ref{fig:Scutum} shows the Scutum Cloud, one of the brightest parts of the Milky Way, compared to the same part of the ESO {\it Milky Way Panorama}. The isophotal map is difficult to read because of the large number of contours (half to a third of the isophotes would have sufficed). Close inspection shows that there is an excellent correspondence between the isophotes drawn and the photograph. Even rather small features can be recognized in the isophotes. This is important to note since this is measured as a background in an image with many out of focus stars that cover a not insignificant part of it. The unique things is that this apparently works even in bright parts of the Milky Way where star fields are relatively crowded.

\begin{figure}[t]
\begin{center}
\includegraphics[width=0.98\textwidth]{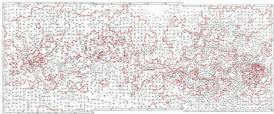}
\end{center}
\caption{\normalsize Isophotal map of the same region in the Milky Way as Figs.\ref{fig:Maps1} and \ref{fig:Maps2} from the photographic survey of the northern Milky Way (Pannekoek, 1933). It runs in old coordinates from $l^{\rm I}$ = 65\degs\ to 350\degs\ , $b^{\rm I}$ = -15\degs\ to +15\degs. Using the conversion tables in Lund Observatory (1961), the four corners are rounded off clockwise at ($l^{\rm II}$,$b^{\rm II}$) respectively (98\degs,+16\degs), (97\degs,-14\degs), (23\degs,+15\degs), (22\degs,-15\degs),(97\degs,-14\degs). The unit is number of stars of magnitude 10 per square degree. To reveal the structure of the isophotes I have emphasized the isophotes at values that are a multiple of 50 in these units with thick, red lines.}
\label{fig:PhotN1}
\end{figure}

To give an impression of the maps on a larger scale I show in Fig.~\ref{fig:PhotN1} the same area from the northern photographic survey as in Figs.~\ref{fig:Maps1}, \ref{fig:Maps2} and \ref{fig:blur}. For this I had to join three individual figures as published in Pannekoek (1933) to generate the same longitude range. The correspondence of the isophotes at the line where these separate figures join is not always perfect, but this may be due at least in part to the printing and reproduction process. To make the presentation appear more orderly I have thickly drawn in red only isophotes every 50 units, so that four out of five of the isophotes have disappeared. Again the comparison with the visual images in Fig.~\ref{fig:Maps1} is very good.
\bigskip

\begin{figure}[t]
\begin{center}
\includegraphics[width=0.98\textwidth]{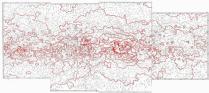}
\caption{\normalsize  The same region of Figs.~\ref{fig:VisS1} and  \ref{fig:VisS3} in the southern Milky Way is represented here in an isophotal map from Pannekoek (1949) by joining a few separate maps (Plates III to IX) together. The unit is number of stars of magnitude 10 per square degree. To reveal the structure in the isophotes I have emphasized those at values that are a multiple of 50 in these units with thick, red lines. The coordinates of the corners are (l$^{\rm I}$, b$^{\rm I}$) clockwise (307\degs, +19\degs), (218\degs, +13\degs), (218\degs, -17\degs) and (307\degs, -19\degs), rounded off (l$^{\rm II}$, b$^{\rm II}$) (274\degs, +20\degs), (185\degs, +14\degs), (185\degs, -18\degs), (274\degs, -20\degs).}
\label{fig:PhotS}
\end{center}
\end{figure}

Obviously Pannekoek wished to complement the survey also to include the southern Milky Way.  When he visited the Bosscha Observatory in 1926, the work on the northern photographic survey was already underway; the preliminary study of the Scutum Cloud had already  appeared  (Pannekoek, 1923c) and the taking of the plates had started. So Pannekoek had a photographic survey of the southern Milky Way as much on his mind as the visual one that he performed during this stay and experimented with two cameras that were  available. The aperture was larger than in the north, so the exposure times less but the field-of-view more restricted. The taking of the survey plates started in 1926, but proceeded very slowly due to other work and only by 1939 reached some level of completeness north of declination -55\degs\ or so. The problem with Lembang is that its latitude (7\degs\ south) is not far enough south for the southernmost part of the Milky Way. The plan had been taken up that when the director Joan Vo\^ute would retire (probably in the 1940s; he would turn 65 in 1944) he would move to Johannesburg, take the camera along and complete the program from there. But then the Second World War intervened. During the Japanese occupation Vo\^ute had been protected from being taken prisoner by the Japanese director, who had argued he needed him to run the Observatory. This is remarkable since Vo\^ute had stepped down in 1939 and Aernout de Sitter (1905–1944), son of Willem de Sitter, had become the director. De Sitter and another Dutch staff member, Willem Christiaan Martin (1910–1945), died in Japanese prisoner camps. Vo\^ute after the War moved via Australia back to the Netherlands. Pannekoek arranged in 1940 with Shapley that the remaining plates would be taken at Harvard’s Boyden station, which in 1927 had moved to Mazelspoort, Bloemfontein in South-Africa. This had been completed in 1946 (for more details again see Tai, 2022, section 1.3).

The measuring and reduction of the plates was mostly performed by David Koelbloed (1905--1977). He had been hired as computer and assistant in 1921 at age 16. He eventually completed physics and astronomy studies in Amsterdam, defending a PhD thesis in 1953 on {\it Line spectra of some giant and dwarf K-type stars}; Herman Zanstra (1894—1972) had succeeded Pannekoek in 1946 and acted as supervisor. Koelbloed was eventually promoted to lecturer. The survey was published  by Pannekoek \&\ Koelbloed (1949). The project had taken 95 plates to complete, but over much of the longitudes  in latitude extending to -/+ 20\degs\ (the northern part extended generally to -/+ 10\degs). The overlap with the northern photographic survey was used to unify the magnitude scales. In Fig.~\ref{fig:PhotS} I show part of the isophotal maps for the same longitude interval as in Fig.~18 and 19. Again because of the excessive number of contours I have traced a selection (every 50 in the same units as before) with red lines.
\bigskip

The conclusion is that in this work Pannekoek has been successful in providing isophotal maps of the Milky Way using the technique of out-of-focus photographic plates. This is highly commendable, but it should also be said that in this kind of work getting the shape of the isophotes right is not the most difficult part; getting the calibration correct is much more difficult. I turn to that in the next section.

\section{Comparison to Bochum and Pioneer}

There was not much material for Pannekoek to compare his maps and his photometric calibration to. He was interested in calibrated measurements of the appearance of the Milky Way, which is different from, although related to, studies of the integrated starlight. This goes back to Kapteyn’s (1906) {\it Plan of Selected Areas}, in which he proposed  to measure this to constrain models for the distribution of stars in the Sidereal System. It resulted in the PhD theses already referred to by Yntema (1909) and van Rhijn (1915, see also 1919). Yntema found evidence for  ‘earth light’ or airglow, and it subsequently became known that measurements of the sky brightness in addition have contributions from zodiacal light, diffuse Galactic light and  extragalactic background light. The last of these is small and diffuse Galactic light is of no concern here, since it was part of Pannekoek’s observations as well. Zodiacal light could have compromised the Pannekoek surveys, although he did try to avoid contamination by it. But no quantitative work was available on this at the time he performed this work

Few photometric studies like Pannekoek’s have been undertaken since he produced his isophotal maps. There are two that are sufficiently extensive to consider; the first is the extensive mapping program of the southern Milky Way by Theodor Schmidt-Kaler and his research group at the Astronomisches Institut der Ruhr-Universit\"at Bochum.They used the Bochum Super-Wide-Angle Camera (Schlosser and Schmidt-Kaler, 1977), which has a concave, spherical mirror of 30 cm diameter. This half-sphere sits on the ground looking up to a camera on a tripod positioned on top of it so that the camera can look down on it and can see the sky in principle from horizon to horizon. This set-up was actually pioneered at Yerkes Observatory by Louis George Henyey (1910--1970) and Jesse Leonard Greenstein (1909--2002) and their camera was used to make very well-known wide-angle photographs of the Milky Way by Osterbrock and Sharpless (1951) and by Code and Houck (1955). The Bochum exposures, taken at the ESO La Silla Observatory in Chile, were measured and reduced to maps that cover a field of view of 145\degs\ and have a resolution of 0\degspt25.  Bright stars up to about magnitudes 7.5 in the color band concerned, which were U, B, V and R,  were deleted  from the maps. The survey is restricted the the southern Milky Way, except in one band in the north but that is U and provides no useful data to compare Pannekoek’s work to. The publications that I will use  are Seidensticker {\it et al.} (1982), which is a UBVR study of the southern Coalsack, Kimeswenger {\it et al.}, (1993), mapping of the southern Milky Way in the B-band, and Hoffmann {\it et al.}, (1998), the same in U, V and R. I will not use the U-band data.

These observations are corrected using standard procedures as good as possible for airglow and zodiacal light, as well as  for extinction in the atmosphere. It is not easy to do this completely reliably unless one goes to space to get rid of the earth-bound contributions. Still, that does not solve the problem of the zodiacal light, which is observable along the full ecliptic and results from scattering of light from the Sun by interplanetary dust. Looking far away from the Sun does not necessarily help, since there is a secondary maximum in the form of the Gegenschein opposite the Sun, which results from back-scattering. The distribution of the dust in interplanetary space is and has been an intense subject of study. The fortunate situation is that here we encounter the reverse of that of the Galactic astronomer studying the Milky Way, namely these researchers would want to accurately map the zodiacal light, but need to  correct their observations for annoying contributions from the Milky Way background.

In the 1970s this gave rise to the the Background Starlight Experiment performed using the Imaging Photopolarimeters aboard the Pioneer 10 and 11 spacecraft that visited the planet  Jupiter (and also Saturn in Pioneer 11’s case). In the following I am concerned only with data from Pioneer 10. This instrument (Principal Investigator was Anton Marie Jacob (Tom) Gehrels (1925--2011), a Dutch-born specialist on asteroids and planets) was designed to scan the image of Jupiter (and Saturn) during encounter. But it was also used to scan the sky  extensively during the spacecraft’s voyage to Jupiter, while it slowly rotated around its axis, to study zodiacal light and dust distribution in the Solar System. When the zodiacal light became negligible at distances beyond 3 A.U. from the Sun, it has been used in this mode to map the undisturbed Galactic background to be able to correct observations of zodiacal light from Earth or spacecraft near to it for the Milky Way contribution. The solar distance of Jupiter is just over 5 AU and 3 AU is roughly the outer radius of the the main asteroid belt, so most of the time to cross this distance was used to build up maps of the Galactic background.

Particularly the results from Pioneer 10 were reduced, analyzed and published, notably  in the PhD thesis of Toller (1981). I became aware of this when I happened to see isophote maps of the Milky Way in  a {\it Sky \&\ Telescope} article by Weinberg (1981). After contacting Weinberg and Toller I was sent a computer print-out with a listing of the reduced full-sky surface brightness in two optical bands at a 2-degree resolution. And I used it to make a model of the stellar luminosity distribution in the disk of our Galaxy (van der Kruit, 1986), similarly to what I had done in extragalactic edge-on spiral galaxies (van der Kruit \&\ Searle, 1981,1982, and further papers in that series). Having this print-out still in my archives after forty years or so, I can make a comparison to Pannekoek’s maps. The surface brightnesses in the Pioneer maps were corrected for stars brighter than apparent V-magnitude 7.5. There is a hole in the distribution of diameter some 20\degs to 30\degs\ around the position of the Sun as seen from Jupiter ($l^{\rm II}$, $b^{\rm II})  \sim$ (205\degs, +40\degs) during encounter in December 1973.  I produced from these data two full-sky contour maps of the Galactic surface brightness measured by Pioneer 10 in the two colors smoothed to an 8-degree resolution (van der Kruit, 1986, 1990). Toller (1990) has presented an interesting overview of the project including a general background of the study of the integrated light from the Galaxy and some implications of the Pioneer studies for its structure. Toller, Tanabe and Weinbeg (1987) show isophotal maps in 15\degs\ areas around the celestial, ecliptic and Galactic poles (the south ecliptic pole is in the Small Magellanic Cloud).

The maps still contain next to the integrated starlight, contributions from diffuse Galactic light and extragalactic background light. These contributions have been studied in much detail by Toller (1981). The extragalactic contribution is very small at most (see also Toller, 1983) and will be of no further concern. Diffuse Galactic light (in the optical mostly from scattering by dust) is not negligible at low latitudes (Toller, 1981, 1983; Leinert {\it et al.}, 1997), but it would have contributed to Pannekoek’s observations as well.
\bigskip

Before making a comparison we have  to look first at wavelength bands and units. The two Pioneer 10 maps are in two wavelength bands: ‘blue’ (3950-4850 \AA) and ‘red’ (5900-6900 \AA). The blue band has an effective wavelength of 4407 \AA\ and is rather similar to the Johnson B-band (effective wavelength 4361 \AA). The red band starts around the sharp blue increase and peak of the R-band of the Johnson-Cousins UBVRI system, but misses the red tail which extends another 1000 \AA\ (Bessel, 2005). The effective wavelengths are similar (Pioneer red has 6417 \AA\ and the R-band 6407\AA). The red band can be treated as similar to the Johnson-Cousins R-band,

The question then is what the bands are for Pannekoek. His visual observations correspond to the response curve of the human eye, which peaks around 5500 \AA. The Johnson V-band  peaks at about 5300 \AA,  so the difference between $m_{vis}$ and V is not great, and for practical purposes we may treat them as equivalent. The photographic (for filter-less, early photographic emulsions) band has an effective wavelength of  about 4200 \AA, while for the Johnson B-band  it is more like 4400 \AA, so in first approximation we may treat $m_{phot}$ and B as equivalent. 
\bigskip

Now for units. Pannekoek’s unit is 1 star of magnitude 0 per square degree, but sometimes 1 star of magnitude 10 per square degree. The former is 17.78 mag arcsec$^{-2}$ and the latter 27.78 in the same units, where we may add the designation {\it  vis} or {\it phot} where appropriate. Most of Pannekoek’s maps employ the second choice.

The Bochum surface photometry is also expressed in the number of stars of magnitude 10 per square degree in the respective wavelength band. These observers designate this unit as S10 and I will also adopt that here, but add the wavelength band, so e.g. S10(vis) or S10(phot). In view of what has just been discussed this may also be labeled V and B respectively. We can then calibrate Pannekoek's steps in the visual observations, which he gave in terms of number of  stars of magnitude 0.0 per square degree,  in terms of magnitudes per square arcsec as listed  in Table~\ref{table:steps}. The step differences change systematically (as he noted himself) and are 0.48, 0.34, 0.29, 0.24, 0.22,  and 0.21 magnitudes.

For Pioneer the unit used is more complicated. What was used as unit was S$_{10}$(V)$_{\rm G2V,band}$ with `band' either `red' or `blue', which is the number of G2V stars (like the Sun) per square degree that are of apparent magnitude 10 in the V-band. In the V-band this is equal to 27.78 V-mag arcsec$^{-2}$, and in this band the Pioneer and S10 units are the same. For other bands we need the appropriate color index of the Sun, which for (B--V) is 0.653 (Ram\`{\i}rez et al., 2012). Then using these latest colors the unit becomes 28.43 B-mag arcsec$^{-2}$. In van der Kruit (1986) a different route was taken, following Toller’s (1981) calibration of the unit in the blue Pioneer band of $1.16 \times 10^{-9}$ erg cm$^{-2}$ sec$^{-1}$ sterad$^{-1}$. Then the unit came out as 28.49 B-mag arcsec$^{-2}$, a difference of 0.06 magnitudes or about 6\%. This is very likely significantly smaller than the accuracy of the calibration and measurement in Pannekoek’s maps. I will adopt the former value for consistency with the red band and then 1  S$_{10}$(V)$_{\rm G2V,blue}$ = 28.43 B-mag arcsec$^{-2}$= 0.283 L$_{\odot,B}$ pc$^{-2}$. For the red band the relevant solar color index (V--R) is 0.356 and then 1 S$_{10}$(V)$_{\rm G2V,red}$  = 27.42 R-mag arcsec$^{-2}$.

For reference, quoting surface brightnesses accurate to 0.1 magnitude, the conversions have been summarized in Table~\ref{table:convs}.
\bigskip

\begin{table}[t]
\caption{\normalsize Calibration of Pannekoek's steps in terms of surface brightness.}
\begin{center}
\begin{tabular}{ccc}
\hline
\hline
Step &  S10(vis) & V-mag \\
number &  & arcsec$^{-2}$ \\
\hline
0 & 0.0112 & 22.58 \\
1 & 0.0172 & 22.11 \\
2 & 0.0236 & 21.77\\
3 & 0.0309 & 21.48\\
4 & 0.0390 & 21.22\\
5 & 0.0480 & 21.00 \\
6 & 0.0579 & 20.79 \\
\hline 
\end{tabular}    
\end{center}
\label{table:steps}
\end{table}

\begin{table}[t]
\caption{\normalsize Calibration of surface brightnesses in stars of magnitude 10 per square degree to magnitudes per square arcsecond to Pioneer units.}
\begin{center}
\begin{tabular}{rcccl}
\hline
\hline
\multicolumn{5}{c}{Blue/photographic:} \\
100 S10(phot) &=& 22.8 B-mag arcsec$^{-2}$ &=& 171 S$_{10}$(V)$_{\rm G2V,blue}$, \\
100 S$_{10}$(V)$_{\rm G2V,blue}$ &=&  23.4 B-mag arcsec$^{-2}$ &=& 59 S10(phot).\\
\multicolumn{5}{c}{Red/visual:}\\
100 S10(vis) &=& 22.8 V-mag arcsec$^{-2}$ &=& 100 S$_{10}$(V)$_{\rm G2V,red}$,\\
100  S$_{10}$(V)$_{\rm G2V,red}$ &=& 22.4 R-mag arcsec$^{-2}$ &=& 100 S10(vis).\\
\hline
\end{tabular}
\end{center}
\label{table:convs}
\end{table}

\begin{figure}[t]
\begin{center}
\includegraphics[width=0.88\textwidth]{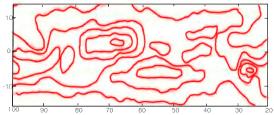}
\includegraphics[width=0.88\textwidth]{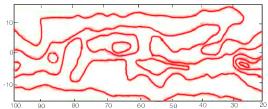}
\end{center}
\caption{\normalsize Isophotal maps of the surface brightness of the Milky Way as determined in the Pioneer 10 Background Starlight Experiment. This figure concerns the northern area of the Milky Way that was also illustrated in Figs.~\ref{fig:Maps1}, \ref{fig:Maps2}, \ref{fig:blur} and \ref{fig:PhotN1}. The resolution is 2\degs. Top: Isophotes in the Pioneer blue band. The faintest  isophote is at 100 S$_{10}$(V)$_{\rm G2V,blue}$ and the contour interval is 50 S$_{10}$(V)$_{\rm G2V,blue}$. The brightest isophote is at 350 S$_{10}$(V)$_{\rm G2V,blue}$. Bottom: Idem in the Pioneer red band. The faintest isophote is at  200 S$_{10}$(V)$_{\rm G2V,red}$, the interval is 100 S$_{10}$(V)$_{\rm G2V,red}$ and the brightest isophote at 600 S$_{10}$(V)$_{\rm G2V,red}$.}
\label{fig:PioneerN}
\end{figure}

\begin{figure}[t]
\begin{center}
\includegraphics[width=0.80\textwidth]{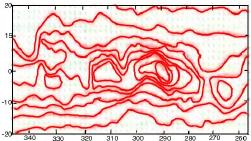}
\includegraphics[width=0.80\textwidth]{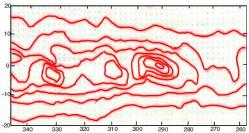}
\end{center}
\caption{\normalsize Isophotal maps of the surface brightness of the Milky Way from Pioneer 10 as in Fig.~\ref{fig:PioneerN}, but now the southern area of the Milky Way that was also illustrated in Figs.~\ref{fig:VisS1}, \ref{fig:VisS3}, and \ref{fig:PhotS}. Top: Isophotes in the Pioneer blue band. The faintest  isophote is at 100 S$_{10}$(V)$_{\rm G2V,blue}$ and the contour interval is 50 S$_{10}$(V)$_{\rm G2V,blue}$ up to  400 S$_{10}$(V)$_{\rm G2V,blue}$, and then  500 and  600 S$_{10}$(V)$_{\rm G2V,blue}$. Bottom: Idem in the Pioneer red band. The faintest isophote is at 200 S$_{10}$(V)$_{\rm G2V,red}$, the interval is 100 S$_{10}$(V)$_{\rm G2V,red}$ and the brightest isophote at 700 S$_{10}$(V)$_{\rm G2V,red}$.}
\label{fig:PioneerS}
\end{figure}

The isophotal maps representing the Pioneer 10 observations of the two regions used above to illustrate Pannekoek’s visual and photographic mapping are presented in Figs.~\ref{fig:PioneerN} and \ref{fig:PioneerS}. The northern field in Fig.~\ref{fig:PioneerN} should be compared to Fig.~\ref{fig:Maps1} for Pannekoek’s visual impression and the corresponding part of the {\it ESO Milky Way Panorama}, and to Fig.~\ref{fig:Maps2} for the visual isophotal maps (to Fig.~\ref{fig:blur} for a gray-scale impression of the visual observations) and to Fig.~\ref{fig:PhotN1} for his photographic isophotes. Similarly, Fig.~\ref{fig:PioneerS} should be compared to respectively Figs.~\ref{fig:VisS1}, \ref{fig:VisS3} and \ref{fig:PhotS}. Obviously the Pioneer resolution of 2\degs\ results in a quite different appearance,  but the correspondence is quite good when allowing for that.

The comparisons to make are the following. Firstly, there are two separate issues, the brightness or magnitude scale and the zero point calibration. For the visual Pannekoek data the magnitude scale was calibrated by observing the Milky Way at various elevations above the horizon and the (uncertain) zero point comes from the integrated starlight determinations of Yntema (1909) and van Rhijn (1919, 1921b). The wavelength band is roughly Johnson V, and comparison is to the Pioneer red band, translated into Johnson-Cousins R. In the photographic work the scale comes from determinations of the characteristic curve and the zero point from surface brightnesses of distributed light of out-of-focus stars.

The most ideal comparison would be by smoothing Pannekoek’s data to the Pioneer resolution and then rebinning to the same grid followed by performing a full regression analysis, such as done for example by Kimeswenger {\it et al.} (1993) or Toller (1989). This, I note in passing, gave reasonably good agreement between Pioneer and Bochum, the ratio between their S10 values differing from unity by 4\%\ (if the regression is forced through the origin). However this requires the Pannekoek and Bochum maps to be turned into digital form.. This is not straightforward in practice; we only have isophotal maps with numerical brightness values on very irregularly distributed points. Although probably theoretically not impossible to turn this into a grid of brightness values, the investment of effort, time and manpower involved would be excessive and wasteful in view of the information to be obtained. The Pioneer listings could be scanned and turned into digital form, but in the absence of numerical information on Pannekoek's data this would not really serve any purpose.

The way I will proceed to compare Pannekoek’s isophotal maps to those produced by Pioneer 10 and the Bochum group is by selecting some specific areas. For example, I will select general brightness levels at certain latitudes in longitude ranges where the contours run more or less parallel to the Galactic equator. I will also examine the smooth brightness over extended areas of stellar ‘clouds’ or of extended areas of dust. Such an exercise will obviously not fully {\it recalibrate} Pannekoek’s isophotal maps; in the unlikely case that one would want to use these for a quantitative analysis this quite extensive, laborious and careful recalibration procedure would need to be performed. My aim here is merely to establish as a general idea how reliable the isophotal values are. I will proceed to go through the comparisons I made; the numerical values derived along the way have been collected in Table~\ref{table:compare}.

\begin{table}[t]
\caption{Comparison of surface brightnesses in the Milky Way, measured visually and photographically by Pannekoek, in a blue and red band by Pioneer 10 and in UBV by the Bochum group. }
\begin{center}
\begin{tabular}{ccccccrcccr}
\hline
\multicolumn{3}{c}{Pannekoek} & \multicolumn{4}{c}{Pioneer} & \multicolumn{4}{c}{Bochum}\\ 
\hline
 Band($\Delta \lambda$) & Step/S10($\Delta \lambda$) & mag   & \ \ \ Band($\Delta \lambda$) & \ \ S$_{10}$(V,)$_{{\rm G2V},\Delta\lambda}$ & mag & Diff. & \ \ \ Band($\Delta \lambda$) & \ \ S10($\Delta \lambda$)\ \ \ &   mag & Diff. \\
&& arcsec$^{-2}$ &&& arcsec$^{-2}$ &&&& arcsec$^{-2}$ & \\
\hline
\multicolumn{11}{c}{{\bf NORTH}}\\
  \multicolumn{11}{c}{{\bf Parallel rectangle} (l$^{\rm II}$, b$^{\rm II}$) = (100\degs\ -- 70\degs, -8\degs $\pm$ 2\degs)}\\
Vis/V & 2.6 & 21.5 & Red/R & 280 & 21.3 & 0.2 &&&& \\
Phot/B & 140 & 22.4 & Blue/B & 210 & 22.6 & -0.2 &&&& \\
  \multicolumn{11}{c}{{\bf Parallel rectangle} (l$^{\rm II}$, b$^{\rm II}$) = (100\degs\ -- 70\degs, -12\degs $\pm$ 2\degs)}\\
Vis/V & 1.5 & 22.0 & Red/R & 185 & 21.7 & -0.3 &&&& \\
Phot/B & 115 & 22.6 & Blue/B & 150 & 23.0 &-0.4 &&&& \\
 \multicolumn{11}{c}{{\bf Cygnus cloud} (l$^{\rm II}$, b$^{\rm II}$) = (76\degs\ -- 64\degs, 2\degs\ -- 6\degs)}\\
Vis/V & 4.3  & 21.2 & Red/R & 450  & 20.8 & 0.4 &&&& \\
Phot/B & 215 & 22.2 & Blue/B & 215 & 22.6 & -0.4 &&&& \\
  \multicolumn{11}{c}{{\bf Dark streak} (l$^{\rm II}$, b$^{\rm II}$) = (28\degs\ $\pm$ 2\degs, 4\degs $\pm$ 2\degs)}\\
Vis/V & 1.1 & 21.8 & Red/R & 165  & 21.9 & -0.1 &&&& \\
Phot/B & \ 95 & 22.9 & Blue/B & \ \ 85 & 23.6 &-0.7 &&&& \\    
\multicolumn{11}{c}{{\bf Scutum cloud} (l$^{\rm II}$, b$^{\rm II}$) = (27\degs\  $\pm$ 2\degs, -3\degs\ $\pm$ 2\degs)}\\
Vis/V & 5.8  & 20.8 & Red/R & 550  & 20.5 & 0.3 &&&& \\
Phot/B & 285 & 21.7 & Blue/B & 350 & 22.0 & -0.3 &&&& \\
\multicolumn{11}{c}{{\bf SOUTH}}\\
  \multicolumn{11}{c}{{\bf Carina} (l$^{\rm II}$, b$^{\rm II}$) = (287\degs\  $\pm$ 1\degs, -2\degs\ $\pm$ 1\degs)}\\
Vis/V & 6.0  & 20.8 &&&&& V & 700 & 20.7 & 0.1 \\
&&& Red/R & 625 & 20.4 & 0.4 & R & 1500 & 19.9 &\\
Phot/B & 600 & 20.9 & Blue/B & 670 & 21.4 & -0.5 & B & 400 & 21.3 & -0.4 \\
  \multicolumn{11}{c}{{\bf Coalsack} (l$^{\rm II}$, b$^{\rm II}$) = (303\degs\  $\pm$ 6\degs, -2\degs\ $\pm$ 6\degs)}\\
Vis/V & 0.0 & 22.6 &&&&& V & 350 & 21.4 & 1.4 \\
&&& Red/R & 750 & 20.2 & 2.4 & R & 900 & 20.4\\
Phot/B & \ 70 & 23.2 & Blue/B & 385 & 22.0 & 1.2  & B & 120  & 22.6  & 0.6 \\
 \multicolumn{11}{c}{{\bf Parallel rectangle} (l$^{\rm II}$, b$^{\rm II}$) = (340\degs\ -- 320\degs, -10\degs -- -13\degs)}\\
Vis/V &  2.5 & 21.6 &&&&& V & 250 & 21.8 & -0.2 \\
&&& Red/R & 330 & 21.1 & 0.5 & R & 900 & 20.1 \\
Phot/B & 245 & 22.4 & Blue/B & 245 & 22.5 & -0.1 & B & 110 & 22.7 &-0.3  \\
\multicolumn{11}{c}{{\bf Parallel rectangle} (l$^{\rm II}$, b$^{\rm II}$) = (340\degs\ -- 320\degs, -13\degs -- -16\degs)}\\
Vis/V &  2.0 & 21.8 &&&&& V & 200 & 22.0 & -0.2  \\
&&& Red/R & 200 & 21.6 & 0.2 & R & 700 & 20.7 \\
Phot/B & 120 & 22.6 & Blue/B & 265 & 22.3 & 0.3 & B & \ \ 90 & 22.9 & -0.3 \\
\multicolumn{11}{c}{{\bf Parallel rectangle} (l$^{\rm II}$, b$^{\rm II}$) = (340\degs\ -- 320\degs, -16\degs -- -19\degs)}\\
Vis/V &  1.5 & 22.0 &&&&& V & 170 & 22.2 & -0.2  \\
&&& Red/R & 190 & 21.7 & 0.3 & R & 450 & 21.2 \\
Phot/B & 110 & 23.3 & Blue/B & 150 & 23.3 & 0.0 & B & \ \ 50 & 23.6 & -0.3\\
\end{tabular}
\label{table:compare}
\end{center}
\end{table}

In the northern field in the Pioneer map (Fig. 26) between (new) longitudes 10\degs\ and 70\degs, the surface brightness is reasonably stable at latitudes of -10\degs\ and lower. I will go through the actual comparison here in detail. The mean Pioneer values are about 210 and 150  S$_{10}$(V)$_{\rm G2V,blue}$ at latitudes -8\degs\  and -14\degs\ and about 280 and 185 S$_{10}$(V)$_{\rm G2V,red}$. The red band could serve for comparison with Pannekoek’s visual maps. I look at the visual Pannekoek observations first. From the extensive, tabular data in Pannekoek (1920, pp.A100 and A101, observations ‘S. + B. +E.+P.’, the mean of four observers) I estimate the corresponding values (old longitudes 70\degs\ to 40\degs) as about step 2.6 and 1.5, which from the calibrations in Table~\ref{table:steps}  correspond to 21.9 and 22.1 V-mag arcsec$^{-2}$. The Pioneer values of 280 and 185 S$_{10}$(V)$_{\rm G2V,red}$ correspond to 21.3 and 21.7 R-mag arcsec$^{-2}$. Assuming for the Milky Way a solar color (V —R) of 0.4  this would be 21.7 and 22.1 V-mag arcsec$^{-2}$. This compares rather well. In the photographic observations of the surface brightness by Pannekoek (1933; see also Fig.~\ref{fig:PhotN1}) I estimate the average surface brightnesses of Pannekoek as 140 and 115 S10(phot), or 22.4 and 22.6 B-mag arcsec$^{-2}$, while the Pioneer values correspond to 22.6 and 23.0 B-mag arcsec$^{-2}$. This points at a small possible zero-point error by Pannekoek and some deviations in the magnitude scale.

 Next I will look at the Cygnus Cloud (see Fig.~\ref{fig:Maps1}, for which I will take the area between longitudes (new) 76\degs\  and 64\degs\ (old 43\degs\ to 31\degs), latitudes 2\degs\  to 6\degs. From the results in the table we see that the comparison is reasonably satisfactory, if we allow for a possible color index (V—R) of 0.4 like the Sun. Note that the color index (B—V) that Pannekoek would have deduced would have been 1.0, which is redder than the Sun and corresponds to spectral type early-K to mid-K. Pioneer gives a (B--R) color of 1.8, typical for a K5- to K-7 star, which is not unreasonable for an old disk population dominated by giants. For the Scutum Cloud Pannekoek  deduced an F-type with more or less the same data (Pannekoek, 1923c).

Next I choose a small area in the dark streak of extinction around (l$^{\rm II}$, b$^{\rm II}$) = (28\degs, +8\degs).  The differences are somewhat larger now, Pannekoek’s surface brightnesses being brighter by about half a magnitude  The color index according to Pannekoek is (B—V) = 1.1, while Pioneer gives (B—R)= 1.7, which would indicate we are looking at reddened stars of spectral type F, G or early K.

And finally I look at the brightest part of the Scutum Cloud (see Fig.~\ref{fig:Scutum}) in an area of some 6\degs\ diameter centered at (l$^{\rm II}$, b$^{\rm II}$) = (27\degs, -3\degs). As can be seen from the Table the agreement is now somewhat less satisfactory. This may be due in part to the different resolutions.

This all is encouraging, but it has not involved yet the Bochum data. In the southern field I start with the brightest part of the Milky Way around (l$^{\rm II}$, b$^{\rm II}$)  = (287\degs, -2\degs), which it the well-known effect of the Carina spiral arm. I now take an area of about 4\degs\ in diameter at the brightest peak at the position indicated. The southern visual observations are in Pannekoek (1928) in step values that from the overlap on the sky have been made consistent with the northern survey. However, this time no listings with numerical values are provided as was the case with the northern Milky Way, but in addition to the isophotes we can use the numbers added to the maps. In the Carina region the brightest part has been shaded and the numbers around it are 4.7, 5.4 and 6.5 (see Fig.~\ref{fig:VisS3}, as before Pannekoek has deleted decimal points). I use as observed value step 6.0. The photographic data are now in Pannekoek (1949; see also Fig~\ref{fig:PhotS}), where we need to read again values from the isophotal map. In the center of the nebula values of 700-1100 are written, but that is more concentrated than in the Pioneer maps. The value I adopt is 600 S10(phot).  This gives for Pannekoek 20.8 V-mag arcsec$^{-2}$ and 20.9 B-mag arcsec$^{-2}$, while for Pioneer the surface brightnesses are 20.4 R-mag arcsec$^{-2}$ and 21.3 B-mag arcsec$^{-2}$. The Bochum maps indicate surface brightnesses of  21.4 B-mag arcsec$^{-2}$, 20.7 V-mag arcsec$^{-2}$ and 19.9 R-mag arcsec$^{-2}$. The authors do indicate that these areas should be treated with caution (because they are bright?), yet a the agreement between Pannekoek and Bochum in B and V and between Bochum and Pioneer in B and R is good, except for Pannekoek’s surface brightness in B being somewhat bright (and Bochum R rather bright). All color indices are fairly blue, which is not surprising since Carina is a star forming area,

Next I look at the southern Coalsack, centered at about (l$^{\rm II}$, b$^{\rm II}$) = (303\degs, -2\degs). This corresponds to  (l$^{\rm I}$ = 270\degs). Pannekoek (1928) wrote at this position in his isophotal map (Fig.~\ref{fig:Oph}) step 0.0, which is the faintest step he has given and corresponds to 22.6 V-mag arcsec$^{-2}$. The photographic isophote map (Fig.~\ref{fig:PhotS}) has in a few degree area as average number of about 70 S10(phot); the very detailed  isophotes would give a similar value if smoothed by eye to a few degree resolution.  This corresponds to 23.2 B-mag arcsec$^{-2}$. The Bochum photometry (Seidensticker, Schmidt-Kaler and Schlosser, 1982) shows isophote maps of the area of the Coalsack. Averaged over a few degrees the surface brightnesses are about 20.9  R-mag arcsec$^{-2}$, 21.4 V-mag arcsec$^{-2}$ and 22.4 B-mag arcsec$^{-2}$. We have to be aware that comparison in this dark cloud maybe affected by the different angular resolutions. The comparison between Pioneer and Bochum in V and R is satisfactory, but Pannekoek had  now obtained much fainter surface brightnesses. It would therefore be interesting to also look at colors. Pannekoek would give (B--V) = 0.6, Pioneer (B—R) = 1.8 and Bochum (B--V) = 1.2, (B--R) = 2.2. Seidensticker, Schmidt-Kaler and Schlosser (1982) give average values for the Coalsack, which constitutes more accurate treatment, of about (B--V) = 0.8 and (V--R) = 1.0, more or less consistent with this. Pannekoek has systematically overestimated the decrease in surface brightness but has the color more or less right.

Finally I take an area outside the plane, like I did in the north, over which in longitude the surface brightness changes little. This consists of three rectangular areas at (l$^{\rm II}$, b$^{\rm II}$) =   (340\degs\ — 320\degs,  -12\degs, -15\degs, -18\degs) (l$^{\rm I}$ = 306\degs\  — 287\degs). The agreement now is generally satisfactory, being in agreement up to a few tenths of a magnitude.
\bigskip

The comparisons are, as stated above, not meant to be a calibration of Pannekoek’s isophotal map, but intended to provide a general impression of the accuracy. The number of entries in the table is too small for a statistical analysis. The following can be said. Firstly, the observations in the Coalsack by Pannekoek are much too faint. The Pioneer and Bochum values are in reasonable agreement, so it is likely that Pannekoek’s analysis has produced some spurious results. The `dark streak’ in the north actually is of lower surface brightness than the Coalsack, according to Pioneer. What might have played a role is that contrary to the northern dark streak the surrounding area of the Milky Way  around the Coalsack is very bright and this may have resulted in the contrast to be overestimated. This may have been a factor in visual observing; in the photographic measurements the possible relatively dark photographic density may have played a role since a heavier exposed part of the characteristic curve was involved.

Taking all comparisons together, except in the Coalsack, results in rather small average differences. In the red the difference (Pannekoek--Pioneer) is 0.3 $\pm$ 0.2 (this is the mean and rms scatter), but using a (V--R) of solar of say 0.4 reduces this to -0.1 magnitudes. For the blue the mean difference is -0.3 $\pm$ 0.3.  Not too much weight should be attached to the scatter since effects of reading from contour maps versus smoothed data can influence this. Doing this for the (fewer) Bochum comparison results in (Pannekoek--Bochum) of -0.1 $\pm$ 0.2 magnitudes in V and -0.3 $\pm$ 0.1 magnitudes in B (together -0.1 $\pm$ 0.2 magnitudes). The Pioneer surface brightnesses minus the Bochum ones differ by -0.1 $\pm$ 0.4 magnitudes in R and -0.1 $\pm$ 0.3 magnitudes in B (together -0.1 $\pm$ 0.3 magnitudes). There is no clear systematic effect in the differences in  Table~\ref{table:compare} with surface brightness (except for the Coalsack), although admittedly the number of comparisons is small.  Altogether this means that the only important effect is a systematic error of 0.3 magnitudes in Pannekoek’s photographic zero point in the sense that Pannekoek’s surface brightnesses are probably about 0.3  magnitudes too faint, except in the southern Coalsack, where Pannekoek is too faint by 1.5 magnitudes or more.  The difference between Pioneer and Bochum shows more consistency confirming that the differences are mainly due to Pannekoek’s zero points and magnitude scales. Translating a difference of 0.3 magnitudes into linear brightness gives a difference of 30\%. However, remember that both the visual impression and response of a photographic emulsion are both intrinsically logarithmic. 

My final conclusion then is that {\it the photometry of Pannekoek is of high quality.} This is a highly commendable achievement.

\section{Discussion}

In the Introduction I announced that in this paper I would address the question of how Pannekoek determined distances of up to 60 kpc for parts of the Milky Way and how well his isophotal maps compared to modern ones, in particular his magnitude scale calibration. This required insight into his methods and ways of interpretation of the data. This evolved into an effectively complete discussion of all Pannekoek’s research concerning the Milky Way and the structure of the Stellar System. I divide this discussion into two parts, his work on the Galactic structure and separately his efforts to produce isophotal maps of the Milky Way.

\subsection{Structure of the Stellar System}
I first turn to the matter of the distances to the  Cygnus and Aquila Clouds. We have seen that Pannekoek could measure such large distances of tens of kpc because  he made an enormous extrapolation. He assumed that the star clouds are more or less isolated structures in space of which he sees only a small fraction of the stars further than some 3.5 standard deviations out in the bright tail of the luminosity distribution, which corresponds to a fraction of 0.0002 or so of the total. This requires extreme confidence in the accuracy with which the distribution (the luminosity function) is known, but also in its universality. Of course, Pannekoek was aware of this (p.506):
\begmarg
As regards the numerical values deduced from observational data (reaching from M = -6 to +8) the curve deserves great confidence, and between these limits the formula adopted is fixed by them [Kapteyn, van Rhijn and Schouten] with great accuracy.
\endmarg

The other, and I think more acute, uncertainty is the counts used. These are far from uniform: the {\it Bonner Durchmusterung} up to magnitudes 9 or so, star counts by Epstein from Frankfurt am Mainz, counts on photographic plates of the {\it Carte du Ciel} and William Herschel’s star gauges. The counts by Epstein have never been published. So the question arises what these really are. According to Pannekoek (1910), Epstein surveyed between 1877 and 1888 about 2700 fields distributed over the  whole sky with a 6-inch (15 cm) telescope. `The whole sky’ surely means the whole northern sky and the telescope probably was his private property. These `gauges’ were never published, and Pannekoek had obtained these data directly from Epstein. He must have heard about it from Cornelis Easton, who had also referred to these data (Easton, 1900).

Theobald Epstein (1836--1928) was a Frankfurt teacher in mathematics and physics at the {\it Philanthropin}, a Jewish gymnasium that had been founded in 1804. He was also interested in astronomy, particularly sunspots, on which he published a number of {\it Astronomische Nachrichten} contributions. He also authored in the {\it Vierteljahrsschrift der Astronomischen Gesellschaft}, but these are not in the {\it SAO/NASA Astrophysics Data System} (ADS) (some can be found on Google Books). He also  used the observatory of the Physical Society, of which he became the head when it was opened in 1908 on the tower of the Paulskirche. This {\it Physikalische Verein} is an old society that was founded in 1824 and still is very active. Epstein used the observatory to provide the local  time service for the regulation of the Frankfurt clocks. It had  a 21-cm refractor that had been installed in 1908. But all this is well after his star gauges, which date from the 1870s and 1880s.

As a point of interest: Epstein actually was the gymnasium teacher of Karl Schwarzschild, whose father knew Epstein well. Karl’s interest in astronomy was developed further under Epstein with the support of father Schwarzschild. Karl was also a friend of Epstein’s son, well-known mathematician Paul (1871--1939), who was only two years his senior.

The limit of the counts performed by Epstein were judged to be magnitude 12.51 (Pannekoek, 1910; based on comparing fields all over the sky to counts in Kapteyn, 1908) and those of Herschel to 13.90 (this value was attributed to Kapteyn). For the {\it Carte du Ciel} plates it depended on the observatory they came from, in this case he used  11.73 and 13.20. This is far from uniform material and the limiting magnitudes will not be very accurately determined (Pannekoek actually corrected them using work of van Rhijn); yet, in calculating the gradients these magnitudes are used without any qualification. The magnitude limit of 13.90, attributed to Kapteyn, is too bright (see van der Kruit, 1986), but surely is also not that uniform across the sky to warrant two decimals. My analysis there, when I  compared counts from Herschel to those in a model for the Galaxy constructed by Bahcall \&\ Soneira (1980) and adapted by me on the basis of the Pioneer data presented above, showed a limiting magnitude of 14.5 or so,  but with a scatter from field to field of about  $\pm$0.5 magnitudes. This is at moderate latitudes, at higher latitudes the limit is more like magnitude 15, which is a full magnitude fainter than Pannekoek’s adopted value. There is absolutely no reason to confidently adopt a value with even a single decimal. In Pannekoek (1923a) the limiting magnitude of the Herschel counts was corrected to 14.77 (and Epstein’s to 12.77), a correction of almost 0.9 magnitudes, which seems closer to the truth, but the use of two decimals remained!

Yet to derive the gradients Pannekoek has taken these limiting magnitudes  as an definite value with an accuracy to two decimals. The same holds for the other sets of counts. He calculated gradients for intervals of 0.8 or 0.9 magnitudes and an error of 0.2 magnitudes  in one of these limiting values that make up the ends of the intervals produces an error of more than 20\%\ in the gradients. Much depends on the gradient out to the limit of Herschel's counts and there the uncertainties are much larger than this. Much criticism could be brought forward on the accuracy of the data used and it seems to me the paper at the present time would never had withstood careful refereeing.

But that is not the main point. What I am concerned with here in the first place is the basics of the method. In order to follow this way of attacking the problem, Pannekoek made assumptions. First, like most astronomers at the time (1919) such as in particular Shapley, von Seeliger and Kapteyn, the star counts were assumed unaffected by interstellar extinction. Secondly, the variations in brightness in the Milky Way were assumed to reflect principally variations in space density of stars rather than the result of intervening dust. And thirdly, the patches in Cygnus and Aquila were assumed to be localized clouds limited in extent not only on the sky, but also along the line-of-sight. The Sidereal System was in Pannekoek’s view primarily a collection of star clouds with in between a rarefied smooth stratum of stars rather then primarily a smooth system of stars  with condensations of enhanced density superimposed on it. And on top of that isolated clouds of dust. All these assumptions  -- and the one of a constant luminosity function across the Galaxy --  have subsequent to Pannekoek’s paper proved to be doubtful or incorrect.

All of this is not to say Pannekoek’s approach was not clever, ingenious or original. It was all of that. The result was that he came to the realization that the Milky Way was a larger stellar body comparable in size with that outlined by Shapley in his system of globular clusters. His assumptions were incorrect, his analysis led to incorrect distances, but the consequences of a large, extended Stellar System were correct and significant. So, what made him make these assumptions? Where did his conviction that the sidereal system consisted primarily of patches, clouds and streams of stars and isolated dust clouds come from? Obviously from his life-long fascination with the appearance of the Milky Way.
\bigskip

To complete the discussion of these matters of the distances of these star clouds, a few words on the question why the counts Pannekoek used indeed showed the effect he was looking for. We now know that in directions of the Milky Way there is significant extinction along each line of sight. The `clusters’ he saw are bright areas on the sky, so with less extinction than their surroundings. So along these lines of sight we look further into the stellar distribution than in the dark area he used for comparison. No wonder he found an excess towards the bright areas at the faint end of the counts. Pannekoek used as primary comparison the counts towards the Galactic poles. The stellar distribution there will likewise be limited in extent, now not due to extinction, but to the flatness of the old disk population, dropping to half the density in the plane in already somewhat less than 1 kpc (van der Kruit \&\ Searle, 1981, 1982). The star counts in both comparison areas will be lower than towards the bright ones and the gradients changing more gradually, producing the effect Pannekoek was looking for.
\bigskip

It is remarkable that the discussions of the 1919 paper at the Amsterdam Symposium  (Tai, van der Steen en van Dongen, 2019) on the distance of the Milky Way do not refer to the sequels Pannekoek (1922b, 1923a) in which he reduced his earlier distance to the Cygnus Cloud  by a factor two and subsequently withdrew the result altogether. It is true that the Pannekoek (1922b) paper and the reduction of the distance was noted and discussed in Tai (1920), but I see no reference to Pannekoek (1923a) there. 
\bigskip

So how would he evaluate this work himself?
Pannekoek (1982) described this work as follows (p.247; my translation):
\begmarg
What I envisioned as a work task was primarily the question of structure of the galaxy. I had strongly sympathized with Kapteyn's work, but always felt it was based on a different point of view. He left out the change with galactic longitude, treated only the stellar distribution in depth and latitude. I had grown up with the aspect of the Milky Way clouds always first in my mind, and did not see in it an outwardly thinning rotating body, but it consisting of its own individual accumulations, similar and as prominent as Kapteyn's local system. So here I saw my own task to be performed according to my own ideas: the spatial structure of the Galaxy as a collection of physical clouds and streams. Later these `clouds' lost much of their individuality, and turned out to be apparent effects, outlined by dark nebulae; but at that time this was not so clear.
\endmarg 

Obviously the fact remains that Pannekoek’s analysis of the Cygnus and Aquila star clouds enticed  him to accept the larger stellar system of Shapley, regarding Kapteyn’s Universe as one of the condensations it consisted of. Note that the Cygnus Cloud measures about 20\degs\ by 7\degs\ on the sky, and the Aquila Cloud 15\degs\ by 5\degs, which at 40 and 60 kpc corresponds to about 15 by 5 kpc, not that much smaller than Kapteyn’s Universe.But the story as usually told up until now is incomplete. I have shown above that earlier researchers  maybe have not carefully read  Pannekoek’s papers and in any case have misrepresented it. His methods have been misunderstood and further publications by him on the subject have been ignored. The distances were based on enormous extrapolations of the luminosity function, and on highly uncertain calibrations of the faint limits of older star counts. As we saw, soon he had to correct the distances by a factor two using better counts and not long after that to withdraw his results in view of non-universality of the luminosity function. However, this did not shake his belief in the larger Galaxy and his early conversion remains admirable.
\bigskip

The question then is what impact this work had on the debate on the scale of the Stellar System and the structure of the Sidereal World. The famous `Great Debate’ between Heber Curtis and Harlow Shapley took place  in 1920, so after Pannekoek published his distances to the star clouds. The printed version of the Debate (Shapley \&\ Curtis, 1921)  does not mention Pannekoek’s paper, while Shapley obviously could have used it to support his case. Instead he resorted to Kapteyn (p.173):
\begmarg
For instance, Professor Kapteyn has found occasion, with the progress of his elaborate studies of laws of stellar luminosity and density, to indicate larger dimensions of the Galaxy than formerly accepted. In a paper just appearing as {\it Mount Wilson Contribution}, No. 188, he finds, as a result of the research extending over some 20 years, that the density of stars along the galactic plane is quite appreciable at a distance of 40,000 light-years- — giving a diameter of the galactic system, exclusive of distant star clouds of the Milky Way, about three times the value Curtis admits as a maximum for the entire galaxy.
\endmarg

Shapley used the feature of Kapteyn’s universe stretching out to 13 kpc as an indication of a {\it large} sidereal system, even though it was significantly smaller than his system of globular clusters, which had its center at 20 kpc. He could have used Pannekoek’s distance to the Milky Way of 40 and 60 kpc as much more forceful support. So either, Shapley missed Pannekoek’s paper, distrusted the result or actually found 60 kpc uncomfortably large. Whatever is the case, Pannekoek’s work failed to make much impression on Shapley. The comprehensive and authoritative discussion by Robert Smith (1982) of the Great Debate also provides no indication that Shapley had welcomed Pannekoek’s results as support for his case. Smith only refers to `historian-astronomer A. Pannekoek’ to quote his description of the emergence of statistical astronomy in Pannekoek (1961).
 In an address for the British Astronomical Association on May 31, 1922 (Shapley, 1923) he likewise did not mention Pannekoek. This was only three days after his meeting in Leiden with Dutch astronomers including Pannekoek that I alluded to above (and see Tai, 1921). Had he been impressed by Pannekoek’s results  he might have mentioned this in this UK address.
\bigskip

What then about other (leading) astronomers? For example Henry Norris Russell (1877–1957) or Arthur Eddington? In his review of sidereal astronomy, Russell (1920) touched upon the subject, but apparently saw no reason to mention Pannekoek. There is no reference in the first half of the 1920s to an Eddington publication in ADS in which he expressed a view on the size of the Galaxy: maybe not surprising since by that time his interest had changed to stellar structure and relativity theory. Another leading astronomer in the Netherlands that could have commented on Pannekoek’s work was Ejnar Hertzsprung; he had assumed his position in Leiden as part of the same reorganization that failed to attract Pannekoek. However, there is also no publication by Hertzsprung that I could find in which he expressed an opinion on the size of the Galaxy. The same holds for Willem de Sitter. Searching in ADS for papers with `galaxy’, `galactic’, `Milky Way’, `Sidereal System’, etc. in the title between 1920 and 1925 did not uncover any papers in which reference was made to Pannekoek. In his comprehensive review of highlights in the study of the  Milky Way up to 1930, Smith (1985) made no mention of Pannekoek’s work. It seems his work on the distance of the Milky Way did not make much impression.

To summarize, Pannekoek’s attempt to derive distances to star clouds like in Cygnus resulted in him being convinced of a large system of stars comparable to Shapley’s globular cluster system. This large system consisted of star clouds and Kapteyn’s universe may be an example of this. Yet only after a few years he had to revise the distance to this and other clouds by a factor two or so and a year later to admit his method had failed to give reliable distances. It seems his support was not welcomed by Shapley.
\bigskip

The work on the distances of the bright stellar clouds of the Milky Way in Cygnus and Aquila was not the end of Pannekoek’s research on the structure of the Galactic System. The series of papers on the subject (Pannekoek, 1924, 1929a, 1929b, 1929c) included among others detailed work on the distribution of stellar clouds and agglomerations in the general neighborhood of the Sun. This was undertaken to investigate the effects of the structure in the Milky Way with longitude that Kapteyn had ignored. My summary of this work is that it seems that it only complicated matters and that no clear unifying picture emerged. Obviously, taking Pannekoek’s point of view of concentrating on details in the distribution of stars on the sky rather than data smoothed in longitude, is a very valid  one, and in the end it did yield some information on the local distribution of stars, but these results turned out rather different depending on the type of star in the data set used. As a result, it did not provide much new insight into the nature of the Milky Way Galaxy, in the absence of an understanding the Hertzsprung-Russell diagram in terms of stellar structure and evolution.

It is interesting to note that in a sense this failure to elucidate the local structure in the Galaxy bears some resemblance to the also unsuccessful work of van Rhijn (1936), who analyzed star counts in {\it Selected Areas} near the Galactic plane. He used average values for the amount of extinction, derived from reddening and distances from Galactic differential rotation, to correct for the effects of dust. The difference with Pannekoek’s work was that he assumed extinction to take place everywhere and rather uniformly in all directions. Like Pannekoek’s work, who instead assumed extinction to be restricted to dark clouds, the outcome of van Rhijn’s studies did not contribute any new insights or understanding (van der Kruit, 2022). Real progress came only when Oort used the data of the {\it Plan of Selected Areas} in Areas, that are situated {\it away from the Galactic plane} at moderate and high latitudes, and wrote the paper that constituted what I consider to be the major outcome of the {\it Plan} (Oort, 1938, see discussion in van der Kruit, 2019, 2021b, 2022). Ignoring the data at low latitudes, that Pannekoek and van Rhijn had concentrated on, and from galaxy counts, provided by Edwin Hubble, estimating the extinction, which now was all produced at small distances along the line of sight, he derived a crosscut through the Galactic disk, that still holds in broad terms.
\bigskip

An important line that Pannekoek followed was his thinking about the  dynamics of the Galaxy. Pannekoek's point was an excellent one when he resolved that following Shapley’s views it was important to see how the Kapteyn Star Streams could be fitted into the concept of a larger stellar system. And following Oort’s discovery of the rotation of the Galaxy to see if he could identify the central mass in the rotation center that was required to provide the gravitational force to correspond to the amount of rotation and the distance of the rotation center. In both inquiries he was not successful and did not produce much new insight. 
\bigskip

An area in which Pannekoek did contribute significantly, but did not get the recognition he deserved, concerned distances of dark clouds. His approach was rigorous and developed to give accurate distances, but required much effort, while the more popular, but quick and dirty approach by Wolf gave less accurate results. Although Pannekoek’s method is mentioned regularly in the literature, at least in the first two decades after he first presented it, modern references are very few. The {\it SAO/NASA Astrophysics Data System} (ADS)  is highly incomplete to citations before the 1950s. Keeping this in mind:  Pannekoek’s four papers on distances and properties of dark nebulae collected a total of 5 citations (discounting van der Kruit, 2022), while the paper by Wolf (1923) has 85! Considering that the comparison of star counts was already used by Barnard (1919)  and by Dyson \&\ Melotte (1919) and that it was Pannekoek (1919a, 1921a) who had devised the rigorous technique, the overwhelming attribution to Wolf (1923) is unjust. As an example take the statement from Reid (1993):
\begmarg
[...] after Barnard and Wolf (1923) convincingly demonstrated the existence of small-scale dark clouds (using star-counts in adjacent comparison fields), [...]
\endmarg
This is an unfortunate wording, since there of course is no such reference (Barnard and Wolf, 1923). Although Barnard is mentioned, albeit without a proper reference (but Dyson and Melotte and Pannekoek are not), the statement implies that the most significant paper in this development is Wolf (1923).

Did Pannekoek suffer from this lack of recognition? In his autobiographical notes (1982) he complained about lack of interest from  colleagues of his work in general. He recalled that when he had read the papers by Saha that `Dr. Groot’,  his successor as teacher in Bussum, had sent him, he had made some calculations on the physical conditions in stellar atmospheres deduced from their spectral lines (Pannekoek, 1922d). He then noted (Pannekoek, 1982; p.152; my translation):
\begmarg
I published a summary of my calculations in the Bulletin of the Astronomical Institutes of the Netherlands No. 19 (1922).  […] The piece in the BAN gave rise to some disappointment in that there was no response to it anywhere; Minnaert was the only one, from whom I received an appreciative letter about it. It often struck me as a disappointment that people took so little note of each other's work. When Publication 1 of Amsterdam appeared, I expected a friendly word in the `Observatory', which would have been important for the Amsterdam authorities, but it was completely ignored. At the time I took it as proof that all that work meant less in the grand scheme of astronomy than I had imagined. Later I understood, that younger people had noticed it; so the reason probably is, that all the older ones, who already had made a name, were so completely focused on their own work, that they had no time to thoroughly take notice of what was outside of it; I realized the same thing happened later on to myself.
\endmarg

On the other hand, when describing the festivities around him being awarded an honorary doctorate at Harvard in 1936, he recalled (p.271):
\begmarg
Here are all the special events and ceremonies of this celebration, the lodging in a student dormitory on Campus, the attendance at all kinds of meetings, the rained-out solemn presentation of all the honorary doctor's degrees, the lodging with one of Boston's notables (before this we, my wife and I, had stayed at the Agassiz Cottage on Oak Ridge, then in a student dormitory on Campus, and then at Menzel's), the many dinners and exhibitions, all of no special astronomical interest, as well as the meetings with [philosopher Rudolph] Carnap and with the Aristotle scholar Werner Jaeger. The only thing worth mentioning is, that the brief justification of my award started with the distance determination of the dark nebulae in Taurus, which I thought had not attracted any attention at all, because my contemporaries kept silent about it.
\endmarg

\subsection{Isophotal maps of the Milky Way}
Pannekoek had attempted to capture in drawings the appearance of the Milky Way ever since 1889 or 1890, when he was sixteen or seventeen years of age. This has become an almost lifelong occupation involving detailed mapping of first the northern Milky Way by visual observations, then extending that southern Milky Way from the Bosscha Observatory and further repeating the whole work using out-of-focus photographic observations.

 Pannekoek did not really state in his papers what possible use he saw for his Milky Way maps. At the end of the southern photographic survey he noted (Pannekoek, 1949; p.26):
\begmarg
 A comparison with the visual representations of the Milky Way […] shows at once what far greater wealth of detail is offered by the photographic method. We might describe the picture as the aspect the Milky Way would present to eyes that were far more sensitive to faint glares of light than ours and at the same time able to distinguish smaller details. A comparison with the focal photographs of Barnard and Ross shows a smoothing out of all sharp detail, thus gaining a true representation of the surface intensity which is lacking there.
\endmarg

This does not refer to any specific scientific question or a research program aimed at a deeper understanding of a cosmic phenomenon, property or structure. He must have done all this mapping work then without any specific scientific question in mind, but simply because the Milky Way is beautiful. He did comment on an extension of this work in the form of providing colors (Pannekoek, 1957), but no further analysis of his calibrated photographic surface photometry of the Milky Way has been published by him or collaborators. 

While doubts could be raised about the reliability of the visual surveys because of subjectivity and dependence on observer, this would seem unwarranted in the case of  his photographic work. The use of out-of-focus stars to calibrate at the same time the characteristic curve and through that the surface brightness  magnitude scale  as well as the zero point, would ensure a result well suited for use by others. There is one aspect of the reduction and calibration procedure that might raise doubts. The use of a wide range of photographic emulsions and developers and determination of the characteristic curve on a different plate, although specified as ‘the same kind of plate as the plate to be measured, {\it if possible’} (my italics) would certainly worry more recent users of photographic plates. The ‘of possible’ suggests that sometimes a plate has been calibrated with a separate plate with a different emulsion. It is not mentioned whether or not the exposure times on the calibration plates are similar to those on the survey plates to eliminate effects of LIRF (low-intensity reciprocity failure). More recent photographic surface photometry required utmost care to provide reliable results, such as uniformity in material and procedures. The outcome of the comparison with the Pioneer and Bochum surveys, however, does not point to major calibration problems, but one would think that there has been an element of luck in this.
\bigskip

I already noted that Pannekoek, although using them as a background -- and an important guide -- to a number of his studies of the local structure of the distribution of stars and dust, he did not analyze his maps in order to arrive at an overall, new picture of the Galaxy. It is difficult to see what more can be done than fitting large-scale distribution parameters of the Galaxy. In fact, this is also the main outcome of the extensive Bochum studies, who derived radial and vertical exponential scalelengths of the Galactic disk (Kimeswenger {\it et al.}, 1993), merely confirming my earlier (van der Kruit,  1986) results based on the Pioneer data.

What about others? I will below say a few words on the issue of usefulness and visibility of Pannekoek’s work as far as evident from bibliometrics, but here I note that in the {\it SAO/NASA Astrophysics Data System} (ADS) there are only a few citations to the four Milky Way maps: fourteen to the visual maps and one to the photographic maps and none of these analyze a significant part of the information. However, there is a problem with the Pannekoek \&\ Koelbloed (1949) paper. I discovered this only because I had at the time it was published been interested in the paper {\it Fluctuations in brightness of the Milky Way and interstellar clouds} by Peters (1970). My vague recollection resulted after much searching in locating this paper. It employs the Pannekoek \&\ Koelbloed (1949) maps of the southern Milky Way to derive structural parameters of the distribution of dust in the Galaxy. So why is this not entered in the ADS? The reason turned out ot be that the title {\it Photographic photometry of the southern Milky Way after negatives chiefly taken at the Bosscha Observatory at Lembang by Dr. J.G.E.G. Vo\^ute} caused the inclusion of Joan Vo\^ute as third author. No other astronomers would of course refer to this paper as Pannekoek, Koelbloed \&\ Vo\^ute! There probably are  (some) more citations that disappeared this way. One is easily found by looking at papers citing Peters' (1970), and that is Matilla \&\ Scheffler (1978) on a quite similar study. Another study using Pannekoek’s maps that is preceding Peters is Limber (1952), based on a theory of fluctuation in the brightness of the Milky Way by Chandrasekhar and M\"unch (1952), but I have not pursued that further.

The conclusion is that this massive effort has produced isophotal maps that were definitely up to standards at the time and are reasonably useful even now,  but that the actual use made of it by Pannekoek himself is minimal and by others very limited.

\begin{figure}[t]
\begin{center}
\includegraphics[width=0.84\textwidth]{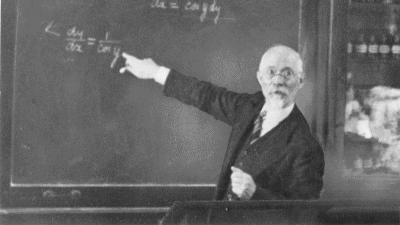}
\end{center}
\caption{\normalsize Pannekoek teaching later in his career. This photograph has been taken around 1939 during his mathematics lectures for students who had mathematics as a minor (physics and chemistry majors, and the like). Only half of Pannekoek’s teaching assignment was for astronomy; the other half concerned mathematics for minor students. From the Oort Archives.}
\label{fig:PkfromJHO}
\end{figure}

\subsection{Concluding remarks}
I first will have a look at some bibliometric statistics, using the {\it SAO/NASA Astrophysics Data System} (ADS). I consulted ADS for this on 29 September 2023. I look first at all Pannekoek's papers, including the ones on stellar atmospheres and astrophysics (see Fig.~\ref{fig:PkfromJHO}). The ADS lists 113 papers by Pannekoek, of which 71 refereed. However ‘refereed’ in the ADS sense leaves out his observatory publications (so Pannekoek’s Milky Way maps) and his KNAW papers. By hand I have left out books in Dutch and popular articles, after which 100 papers remain. These papers have collected  255 citations, mostly after the early 1950s, when journals started to collect references at the ends of papers rather than in footnotes. This is 2.5 per paper.  Since about the year 2000 ADS started to count ‘reads’ and ‘downloads’, a read being someone accessing the bibliographic record and a download an actual download of the scans or other full-text file of the paper. This took at number of years to pick up.  These Pannekoek papers collected 4409 reads and  2457 downloads.    Averages per paper  then are 44.1 reads  and 24.6 downloads per paper. These values return in Table~\ref{table:biblio}. In that table I collect the numbers for the same properties separated out into three categories: `Milky Way Galaxy', which stands for statistical astronomy and Galactic structure; `Astrophysics' for stellar (and solar) atmospheres and spectrophotometry; and `Stellar photometry' for photometry and studies of variable stars.

\begin{table}[t]
\caption{Bibliometric statistics of Pannekoek’s publications according to the \textit{SAO/NASA Astrophysics Data System} (ADS).}
\begin{center}
\begin{tabular}{lccccccc}
\hline
 Category & papers & citations & citations & reads & reads  & downloads & downloads \\
&&& per paper && per paper && per paper \\
\hline
All papers & 131\ \ \  & 262\ \  & 1.8 & 5596 & 49.5 & 2981\ \  & 26.3 \\
Scientific papers & 100\ \ \  & 255\ \  & 2.5 & 4409  & 44.1 & 2457\ \  & 24.6 \\
Milky Way Galaxy  & 31 & 35 & 1.1 & 1613 & 52.0 & 973 & 31.4 \\
Astrophysics & 32 & 144\ \   & 4.5 & 1528 & 47.8 & 893 & 35.2 \\
Stellar photometry  & 20 & 23 & 1.1 & \ \ 543 &  27.1 & 311 & 15.6 \\
\hline
\end{tabular}
\end{center}
\label{table:biblio}
\end{table}

The paper with the highest citation  count is the first paper on the application of Saha’s equation (Pannekoek, 1922d), which received 82 citations since 1966. On average it still has two citations per year, and therefore is a real classic, being still cited after more than a century. It has 263 reads (actually at present still 10-15 per year) and a total of 124 downloads (so more than 5 downloads annually). The second highest citation count is for the English translation of his book {\it De groei van ons wereldbeeld: Een geschiedenis van de sterrekunde} (Pannekoek, 1951), which appeared as {\it A history of astronomy} (Pannekoek, 1961). It still has on average one citation per year; there are a few reads per year (only to find out it is not on the Web in electronic form, I suppose), and of course no downloads. The most cited paper on Galactic studies is Pannekoek (1926) on the visual maps of the northern Milky Way (Pannekoek, 1920) on place 6; it has 9 citations, 157 reads and 87 downloads (currently still about 10 per year). The ADS has the year of publication wrong of this publication; it has 1929. I suspect that the southern photographic survey (Pannekoek and Koelbloed, 1949) would have ended much higher than now, had the ADS not added Vo\^ute as the third author; it has only 1 citation. There are 89 and 68 reads and downloads, presumably by persons that have found it as one of the publications of Pannekoek. The whole set of four papers with visual and photographic isophotal maps has collectively only 15 citations, but in recent years on average 30 reads and presumably 15-20 downloads annually. However, this is affected by the error in authorship made by ADS in  the southern photographic survey publication,

Pannekoek's papers on stellar atmospheres collect more than half his citations, but it may come as a surprise to some, that  Pannekoek's legacy, measured in terms ADS reads and ADS downloads, in the fields of Galactic mapping and structure on the one hand and of stellar atmospheres and astrophysics on the other is comparable. It is a tribute to Pannekoek's efforts that still his publications are looked up on ADS ('reads') 250 times each year and (if indeed proportional) every year on average about almost a century a full electronic text of his publications is downloaded more than 100 times!
\bigskip

To conclude this paper I quote from Pannekoek's autobiographical notes (Pannekoek, 1982, p.273): 
\begmarg
When I look back at all this, it is clear in hindsight that there has been quite some research of details and introduction of new methods, but not a single problem has been solved, no discovery made, and no new direction identified. This is probably so because we are living in a time of social crisis, which has strongly affected me, and reflecting on this has always formed the deep basis of all my  thinking. I did see the value of scientific and intellectual work; but the persons that devote their intellectual power to it is very large, and attracts the most talented and brilliant minds. Those that study the social issues of today at a similar scientific level is small; also because everything there is so much more obscure. 
\endmarg

The first sentence is in my view much too harsh a judgment, even when one only considers his Galactic research. The justification that he has devoted much time to social issues, while so few others do likewise, is therefore unnecessary. 

In response to one of a referee’s remarks I summarize some salient questions. Why did Pannekoek make the choices he did? His approach to the Galaxy as an object for study has been heavily biased by his fascination and very thorough and detailed mapping the of the Milky Way. His view of the Sidereal System was primarily its appearance to the visual observer of the Milky Way, contrary to Kapteyn whose approach related to star counts. How did these choices fit into the ways of working of his peers? The answer can be short: it didn’t. Almost all astronomers, especially during the 1920s did not share his view that the Sidereal System was primarily made up of clouds of stars like his Cygnus Cloud with a low-density smooth underlying stratum but rather of an extended distribution with condensations or density enhancements superposed on it. Why did other astronomers pay so little attention? They did pay attention, but probably did not find his results sufficiently convincing to accept them and follow them up. 

In a double obituary in {\it Sky \&\ Telecope} for both Pannekoek and Pieter van Rhijn, who died days from each other, Bart Bok, himself an authority on Galactic structure, wrote about Pannekoek among others  (Bok, 1960, p.74):
\begmarg
	During the first 40 years of his life, with Pannekoek's political activities seeming to overshadow his scientific work, there was no clearly defined line to his research efforts. But at the early age of 18, he had discovered the variability of Polaris from visual estimates. [...] Toward the beginning of his association with Amsterdam, Pannekoek’s plans for an over-all investigation of our Milky Way system developed. More than any other astronomer, he stressed the importance of accurate spectrographic classification for the study of the nearer parts of the Milky Way system, and his paper on the distribution of A and B stars, in Amsterdam Publications, Vol. 2, had a profound effect on our thinking about galactic structure. He developed, for the first time, the approach used so successfully in recent years by V. A. Ambarzumian and by W. W. Morgan in studying the spiral structure and grouping of stars in the galaxy.

	Pannekoek was above all a man in love with the beauty of the heavens. He began his observations of the Milky Way at the age of 16. On his travels, especially in the tropics (to Sumatra for the eclipse of January 14, 1926), he took every opportunity to draw, photograph, and measure the Milky Way's brightness distribution. Astronomers continue to consult his papers on the subject with pleasure and profit. 
 \endmarg

Bok then went on to  describe his work on stellar atmospheres. I can subscribe to Bok’s remarks, but would add Pannekoek’s method of deriving distances (but not the total extinction) to dark clouds, which Bok (1931) himself advocated as better than Wolf’s. These three things --the mapping of the structure of the nearby part of the Galaxy, the distances of dark clouds, and isophotal maps of the Milky -- constitute the legacy of Pannekoek in the field of Galactic research. Other investigations turned out inconclusive or wrong as a result of his conviction, based on his many years of observing and mapping the Milky Way on the sky, that the nearby structure of the Galaxy is characterized by more or less isolated clouds of stars and that dust is restricted to separate dark clouds.  But that does not take away from the devotion, perseverance and insightfulness evident from his total effort based on his desire to understand his beloved the Milky Way and the stellar system it represents.

\vspace{1cm}

\noindent
{\LARGE {\bf Appendix A: In which constellation is/was the Galactic center located?}}
\bigskip

The position of the Galactic center and the radio source associated with it is uniformly quoted as being in the constellation Sagittarius, but I suspect many people do not realize this is so only just; it is very close to the border with Ophiuchus as defined by the IAU. Sagittarius extends over about 2 hours 50 minutes in Right Ascension or some 37\degs, while the center of the Galaxy lies only less than about half a degree from the border with Ophiuchus to the west. Towards the south it is about one degree from Scorpius.  For a nice map showing the location see ESO (2020). The Galactic anti-center, contrary to the center, does not lie in a constellation of the zodiac; it is situated in  Auriga (the ecliptic misses this constellation by about 4\degs), but almost on the border with zodiacal Taurus and  approximately 2\degs\ west of the corner of Gemini.

How was this when Shapley determined the center of his globular cluster system? First, we need to know what definition of galactic coordinates was used by Shapley. The paper, in which the current system has been defined (Blaauw {\it et al.}, 1960) has a brief summary of the history of the subject. From 1932 the system defined by Lund Observatory (Ohlsson, 1932) had been in fairly uniform use, based on a Galactic pole  at ($\alpha, \delta$) = (12\hs\ 40\ms, +28\degs, 1900.0). Before this a number of different systems were used. Shapley (1918) does not specify what system he used, but it seems to be a reasonable assumption that he used the Harvard system of Pickering (1912), which is actually the one later also adopted by Ohlsson for the Lund system. In both cases the longitude zero-point (l$^{\rm I}$, b$^{\rm I}$) = (0\degs, 0\degs) was one of the interceptions of the Galactic equator with the celestial one, which with the definition of the Galactic pole then has coordinates  ($\alpha, \delta$) = (18\hs\ 40\ms, +0\degs, 1900.0). With this it is straightforward to calculate that Shapley's position  (l$^{\rm I}$, b$^{\rm I}$) = (325\degs, 0\degs) corresponds to  ($\alpha, \delta$) = (17\hs\ 27\mspt3, -30\degs 26\mins, 1900.0) or  after correction for precession ($\alpha, \delta$) = (17\hs\ 33\mspt7, -30\degs 30\mins, J2000.0). Similarly Oort’s position  (l$^{\rm I}$, b$^{\rm I}$) = (323\degs, 0\degs) corresponds to  ($\alpha, \delta$) = (17\hs\ 22\mspt1, -32\degs 56\mins, 1900.0) or ($\alpha, \delta$) = (17\hs\ 28\mspt 5, -32\degs 11\mins, J2000.0). Comparing this with the IAU constellation maps gives the surprising result that the Shapley’s position in fact is in Scorpius but close to the border with Ophiuchus, and Oort’s well (a little over two degrees) inside Scorpius. However, these constellation borders have only been adopted by the IAU in 1928 at the third  General Assembly in Leiden (IAU, 2022);  the constellation names were adopted at the first General Assembly in Rome in 1921) and published by Eug\`ene Delporte (1930a,b) of the Brussels Royal Observatory. So both Shapley’s and Oort’s remarks on a location in Sagittarius, as echoed by Pannekoek, preceded the definition of constellation borders. Compared to the brighter stars in both constellations, the positions of Shapley and Oort to the sky watcher lie in between Sagittarius and Scorpius (locate Shapley’s and Oort's position above in the ESO (2020) map to appreciate this). The brighter stars of Ophiuchus lie further north than those of Sagittarius or Scorpius, so this constellation would not qualify, but in Shapley’s and Oort's days Scorpius could have served equally well as Sagittarius as the constellation that harbors the Galactic center. 
\vspace{1cm}

\noindent
{\LARGE {\bf Appendix B: Some one-paragraph Curriculae Vitae}}
\bigskip

For those readers unfamiliar with some Dutch astronomy PhDs that received their degree early in the twentieth century and figure in this paper, I present brief summaries of their careers, some of which are shorter versions of the descriptions in my Kapteyn biography (van der Kruit, 2015, 2021). I start with Easton, and then the years of their thesis defense are arranged chronologically 
\bigskip

\noindent
{\bf Cornelis Easton} (1864--1929) was primarily a journalist, but in addition was an accomplished amateur astronomer. After his secondary education, he studied various subjects at what is now the Technical University of Delft, dropped out and obtained the qualifications for a secondary school teacher in French. However, he took on jobs as a newspaper journalist, eventually as editor (in-chief) of some daily or monthly newspapers and magazines. He had developed a strong interest already as a child in astronomy and as a young man started producing drawings of the Milky Way, some of which he published. Easton had been intrigued by spiral nebulae and became convinced that our Galaxy had to have spiral structure as well. He developed a model for the Galactic spiral structure from the appearance of the Milky Way with the remarkable, but in hindsight correct conclusion that the Sun was well away from the center and in areas in the Milky Way that were relatively bright we were looking along spiral arms and in darker areas in between such arms. In 1900 he published his ideas in a paper titled {\it A new theory of the Milky Way} in the Astrophysical Journal (Easton, 1900). He eventually improved this using photographs of the Milky Way from many sources (Easton, 1913). Although Kapteyn did not subscribe to all Easton’s ideas, he must have felt sufficiently impressed by Easton to arrange he would be awarded a honorary doctorate in 1903, very likely to mark his jubilee of 25 years as Professor in Groningen (van der Kruit, 2021c). He died at age 64, less than a year before his retirement. Mrs. Easton has been rather unhappy with him spending so much of his spare time on astronomy.
\bigskip 

\noindent
{\bf Lambertus Yntema} (1879--1932) produced in 1909 a thesis under the supervision of Kapteyn with the title {\it On the brightness of the sky and the total amount of starlight — an experimental study}. Yntema was born in a very small village on the west coast of Friesland. After obtaining his PhD, Yntema left the university and astronomy. The records state that he and his family moved to Leeuwarden, the Frisian capital, in 1918 from Breda in the south of the country and that he was registered in Leeuwarden as director of a Christian high school of the type HBS (Higher Citizen's School; unlike in a gymnasium or grammar school no Greek and Latin was taught at an HBS, but there was more emphasis on natural sciences; the science teachers held PhDs, so it provided excellent preparation for university studies in natural sciences). He moved from there to Bussum, not far from Amsterdam, in 1931 to become director of a Christian Lyceum, a secondary school that combined a Gymnasium and an HBS. He died the next year at the relatively young age of 53 years.
\bigskip

\noindent
{\bf Isidore Henri Nort} (1872--1960) defended a thesis in 1917 in Utrecht under Albertus Antonie Nij\-land (1868--1936) on {\it The Harvard Map of the sky and the Milky Way}. According to the Introduction the research was performed following a suggestion by Nijland to look into the major differences in the magnitude scales among brighter stars of Kapteyn and of Pickering and others. Kapteyn did not figure in the acknowledgments, so apparently he had little or nothing to do with this work. After his doctorate, Nort left active astronomical research, becoming a Gymnasium and HBS teacher in Zutphen and Gouda and author of text- and popular books. He published four further astronomical papers up to 1950 through Utrecht Observatory. He was of good health; in a newspaper announcement of his death at age 87, his second wife, whom he married at age 58 while she was only 22, described his surmise as ‘still rather unexpected’
\bigskip

\noindent
{\bf Willem Johannes Adriaan Schouten} (1893--1971), son of a clergyman, studied astronomy under Kapteyn and wrote a thesis {\it On the determination of the principal laws of statistical astronomy}, in 1918. After that he became ‘privaatdocent’ in  astronomy (private lecturer, an in principle unpaid, usually five-year position to teach in a discipline that is not formerly covered) with Pannekoek at the University of Amsterdam, before he became high school (Lyceum, comprising an HBS and a Gymnasium streams) teacher in Arnhem.  He wrote nineteen books and translated one, all popular ones on astronomy, physics and physicists. He died in 1971 at age 78 after having been widowed twice, in 1947 and  again in 1965.
\bigskip

\noindent
{\bf Egbert Adriaan Kreiken} (1896–1964) had a special link with Kapteyn, because his parents had bought the boarding school in Barneveld from Kapteyn’s mother, after Kapteyn’s father had died. So Kreiken and Kapteyn were born in the same house! Kreiken studied with Kapteyn and also started his PhD research under his supervision; Kreiken always considered Kapteyn his real teacher, although he finally defended his thesis with Van Rhijn in 1923, when Kapteyn had already died.  The thesis was entitled {\it On the colour of the faint stars in the Milky Way and the distance of the Scutum group}; it concerned the colors of stars, based on photographic material obtained by Ejnar Hertzsprung at Mount Wilson. Kreiken continued to work in astronomy; eventually he founded the Ankara Observatory in Turkey, which is now named after him. After his thesis he taught at an HBS in Amsterdam until 1928, from 1926 also as privaat-docent at the University of Amsterdam, succeeding Schouten.  After this he moved to the Bosscha Observatory at Lembang in the Dutch East Indies and subsequently to Liberia and Turkey. For a detailed biography see Omay (2011).
\bigskip

\noindent
{\bf Broer Hiemstra} (1911--1994) was born a village in the west of Friesland as a son of an elementary school teacher. He had passed his doctoral exam in Groningen in 1936, but could not be appointed Assistant by van Rhijn, because this position was occupied by a P.P. Bruna, who never completed a PhD thesis. Hiemstra defended a thesis in 1938 on {\it Dark clouds in Kapteyn’s Special Areas 2, 5, 9 and 24 and proper motions of the stars in these regions} (Hiemstra, 1938). He married the same year; in the certificate of his marriage in Dokkum (Friesland) he is listed as having the profession ‘teacher’ (according to a newspaper clipping they were engaged to marry already in 1931!), so  he probably supported  himself as part-time teacher while working on his thesis. Maybe this was turned into a full-time appointment after obtaining the PhD degree, which made the marriage possible (he was 28 years by then and his bride, also an elementary school teacher, 29). They settled in Amsterdam where the young Mrs Hiemstra tragically died in 1940, one would think probably in child birth. Hiemstra remarried and in 1976 at the age of 65 retired as Director of a large secondary school conglomerate in the Hague. He died at age 83.

\vspace{1cm}

}

\noindent
{\bf Acknowledgments} 
{\normalsize
I am grateful to Ed van den Heuvel and Chaokang Tai for email correspondence while preparing this paper, and their willingness to read a draft. I thank Ed for permission to use his photograph of the Pannekoek residence at the Bosscha Observatory. I remain very grateful to Drs. J.L. Weinberg and G.N. Toller for providing me almost forty years ago with the Pioneer 10 results. I thank one of the three referees for making a few constructive remarks which resulted in including a bit more historical context and references to existing literature and another one of these for suggesting some improvements and providing a few links on the Web that I had missed. I thank the staff of the Kapteyn Astronomical Institute for support and help and the Director, Professor L\'eon Koopmans, for hospitality extended to an Emeritus Professor as Guest Scientist. This research made extensive use of the {\it SAO/NASA Astrophysics Data System} (ADS). 
\vspace{0.7cm}
}

\section{References}
{\normalsize

Note: Pannekoek’s publications in the reports and proceedings of the Royal Netherlands Academy of Sciences (KNAW) are not downloadable through links to electronic versions in the {\it SAO/NASA Astrophysics Data System} (ADS), but these are available through {\it The Digital Web Centre for the History of Science (DWC)} of the KNAW at dwc.knaw.nl?pagina=0\&s=Pannekoek\&gender=all\&type=publication\_docum\linebreak[4]
ents\&publicaties.
\bigskip

Abbot, C.G., 1914. The direct and scattered radiation of the sun and stars. {\it Astronomical Journal}, 28, 129-135.

Argelander, F., 1844. Aufforderung an Freunde der Astronomie, zur Anstellung von eben so interessanten und n\"utzlichen, als leicht auszuf\"uhrenden Beobachtungen \"uber mehrere wichtige Zweige der Himmelskunde. {\it Schumacher's Jahrbuch f\"ur 1844}, ed. H.C. Schumacher, Stuttgart \&\ T\"ubingen: J.G. Cotta, pp. 122-254. Translated in 1855 as {\it Handleiding voor vrienden der sterrekunde, tot het volbrengen van belangrijke waarnemingen, die geene werktuigen vorderen, uit het Hoogduitsch  vertaald door W.F. Kaiser, met eene voorrede, aanteekeningen en bijvoegsels van F. Kaiser}. Zwolle, der erven Tijl.

Bahcall, J.N. \&\  Soneira, R.M., 1980. The universe at faint magnitudes. I. Models for the Galaxy and the predicted star counts. {\it  Astrophysical Journal, Supplement Series}, 44, 73–110.

Barnard, E.E., 1919. On the dark markings of the sky, with a catalogue of 182 such objects. {\it Astrophysical Journal}, 49, 1-24.

Barnard, E.E., 1927. {\it A photographic atlas of selected regions of the Milky Way}. Washington, Carnegie Institution of Washington.

Bessel, M.S., 2005. Standard photometric systems. {\it Annual Review of Astronomy \&\ Astrophysics}, 43, 293-336.

Blaauw, A., Gum, C. S., Pawsey, J. L., \&\ Westerhout, G., 1960. The new I.A.U. system of galactic coordinates (1958 revision). {\it Monthly Notices of the Royal Astronomical Society}, 121, p.123-131. 

Boeddicker, O., 1892. {\it The Milky way from the North pole to 10 of south declination drawn at the Earl of Rosse's observatory at Birr Castle}, London, New York, Longmans, Green, and co.

Bok, B.J., 1931. The analysis of star counts. {\it Harvard College Observatory Circular}, vol. 371, 1-40.

Bok, B.J., 1960. Two famous Dutch astronomers. {\it Sky and Telescope}, 20, 74-76.

Br\"unnow, F.F.E., 1951. {\it Lehrbuch der sph\"arischen Astronomie}, Berlin, F. Dummler, 1851.

Cannon, A. J. \&\ Pickering, E. C. (1918-1924). The Henry Draper Catalogue. {\it Annals of Harvard College Observatory}, 92-99.

Chandrasekhar, S. \&\  M\"unch, G., 1952. The theory of the fluctuations in brightness of the Milky Way. V.  {\it Astrophysical Journal},  115, 103-123.

Code A.D. \&\ Houck, T.E., 1955. Wide-angle infrared photograph of the southern Milky Way. {\it Astrophysical Journal}, 121, 553-554. 

de Vaucouleurs, G., 1979. The first half century of galaxy photometry: Methods and results. In: {\it Photometry, kinematics and dynamics of galaxies}. Ed. D.S. Evans. Austin, Department of Astronomy, University of Texas at Austin, p.1-15.

Delporte, E., 1930a. {\it Delimitation scientifique des constellations (tables et cartes)}, Cambridge University Press.

Delporte, E., 1930b. {\it Altas celeste}, Cambridge University Press.

Dyson, F.W. \&\ Melotte, P.J., 1919. The region of the sky between R.A. 3\hs   and 5\hs 30\ms and N. Dec. 20\degs\ to 35\degs, {\it Monthly Notices of the Royal Astronomical Society}, 80, 3-7.

Easton, C., 1893. {\it La  Voie Lact\'ee dans l'hemisphere boreal: cinq planches lithographi\'ees, description d\'etaill\'ee, catalogue et notice historique}, Dordrecht, Blusse et cie; Paris, Gauthier Villars et fils.

Easton, C., 1900. A new theory of the Milky Way. {\it Astrophysical Journal}, 12, 136-158.

Easton, C., 1903. La distribution de la lumi\`ere Galactique, compar\'ee \`a la distribution des \'etoiles catalogu\'ees, dans la Voie Lact\'ee bor\'eal. {\it Verhandelingen der Koninklijke Akademie van Wetenschappen te Amsterdam}, VIII, 1-46.

Easton, C, 1913. A photographic chart of the Milky Way and the spiral theory of the Galactic System. {\it Astrophysical Journal}, 37, 105-118.

Easton, C., 1921. On the distance of the Galactic star-clouds. {\it Monthly Notices of the Royal Astronomical Society}, 81,  215-226. 

Easton, C., 1922. Correlation of the distribution of bright stars and Galactic light in Cygnus, {\it Bulletin of the Astronomical Institutes of the Netherlands}, 1, 157-159.

Easton, C., 1928. A photographic chart of the northern Milky Way. {\it Monthly
Notices of the Royal Astronomical Society}, 89, 207-209.

Eddington, A.S., 1914. {\it Stellar movements and the structure of the universe}, London, Macmillan.

Eddington, A.S., 1918. The dynamical problems of the stellar system. {\it The Observatory}, 41, 132-137.

Eddington, A.S., 1926. Bakerian Lecture. Diffuse matter in interstellar space,. {\it Proceedings of the Royal Society of London,  Series A}, 111, 424-456.

ESO (European Southern Observatory), 2009. {\it The Milky Way panorama}. www.eso.org/public/images/\linebreak[4]
eso0932a/.

ESO (European Southern Observatory), 2020. {\it Location of the Galactic centre in the night sky}. See www.eso.org/public/images/eso1920c/.

Evans, M.R., Schaefer, G.H., Bond, H.E., Bono, G., Katovska, M., Nolan, E.m Sasselov, D., \&\ Madonna B.D., 2008. Direct detection of the close companion of Polaris with the Hubble Space Telescope. {\it The Astronomical Journal}, 136, 1137-1146.

Fath, E.A., 1909. The spectrum of the zodiacal light. {\it Lick Observatory Bulletin}, no. 165, 141-143.

Fath, E.A., 1912. The integrated spectrum of the Milky Way. {\it Astrophysical Journal}, 36, 362-367.

Graff, K., 1920. Die Umrisse und Helligkeitsverh\"altnisse der Milchstra\ss e n\"ordlich von 25 s\"udlicher Deklination. {\it Astronomische Abhandlungen der Hamburger Sternwarte},  2, nr.5, 41-49.

Habison, P., 1999. Karl Schwarzschild's investigations of ‘out-of-focus photometry’ between 1897-1899 and his contribution to photographic photometry. {\it Astronomische Gesellschaft Abstract Series}, No. 15, p. 81, K04. Abstract only, see ADS: 1999AGAb…15…81H.

Hanner, M.S., Weinberg, J.L., DeShields, L.M., Green., BA., \&\ Toller, G.N, 1974. Zodiacal light and the asteroid belt: The view from Pioneer 10, {\it  Journal of Geophysical Research}, 79, 3671-3675.

Hartmann.J., 1899. Apparatus and method for the photographic measurement of the brightness of surfaces. {\it Astrophysical Journal},  10, 321-332.

Hearshaw, J.B., 1996. {\it The measurement of starlight: Two centuries of astronomical photometry}. Cambridge, Cambridge University Press (ISBN 0-521-40393-6).

Heis, E, 1872. {\it  Atlas Coelestis Novus}, Cologne: DuMont-Schauberg, xiii+181; www.lu.lv/en/muzejs/pet\linebreak[4]
nieciba/izstades/zemes-miti-debesis-zvaigznu-atlantu-virtuala-izstade/9-eduards-heiss-1872-atlas-coeles\linebreak[4]
tis-novus-neuer-himmels-atlas-jaunais-debess-atlants/.

Henshaw, C., 2014. {\it On the visual appearance of the Milky Way}. www.academia.edu/31961970/On\_the\_\linebreak[4]
Visual\_Appearance\_of\_the\_Milky\_Way.

Herschel, W., 1785. On the construction of the heavens. {\it Philosophical Transactions of the Royal Society of London}, 75, 213-266.

Hertzsprung, E.,1910. Vorschlag zur Festlegung der photographischen Gr{\"o}{\ss}enskala bei au{\ss}erfokalen Aufnahme. {\it Astronomische Nachrichten}, 186, 177-184.

Hertzsprung, E., 1911a. Nachweis der Ver\"anderlichkeit von $\alpha$ Ursae minoris. {\it Astronomische Nachrichten}, 189, 89-104.

Hertzsprung, E., 1911b.   Bearbeitung der J.F.J. Schmidtschen Beobachtungen von 68 u  Herculis. {\it Astronomische Nachrichten}, 189, 245-256.

Hertzsprung, E., 1911c. Photographische Beobachtung eines Hauptminimums von 68 u Herculis. {\it Astronomische Nachrichten}, 189, 255-258.

Hertzsprung, E., 1911d. \"Uber die photographische Schw\"arzungskurve. {\it Astronomische Nachrichten}, 190, 119-120.

Hiemstra, B., 1938. Dark clouds in Kapteyn’s Special Areas 2, 5, 9 and 24 and proper motions of the stars in these regions. {\it Publications of the Astronomical Laboratory at Groningen}, 48, 1–74.

Hoffmann, B., Tappert, C., Schlosser, W., Schmidt-Kaler, Th., Kimeswenger, S., Seidensticker, K., Schmidtobreick, L., a
nd Hovest, W., 1998. Photographic surface photometry of the Southern Milky Way. VIII. High-resolution U, V and R surface photometries of the Southern Milky Way. {\it Astronomy and Astrophysics Supplement Series}, 128, 417-422.
  
Houzeau, J.-C., 1878. {\it Uranometrie Generale}, Annales de l'Observatoire Royal de Bruxelles. Astronomie; nouveau series I,  Bruxelles : F. Hayez,  x+117, see ui.adsabs.harvard.edu/abs/1878AnOBN\linebreak[4]
...1....5H/abstract.

Hubblesite, 2006. {\it There's more to the North Star than meets the eye}, 
hubblesite.org/contents/news-releases/2006/news-2006-02.html.

IAU {International Astronomical Union}, 2022. {\it The constellations}, version 8. www.iau.org/public/themes/\linebreak[4]
constellations/\#n8.

Jeans, J.H., 1916. On the theory of star-streaming and the structure of the universe, {\it Monthly Notices of the Royal Astronomical Society}, 76,.552-572.

Jeans, J.H., 1919. {\it Problems of cosmogony and stellar dynamics}. Cambridge, University Press.

Jeans, J.H., 1922. The motions of stars in a Kapteyn-Universe. {\it Monthly Notices of the Royal Astronomical Society}, 82, 122-132.

Kapteyn, J.C., 1902. On the luminosity of the fixed stars. {\it Publications of the Astronomical Laboratory Groningen}, 11, 3-32.

Kapteyn, J.C. 1904. Statistical methods in stellar astronomy. Presentation at the Louisiana Purchase Exposition (also known as the fourteenth World’s Fair), held at St. Louis, Missouri, {\it International Congress of Arts and Sciences, Vol. VIII: Astronomy and Earth Sciences}, 396–425.

Kapteyn, J.C. 1906. {\it Plan of Selected Areas}. Groningen, Hoitsema Brothers (Astronomical Laboratory at Groningen).

Kapteyn, J.C., 1908. On the number of stars of determined magnitude and determined galactic latitude. {\it Publications of the stronomical Laboratory Groningen}, 18, 1-54.

Kapteyn, J.C. 1909a. On the absorption of light in space. Second paper. {\it Astrophysical Journal}, 30, 284-317.

Kapteyn, J.C., 1909b. Correction to Professor Kapteyn's article in the November number. {\it Astrophysical Journal}, 30, 398-399.

Kapteyn, J.C. 1914. On the individual parallaxes of the brighter Galactic helium stars in the southern hemisphere, together with considerations on the parallax of stars in general. {\it Astrophysical Journal}, 40, 43–126.

Kapteyn, J.C. 1918a,b,c. On the parallaxes and motion of the brighter Galactic helium stars between Galactic longitudes 150\degs\ and 216\degs. {\it Astrophysical Journal}, 47,  104-133, 146–178 and 255-282.

Kapteyn, J.C.\&\ van Rhijn, P.J., 1920. On the distribution of the stars in space especially in the high galactic latitudes. {\it Astrophysical Journal}, 52, 23–38.

Kapteyn, J.C., 1922. First attempt at a theory of the arrangement and motion of the Sidereal System. {\it Astrophysical Journal}, 55, 302–328.

Kimeswenger, S., Hoffmann  B., Schlosser, W. amd Schmidt-Kaler, T., 1993. Photographic surface photometry of the Milky Way. VII. High-resolution B surface photometry of the southern Milky Way. {\it Astronomy and Astrophysics, Supplement Series}, 97, 517-526.

Kreiken, E.A., 1926. On the centre of the local star-system. {\it Monthly Notices of the Royal Astronomical Society},  86, 665- 686.

Kuijken, K.H. \&\ Gilmore, G., 1989. The mass distribution in the Galactic disc. III - The local volume mass density. {\it Monthly Notices of the Royal Astronomical Society}, 239, 651-664.

Lattis, J., 2014. The Stebbins Galaxy: The origins of interstellar medium studies in the shrinking supergalaxy. {\it Journal of Astronomical History and Heritage}, 17, 240-257.

Leiden Observatory, 1908.{\it Album Amicorum}, presented to H.G. van de Sande Bakhuyzen. See collection\linebreak[4]
guides.universiteitleiden.nl/archival\_objects/0696abf26c3f8a432ce29fa15914ece0.

Leinert, Ch., Leinert, Bowyer, S., Haikala, L.K., Hanner, M.S., Hauser, M.G., Levasseur-Regourd, A.-Ch., MannI.,  Mattila, K., Reach, W.T., Schlosser, W.,  Staude, H.J., Toller, G.N., Weiland, Weinberg, J.L., \&\ Witt, A.N., 1997. The 1997 reference of diffuse night sky brightness, {\it Astronomy \&\ Astrophysics Supplement Series}, 127, 1-99.

Limber, D.N.,  1952. Fluctuations in brightness of the Milky Way. {\it Astrophysical Journal}, 117, 145-168.

Lund Observatory, 1961. Lund Observatory table for conversion of
galactic coordinates. {\it Annals of the Observatory of Lund}. No. 17.

Maass, A., 1926. Sternkunde und Sterndeuterei im malaiischen Archipel, published in three installments: {\it Tijdschrift voor Indische Taal-, Land- en Volkenkunde}, 64 (1924) 1-172 \&\ 347-459, and 66 (1926), 618-670. The journal is online af kitlv-docs.library.leiden.edu/open/Meta\linebreak[4]
morfoze/TBG/tbg.html.

Matilla, K. \&\ Scheffler. H., 1978. Fluctuations in the brightness of the diffuse galactic light and in the brightness of the Milky Way. {\it Astronomy and Astrophysics}, 66, 211-224.

Miller, F.D., 1937. The analysis of general star-counts in obscured regions, {\it Astronomical Journal}, 46, 165-169 (1937).

Morgan, W.W., Whitford, A.E. \&\ A.D.Code, A.D., 1953.  Studies in Galactic structure. I. A preliminary determination of the space distribution of the blue giants. {\it Astrophysical Journal}, 118, 318-322.

Nort, I.H., 1917. The Harvard map of the sky and the Milky Way, {\it Recherches Astronomiques de l'Oservatoire d'Utrecht}, 7, 1-179, Plate I-IV.

Ohlsson, J., 1932. Tables for the conversion of equatorial coordinates into galactic coordinates. {\it Annals of the Observatory of Lund}, 3, 1-151.

Omay, C.G., 2011. {\it Prof. Dr. Egbert Adriaan Kreiken: Founder of the Ankara University Observatory and a volunteer of education}, with contributions by Juus Kreiken. Ankara University, ISBN : 978-605-87419-0-4. See astronomy.science.ankara.edu.tr/wp-content/uploads/sites/124/2022/05/EAK\_Life\_Story\_\linebreak[4]
Book\_CGO.pdf.

Oort, J.H., 1927. Observational evidence confirming Lindblad's hypothesis of a rotation of the Galactic system. {\it Bulletin of the Astronomical Institutes of the Netherlands}, 3, 275-282.

Oort, J.H., 1932. The force exerted by the Stellar System in the direction perpendicular to the Galactic plane and some related problems. {\it Bulletin of the Astronomical Institutes of the Netherlands}, 6, 249–287.

Oort, J.H., 1938. Absorption and density distribution in the Galactic System. {\it Bulletin of the Astronomical Institutes of the Netherlands}, 8, 233–264.

Oort, J.H., 1940. Some problems concerning the structure and dynamics of the
Galactic System and the elliptical nebulae NGC 3115 and 4494. {\it Astrophysical
Journal}, 91, 273-306.

Oort, J.H., 1981. Some notes on my life as an astronomer. {\it Annual Review of Astronomy and Astrophysics}, 19, 1-5.

Osterbrock, D.E. \&\ Sharpless, S., 1951. Photographs with the Henyey-Greenstein
wide- angle camera. {\it Astrophysical Journal}, 113, 222.

Pannekoek, A., 1897. On the necessity of further researches on the Milky Way. {\it Popular Astronomy}, 5, 395-399.

Pannekoek, A., 1898a. II. New charts for inserting the Milky Way. {\it Popular Astronomy}, 5, 485-488; also {\it Journal of the British Astronomical Association}, 8, 80-82 (1897).

Pannekoek, A., 1898b. III. On the best method of observing the Milky Way. {\it Popular Astronomy}, 5, 524-528.

Pannekoek, A., 1906. The luminosity of stars of different types of spectral types. {\it Koninklijke Nederlandsche Akademie van Wetenschappen Proceedings}, 9, 134-148.

Pannekoek, A., 1910. Researches into the structure of the Galaxy. {\it Proceedings of the section of sciences of the Royal Netherlands Academy of Sciences}, 13, 239-258.

Pannekoek,A., 1912. A photographical method of research into the structure of the Galaxy.  {\it Proceedings of the section of sciences of the Royal Netherlands Academy of Sciences}, 14, 579-584.

Pannekoek, A, 1913. Die Ver\"anderlichkeit des Polarsterns. {\it Astronomische Nachrichten}, 194, 359-362.

Pannekoek, A., 1919a. Investigation of a Galactic cloud in Aquila. {\it Proceedings of the section of sciences of the Royal Netherlands Academy of Sciences}, 21, 1323-1337 

Pannekoek, A.,1919b. The distance of the Milky Way. {\it Monthly Notices of the Royal Astronomical Society}, 79, 500-507.

Pannekoek, A., 1919c. On the distribution of the stars down to the eleventh magnitude (A reply to Dr Nort). {\it Monthly Notices of the Royal Astronomical Society}, 80, p.198-200.

Pannekoek, A., 1920. Die N\"ordliche Milchstrasse, {\it Annalen van de Sterrewacht te Leiden}, 11, 1-89, A90-A114, Tafel I-IX. (ADS has year incorrectly as 1929).

Pannekoek,A., 1921a.The distance of the dark nebulae in Taurus. {\it Proceedings of the section of sciences of the Royal Netherlands Academy of Sciences}, 23, 707-719.

Pannekoek, A., 1921b. Further remarks on the dark nebulae in Taurus. {\it Proceedings of the section of sciences of the Royal Netherlands Academy of Sciences}, 23, 720-726.

Pannekoek, A., 1921c. Die absolute Helligkeit des Milchstrassenlichtes. {\it Astronomische Nachrichten},  214, 389-392.

Pannekoek, A., 1922a. Systematic errors in the `Durchmusterung of Selected Areas'. {\it Bulletin of the Astronomical Institutes of the Netherlands}, 1, 53.

Pannekoek, A., 1922b. The distance of the Galaxy in Cygnus. {\it Bulletin of the Astronomical Institutes of the Netherlands}, 1, 54-56.

Pannekoek,A., 1922c. The local star system, {\it Proceedings of the section of sciences of the Royal Netherlands Academy of Sciences}, 24, 56-63.

Pannekoek, A., 1922d. Ionization in stellar atmospheres.  {\it Bulletin of the Astronomical Institutes of the Netherlands}, 1, 107-118.

Pannekoek, A., 1923a. Luminosity function and brightness for clusters and Galactic clouds. {\it Bulletin of the Astronomical Institutes of the Netherlands}, 2, 5-12.
 
Pannekoek, A., 1923b. Dessins de la Voie Lact\' ee faites \`a Ath\`enes par J.Fr. Julius Schmidt dans les ann\'ees 1864-1876. {\it Annalen van de Sterrewacht te Leiden}, 14, 1-8, IA-IIB.

Pannekoek, A., 1923c. Photographic photometry of the Milky Way and the colour of the Scutum cloud. {\it Bulletin of the Astronomical Institutes of the Netherlands}, 2, 19-24.

Pannekoek, A., 1924. Researches on the structure of the Universe. 1. The local star system deduced from the Durchmusterung Catalogues. {\it Publications of the Astronomical Institute of the University of Amsterdam}, 1, 1-120.

Pannekoek, A., 1925. Some remarks on the relative intensity of the two sides of the Milky Way. {\it Bulletin of the Astronomical Institutes of the Netherlands}, 3, 44-46.

Pannekoek, A. 1927. On the possible existence of large attracting masses in the centre of the Galactic system, {\it Bulletin of the Astronomical Institutes of the Netherlands}, 4, 39-40.

Pannekoek, A., 1928. Die s\"udliche Milchstrasse. {\it Annalen van de Bosscha-Sterrenwacht, Lembang, Java}, 2, part 1, A3-A73. Tafel I-VI.

Pannekoek, A., 1929a. Researches on the structure of the Universe. 2 The space distribution of stars of classes A, K and B, derived from the Draper catalogue. {\it Publications of the Astronomical Institute of the University of Amsterdam}, 2,1-70.

Pannekoek, A., 1929b. Researches on the structure of the Universe. 3 The Cape photographic Durchmusterung. {\it Publications of the Astronomical Institute of the University of Amsterdam}, 2, 71-87.

Pannekoek, A. 1929c. Researches on the distribution of the stars in space.  {\it Proceedings of the section of sciences of the Royal Netherlands Academy of Sciences}, 33, 2-14.

Pannekoek, A.,1933. Photographische photometrie der n\"ordlichen Milchstrasse nach negativen auf der Sternwarte Heidelberg (K\"onigsstuhl) aufgenommen von Max Wolf. {\it Publications of the Astronomical Institute of the University of Amsterdam}, 3, 1-71, Tafel I-III, Karte I-VIII.

Pannekoek, A., 1942. Investigations on dark nebulae. {\it Publications of the Astronomical Institute of the University of Amsterdam}, 7, 1-74.

Pannekoek, A. \&\ Koelbloed, D., 1949. {Photographic photometry of the southern Milky Way after negatives chiefly taken at the Bosscha Observatory at Lembang by Dr. J.G.E.G. Vo\^ute. {\it Publications of the Astronomical Institute of the University of Amsterdam},  9, 1-28, plate I-III, chart I-XV.

Pannekoek, A., 1951. {\it De groei van ons wereldbeeld: Een geschiedenis van de
sterrekunde.} Amsterdam : Wereld-bibliotheek.

Pannekoek, A., 1957. Colour differences in the Milky Way? {\it The Observatory}, 77, 241-242. 

Pannekoek, A., 1961. {\it A history of astronomy}. New York, Interscience Publishers; London: Allen \&\ Unwin. ISBN 087-4-7136-5X, 978-0-8747-13657.

Pannekoek, A., 1982. {\it Herinneringen: Herinneringen uit de arbeidersbeweging, Sterrenkundiger herinneringen}, met bijdragen van B.A. Sijer en E.P.J. van den Heuvel, Eds. B.A. Sijer, J.M. Welcker and J.R. van der Leeuw. Amsterdam, van Gennep, ISBN 90-6012-350-6.

Paul, E.R.,  1993. {\it The Milky Way Galaxy and statistical cosmology, 1890–1924}. Cambridge University Press, ISBN 052-1-3536-37.

Peters, G., 1970. Fluctuations in brightness of the Milky Way and interstellar clouds. {\it Astronomy and Astrophysics},  4, 134-143.

Ram\`{\i}rez, I., Michel, R., Sefako, R., Tucci Maia, M., Schuster, W.J., van Wyk, F., Mel\`endez, J., Casagrande, L., and Castilho, B.V., 2012. The UBV(RI)$_{\rm c}$ colors of the Sun. {\it Astrophysical Journal}, 752, id.5, 13 pp.

Reid, N., 1993. Starcounts as a probe of Galactic structure. In: {\it Galaxy evolution. The Milky Way perspective}, Editors, S.R. Majewski, Astronomical Society of the Pacific Conference Series, 49, 37-51.

Reynolds, J.H., 1913. The light curve of the Andromeda nebula (NGC 224). {\it Monthly Notices of the Royal Astronomical Society}, 74, 132-136.

Russell, H.N., 1920. Some problems in sidereal astronomy. {\it Popular Astronomy}, 28, 212-224 and 264-275.

Pickering,  E.C., 1912. Distribution of stellar spectra. {\it Annals of the Astronomical Observatory of Harvard College}, 56, 1-26.

Schouten, W.J.A., 1918. {\it On the determination of the principal laws of statistical astronomy}. Ph.D. thesis, University of Groningen.
www.astro.rug.nl/JCKapteyn/publications/1918Schouten.pdf.

Schlosser, W. \&\ Schmidt-Kaler, T., 1977. A four-colour photographic atlas of the sky. {\it Vistas in Astronomy}, 21, 447-466.

Schmidt, M., 2000. Kapteyn’s (m — log r) table and cosmology, In: van der Kruit \&\ van Berkel, 2000: {\it The  Legacy  of J.C. Kapteyn}, p.325-336.

Schwarzschild, K., 1900. On the deviations from the law of reciprocity for bromide of silver gelatine. {\it Astrophysical Journal}, 11, 89-91.

Seeley, D. \&\ Berendzen, R., 1972. The development of research in interstellar absorption, c. 1900-1930. {\it Journal for the History of Astronomy} 3, 52-64 \&\ 75-86.

Seidensticker, K.J., Schmidt-Kaler, T. \&\ Schlosser, W., 1982. Photographische Fl\"achenphotometrie des Milchstrasse. II. Flachenphotometrie im Gebiet des Dunkelwolke `Kohlensack’ in U, B, V, R. {\it Astronomy and Astrophysics}, 114, 60-70 (1982).

Shapley, H., 1916. Studies of magnitudes in star clusters. I. On the absorption of light in space., {\it Proceedings of the National Academy of Sciences of the United States of America}, 2, 12–15.

Shapley, H., 1917. Studies of magnitudes in star clusters. V. Further evidence of the absence of scattering of light in Space. {\it Proceedings of the National Academy of Sciences of the United States of America,} 3, 267-270.

Shapley, H., 1918. Studies based on the colors and magnitudes in stellar clusters. VII. The distances, distribution in space, and dimensions of 69 globular clusters. {\it Astrophysical Journal}, 48, 154-181.

Shapley, H. \&\ Shapley,  M.B., 1919. Studies based on the colors and magnitudes in stellar clusters. Twelfth paper: Remarks on the arrangement of the Sidereal Universe. {\it Astrophysical Journal}, 49, 311-336.

Shapley, H. \&\ Curtis, H.D., 1921. The Scale of the Universe. {\it Bulletin of the National Research Council}, 2, 171-217.

Shapley, H., 1923. The Galactic System. {\it Popular Astronomy}, 31, 316-327.

Smith, R.W., 1982. {\it The expanding universe - Astronomy's `Great Debate’ 1900-1931}. Cambridge University Press, ISBN 978-05-2123-212-8 (reprinted 2010).

Smith, R.W., 1985. Studies of the Milky Way 1850-1930 - Some highlights. In: {\it The Milky Way Galaxy}, IAU Symposium 106., H. van Woerden, R.J. Allen, \&\  W.B. Burton (Ed’s.), Dordrecht: D. Reidel, 43-58.

Smith, R.W., 2019. Astronomy in the time of Pannekoek and Pannekoek as an astronomer of his times? In: Tai {\it et al.} (2019), p.109-136.

Steinicke, W., 2016. William Herschel's 'Hole in the Sky' and the discovery of dark nebulae. {\it Journal of Astronomical History and Heritage}, 19, 305-326.

Tai, C., 2019. The Milky Way as optical phenomenon: Perception and photography in the drawings of Anton Pannekoek. In: Tai {\it et al.} (2019), 219-248. 

Tai, C., van der Steen, B, \&\ van Dongen, J. (editors), 2019. {\it  Anton Pannekoek: Ways of viewing science and society}. Amsterdam University Press, ISBN  978-94-6298-434-9, e-ISBN 978-90-4853-500-2 (pdf).

Tai, C., 2021. {\it Anton Pannekoek, Marxist astronomer: Photography, epistemic virtues, and political philosophy in early twentieth-century astronomy.} PhD thesis, University of Amsterdam.

Toller, G.N., 1981. {\it A study of Galactic light, extragalactic light, and Galactic structure using Pioneer 10 observations of background starlight}. PhD Thesis, Stony Brook University, 1981.

Toller, G.N., 1983. The extragalactic background light at 4400 \AA. {\it Astrophysical Journal}, 266, L79-L82 (1983).

Toller, G.N.,  Tanabe, H. \&\ Weinberg, J.L., 1987. Background starlight at the north and south celestial, ecliptic, and galactic poles.  {\it Astronomy and Astrophysics}, 188, 24-34..

Toller, G.N., 1990. Optical observations of Galactic and extragalactic light - Implications for Galactic structure. In: {\it The Galactic and extragalactic background radiation}. IAU Symp. 139, eds. S.  Bowyer \&\ C. Leinert. Dordrecht, Kluwer, 21-34.

Trimble, V., 1995. The 1920 Shapley-Curtis discussion: Background, issues, and aftermath. {\it Publications of the Astronomical Society of the Pacific}, 107, 1133-1144.

van den Heuvel, E.P.J., 2019. Anton Pannekoek’s astronomy in relation to his political activities, and the founding of the Astronomical Institute of the University of Amsterdam. In: Tai {\it et al.} (2019), 25-50.

van der Kruit, P.C. \&\ Searle, L., 1981. Surface photometry of edge-on spiral galaxies. I - A model for the three-dimensional distribution of light in galactic disks. {\it Astronomy and Astrophysics}, 95, 105-115.

van der Kruit, P.C. \&\ Searle, L., 1982. Surface photometry of edge-on spiral galaxies. III - Properties of the three-dimensional distribution of light and mass in disks of spiral galaxies. {\it Astronomy and Astrophysics},110, 61-78.

van der Kruit, P.C. \&\ Freeman, K.C., 1984. The vertical velocity dispersions of the stars in the disks of two spiral galaxies. {\it Astrophysical Journal}, 278, 81-88.

van der Kruit, P.C. \&\ Freeman, K.C., 1986, Stellar kinematics and the stability of disks in spiral galaxies. {\it Astrophysical Journal}, 303, 556-572.

van der Kruit. P.C., 1990. The surface brightness of our Galaxy and other spirals,   In: {\it The Galactic and extragalactic background radiation}. IAU Symp. 139, eds. S.  Bowyer \&\ C. Leinert. Dordrecht, Kluwer, 85-97.

van  der  Kruit,  P.C. \&\  van  Berkel,  K.,  2000.  {\it The  Legacy  of  J.C.  Kapteyn:  Studies  on  Kapteyn  and  the Development of Modern Astronomy}. Dordrecht, Kluwer (ISBN 978-94-010-9864-9).

van der Kruit, P.C., 2015. {\it Jacobus Cornelius Kapteyn: Born investigator of the Heavens}. Springer, ISBN 978-3-319-10875-9, accompanying Webpage www.astro.rug.nl/JCKapteyn.

van der Kruit, P.C., 2019. {\it Jan Hendrik Oort: Master of the Galactic System}. Springer, ISBN 978-3-319-10875-9, accompanying Webpage
www.astro.rug.nl/JHOort.

van der Kruit, P.C., 2021a. {\it Pioneer of Galactic  Astronomy: A biography of Jacobus Cornelius Kapteyn}. Springer Publishers. ISBN 978-3-030-55422-4.

van der Kruit, P.C., 2021b. {\it Master of Galactic Astronomy: A biography of Jan Hendrik Oort}, Springer, ISBN 978-3-030-55547-4.

van der Kruit, P.C., 2021c. Karl Schwarzschild, Annie J. Cannon and Cornelis Easton: PhDs Honoris Causa of Jacobus C. Kapteyn. {\it Journal for Astronomical History and Heritage}, 24, 521–543.

van der Kruit, P.C., 2022. Pieter Johannes van Rhijn, Kapteyn’s Astronomical Laboratory and the Plan of Selected Areas. {\it Journal of Astronomical History and Heritage}, 25, 341-438.

van Rhijn, P.J., 1919. On the brightness of the sky at night and the total amount of starlight. {\it Astrophysical Journal}, 50, 356–375.

van Rhijn, P.J., 1921a. Bemerkung zu Dr. H. Seeligers letzten Untersuchungen \"uber das Sternsystem. {\it Astronomische Nachrichten}, 213, 45-48.

van Rhijn, P.J., 1921b. On the brightness of the sky at night and the total amount of starlight. {\it Publications of the Astronomical Laboratory at Groningen}, 31, 1–83. 

van Rhijn, P.J., 1936. The absorption of light in interstellar Galactic space and the Galactic density distribution. {\it Publications of the Astronomical Laboratory at Groningen}, 47, 1–34.

Weinberg, J.L., 1981. Measuring background starlight. {\it Sky and Telescope}, 61, 114-116.

Wolf, M., 1923. \"Uber den dunklen Nebel NGC 6960, {\it Astronomische Nachrichten}, 219, 109-116.

Yntema, L. 1909. On the brightness of the sky and the total amount of starlight. {\it Publications of the Astronomical Laboratory at Groningen} , 22, 1–55.

}
\newpage 

}

\end{document}